\documentclass[aps,pra,twocolumn,superscriptaddress]{revtex4}
\usepackage{graphicx,subfigure}
\usepackage{amsmath,amssymb,amsthm,physics,color,enumerate,tabularx,makecell,mathrsfs}

\allowdisplaybreaks 

\renewcommand{\arraystretch}{2}

\usepackage{algpseudocode}
\usepackage{algorithm}
\usepackage{setspace}
\newcommand*{\diff}{\mathop{}\!\mathrm{d}}
\usepackage{multirow}

\usepackage{tikz,calc}
\newlength{\mydiam}
\newcommand{\whitecirc}[1][1.]{
\begin{tikzpicture}[scale=1.]
\draw[black,fill=white] (0,0) circle (\mydiam/2);
\end{tikzpicture}
}
\newcommand{\blackcirc}[1][1.]{
\begin{tikzpicture}[scale=1.]
\draw[black,fill=black] (0,0) circle (\mydiam/2);
\end{tikzpicture}
}
\newcommand{\whitecircdot}[1][1.]{
\begin{tikzpicture}[scale=1.]
\draw[black,fill=white] (0,0) circle (\mydiam/2);
\draw[black,fill=black] (0,0) circle (\mydiam/10);
\end{tikzpicture}
}
\newcommand{\blackcircdot}[1][1.]{
\begin{tikzpicture}[scale=1.]
\draw[black,fill=black] (0,0) circle (\mydiam/2);
\draw[white,fill=white] (0,0) circle (\mydiam/10);
\end{tikzpicture}
}

\begin{document}
\title{Continuous-time quantum walks on planar lattices and the role of the magnetic field}
\author{Luca Razzoli}
\email{luca.razzoli@unimore.it}
\affiliation{Dipartimento di Scienze Fisiche, Informatiche e 
Matematiche, Universit\`{a} di Modena e Reggio Emilia, I-41125 Modena, 
Italy}
\author{Matteo G. A. Paris}
\affiliation{Quantum Technology Lab, Dipartimento di 
Fisica {\em Aldo Pontremoli}, Universit\`{a} degli Studi di Milano, I-20133 Milano, Italy}
\author{Paolo Bordone}
\affiliation{Dipartimento di Scienze Fisiche, Informatiche e 
Matematiche, Universit\`{a} di Modena e Reggio Emilia, I-41125 Modena, 
Italy}
\affiliation{Centro S3, CNR-Istituto di Nanoscienze, I-41125 Modena, Italy}
\begin{abstract}
We address the dynamics of continuous-time quantum walk (CTQW) on planar 
2D lattice graphs, i.e. those forming a regular tessellation of the 
Euclidean plane (triangular, square, and honeycomb lattice graphs). We 
first consider the free particle: on square and triangular lattice graphs 
we observe the well-known \textit{ballistic} behavior, whereas on the 
honeycomb lattice graph we obtain a \textit{sub-ballistic} one, although still faster than the classical \textit{diffusive} one. We impute this difference 
to the different amount of coherence generated by the evolution and, in turn, 
to the fact that, in 2D, the square and the triangular lattices are Bravais lattices, whereas the honeycomb one is non-Bravais. From the physical point of view, this means that CTQWs are not universally characterized by the \textit{ballistic} spreading. We then address the dynamics in the presence of a perpendicular uniform magnetic field and study the effects of the field by 
two approaches: (i) introducing the Peierls phase-factors, according to which the tunneling matrix element of the free particle becomes complex, or (ii) spatially discretizing the Hamiltonian of a spinless charged particle in the presence of a magnetic field. Either way, the dynamics of an initially
localized walker is characterized by a lower spread compared to the free 
particle case, the larger is the field the more localized stays the walker.
Remarkably, upon analyzing the dynamics by spatial discretization of the Hamiltonian (vector potential in the symmetric gauge), we obtain that the variance of the space coordinate is characterized by pseudo-oscillations, 
a reminiscence of the harmonic oscillator behind the Hamiltonian in the continuum, whose energy levels are the well-known Landau levels.
\end{abstract}
\date{\today}
\maketitle
\section{Introduction}
Quantum walks (QWs) \cite{kempe2003quantum,venegas2012quantum} are the quantum counterpart of classical random walks, and describe the stochastic propagation of one or more quantum \textit{walkers} on a discrete $n$-dimensional graph. QWs can be either \textit{discrete} (DTQW) \cite{aharonov1993quantum,meyer1996quantum,aharonov2001quantum,ambainis2001one} or \textit{continuous} (CTQW) \cite{farhi1998quantum,childs2002example} in time. In the former case the evolution operator of the system is given by the product of two unitary operators - a ``coin flip'' operator and a conditional shift operator - and it is applied only in discrete time steps, while in the latter case, the evolution operator involves the Hamiltonian of the system, it can be applied at any time and no coin is involved. QWs show a \textit{ballistic} spreading, faster than their classical analogous, characterized by a \textit{diffusive} spreading. This is usually observed on a line, but it has been also proved for DTQWs in a higher number of spatial dimensions (the particle moves by one unit in every dimension), revealing the universal feature of a quadratic gain over the classical random walk \cite{mackay2002quantum}.
\par
The dynamical features of QWs make them promising candidates for implementing fast and efficient quantum algorithms \cite{ambainis2003quantum,venegas2008quantum,portugal2013quantum,wang2013physical}, e.g. search algorithms \cite{shenvi2003quantum,childs2004spatial,chakraborty2016spatial,wong2016laplacian} even on graphene \cite{abal2010spatial,foulger2015quantum} and crystal \cite{childs2014spatial} lattices. A simple QW on a sparse graph has been proven to be universal for quantum computation \cite{childs2009universal} and, recently, it has been shown that quantum logic gates can be implemented by multi-particle CTQWs in 1D \cite{lahini2018quantum}. Moreover, the possibility of using graphene armchair and zigzag nanoribbons to implement quantum gates by means of DTQWs has been investigated in Ref. \cite{karafyllidis2015quantum}. QWs provide an important framework also for modeling phenomena of quantum transport \cite{mulken2011continuous,novo2015systematic}, e.g. in biological system \cite{mohseni2008environment,plenio2008dephasing} and on graphene structures \cite{bougroura2016quantum}, state transfer \cite{christandl2004perfect,kendon2011perfect,alvir2016perfect}, and for characterizing the behavior of many-body systems \cite{lahini2012quantum,ahlbrecht2012molecular,crespi2013anderson}. Hence the interest in considering general graphs \cite{kendon2006quantum} or in increasing the number of spatial dimensions of the lattice. 
Experimentally, 2D DTQWs have been implemented for a neutral atom in an array of optical microtraps or an optical lattice \cite{eckert2005one} and for photons by using an optical fiber network \cite{schreiber20122d,jeong2013experimental}. On the other hand, 2D CTQWs have been implemented by using the external geometry of photonic waveguide arrays, e.g. for a square lattice (showing a ballistic spreading) \cite{tang2018experimental} and for a hexagonal graph mapped into a photonic chip (demonstrating quantum fast hitting) \cite{tang2018hexagonal}. 
\par
In the present work we study CTQWs on planar lattice graphs, i.e. those forming a regular tessellation of the Euclidean plane, and we examine the spreading dynamics of the walker by means of the variance of the space coordinates and the maps of probability distribution. The choice of these geometries allows us to go beyond the CTQW on a line, introducing some degree of arbitrariness while avoiding the complexity of higher dimensional lattices. An analogous problem, the CTQW on root lattice $A_n$ (triangular lattice for $n=2$) and honeycomb one, has been investigated by using the spectral distribution method in Ref. \cite{jafarizadeh2007investigation}, and DTWQs on the honeycomb and triangular lattices have been proved to have, as continuum limit, the Dirac equation \cite{arrighi2018dirac}. The basic CTQW on a graph is defined from the graph Laplacian, which, in turn, is defined from the adjacency matrix, which encodes the connectivity of the graph. In principle, any Hamiltonian (or, generally, any hermitian operator) which respects the topology of the graph defines a CTQW \cite{wong2016laplacian,dalessandro2010connection}. Indeed, the graph Laplacian plays the role of the free particle energy, but, in addition to this kinetic term, the Hamiltonian may also include noise \cite{benedetti2016non,siloi2017noisy}, potentials or interaction terms \cite{wang2014quantum,beggi2018probing}.
\par
Recently, DTQWs on square lattices under artificial magnetic fields have been considered \cite{yalccinkaya2015two}. An artificial or synthetic magnetic field can be simulated as follows \cite{buluta2009quantum}: instead of using charged particles in an actual magnetic field, one typically uses neutral particles upon which the effects of a fictitious magnetic field are imposed, e.g. Raman-laser-induced Berry phases \cite{dalibard2011colloquium}. Another approach to realize DTQWs in synthetic gauge fields is to use integrated photonic circuits \cite{boada2017quantum}. It has been shown recently that 2D DTQWs can simulate the coupling of a Dirac fermion to a constant uniform magnetic field \cite{arnault2016landau}.
\par
To the best of our knowledge, 2D CTQWs in the presence of a magnetic 
field have not been yet investigated. We address the problem 
in two ways: (i) by introducing the Peierls phase-factors, according 
to which the tunneling matrix element of the free particle 
becomes complex \cite{yalccinkaya2015two}, and (ii) by 
spatially discretizing the original Hamiltonian in the continuum 
by means of finite difference formulae, which is the way lattice 
quantum magnetometry has been introduced \cite{razzoli2019lattice}.
Whereas the Peierls model is fundamentally based on the graph 
Laplacian, the spatial discretization of the Hamiltonian requires 
also the discrete analog of the first-order differential operator, 
since the linear momentum is now present even at the first order, 
due to the cross-terms with the vector potential. In turn, the spatial 
discretization of differential operators for non-square lattices 
is non-trivial.
\par
The paper is organized as follows.
In Sec. \ref{sec:CTQW_on_plg} we introduce the CTQW of a free particle on a graph, the definition of planar lattice graph, and, after defining the CTQW Hamiltonian on each planar lattice graph, we show the results of the numerical simulations. 
In Sec. \ref{sec:charge_ptc_B} we introduce the Hamiltonian of a spinless charged particle in the presence of a magnetic field. Then, we address the definition of the corresponding CTQW according to two approaches:
in Sec. \ref{sec:peierls_model} introducing the Peierls model, and in Sec. \ref{sec:spatial_disc_H_fd} spatially discretizing the original Hamiltonian in the continuum. Results of the numerical simulations are respectively shown in each section.
In Sec. \ref{sec:conclusions}, which closes the body of the paper, we summarize our results. This is followed by some appendices. In Appendix \ref{app:math_discretization} we deepen the issue of the discretization of the space, how differential operators act on such a space, and we provide further details about the derivation of the CTQW Hamiltonian. In Appendix \ref{app:comp_detail} we introduce the computational details about mapping the probability distribution, indexing of vertices on the planar lattice graphs, how to restore the corresponding $(x,y)$ coordinates, and some remarks about the boundary conditions. In Appendix \ref{app:units} we show how the 
system of units is redefined after setting some characteristic 
parameter of our system to 1 (adimensional).
\section{CTQW on planar lattice graphs}
\label{sec:CTQW_on_plg}
\subsection{CTQW on a graph}
\label{subsec:CTQW_graph}
The CTQW on a graph is defined in direct analogy to a continuous-time classical random walk \cite{farhi1998quantum} and it defines a process on continuous time and discrete space. Given an undirected graph \footnote{A \textit{directed graph} or \textit{digraph} $G$ is a triple consisting of a vertex set $V(G)$, an edge set $E(G)$, and a function assigning each edge an ordered pair of vertices. The first vertex of the ordered pair is the \textit{tail} of the edge, and the second is the \textit{head}; together, they are the endpoints. We say that an edge is an edge from its tail to its head \cite{west2001introduction}.} $G$ with $N$ vertices and no self-loops, we define the \textit{adjacency matrix}
\begin{equation}
A_{jk}=
\begin{cases}
1 & \text{if } (j,k)\in G\,,\\
0 & \text{otherwise}\,,
\end{cases}
\label{eq:adj_matr}
\end{equation}
which describes the connectivity of $G$: the matrix element is non-zero iff vertices $j$ and $k$ ($j,k=1,\ldots,N$) are connected by an edge. In terms of this matrix, we can also define the \textit{graph} (or \textit{discrete}) \textit{Laplacian}
\begin{equation}
L=A-D\,,
\label{eq:graphL}
\end{equation}
where $D$ is the \textit{diagonal degree matrix} with
\begin{equation}
D_{jj}=\operatorname{deg}(j)
\label{eq:deg_matr}
\end{equation}
the degree of vertex $j$, i.e. the number of incident edges \cite{west2001introduction}. The continuous-time random walk on $G$ is a Markov process with a fixed probability per unit time $\gamma$ of jumping to an adjacent vertex. This process can be described by the first-order, linear differential equation
\begin{equation}
\dv{p_j(t)}{t}=\gamma\sum_k L_{jk}p_k(t),
\label{eq:Markov_diff_eq}
\end{equation}
where $p_j(t)$ is the probability of being at vertex $j$ at time $t$. The probability is conserved since the columns of $L$ sum to zero. Indeed, to be a valid probability-conserving classical Markov process, Eq. \eqref{eq:Markov_diff_eq} requires $\sum_j L_{jk}=0$.

The CTQW on a graph takes place in a $N$-dimensional Hilbert space spanned by states $\vert j \rangle$, where $j$ is a vertex in $G$. Due to this choice of basis, we can write a general state $\vert \psi(t) \rangle$ in terms of the $N$ complex amplitudes $q_j(t)=\langle j \vert \psi (t)\rangle$. If the Hamiltonian is $\mathcal{H}$, then the dynamics of the system is determined by the Schr\"{o}dinger equation
\begin{equation}
i\dv{q_j(t)}{t} = \sum_k \mathcal{H}_{jk} q_k(t),
\label{eq:schro_diff_eq}
\end{equation}
in the units in where $\hbar=1$. In the light of the similarities between Eq. \eqref{eq:Markov_diff_eq} and Eq. \eqref{eq:schro_diff_eq}, the CTQW is defined by letting $\mathcal{H}=-\gamma L$ \footnote{Here the sign is chosen so that the Hamiltonian is positive semidefinite. We have defined $L=A-D$ so that for a lattice, $L$ is a discrete approximation to the continuum operator $\nabla^2$. A free particle in the continuum has the positive semidefinite Hamiltonian $\mathcal{H}=
-\nabla^2$ (in appropriate units).}. As an aside, not only the graph 
Laplacian, but any Hermitian operator $\mathcal{H}$ that respects the locality of the graph defines a CTQW. Indeed, being the time-evolution 
operator $\exp{-i\mathcal{H}t}$, Eq. \eqref{eq:schro_diff_eq} 
requires $\mathcal{H}=\mathcal{H}^\dagger$ to be a valid unitary 
quantum process \cite{childs2004spatial}. On the contrary, 
the dynamics of a classical walker is that of an open system, and in turn 
is inherently diffusive.

The graph Laplacian has its roots in the discretization of the space. The Hamiltonian characterizes the total energy of the system, and, for a particle of mass $m$, it includes a kinetic energy term
\begin{equation}
T=-\frac{1}{2m}\nabla^2,
\label{eq:kin_term_free}
\end{equation}
where $\nabla^2=\partial_x^2+\partial_y^2+\partial_z^2$ is Laplace's operator (in 3D Euclidean space). If the particle is confined to discrete spatial locations, then $\nabla^2$ is replaced by the graph Laplacian of Eq. \eqref{eq:graphL}. For example, for a 1D grid with lattice spacing $a$, note the similarities between the continuous-space Laplacian
\begin{equation}
\nabla^2 \psi = \partial_x^2 \psi = \lim_{a \rightarrow 0} \frac{\psi(x+a)+\psi(x-a)-2\psi(x)}{a^2}
\label{eq:Lapl_central_1D}
\end{equation}
and the discrete-space analogue
\begin{equation}
L\psi = (A-D)\psi = \psi_{x+1}+\psi_{x-1}-2\psi_{x}.
\end{equation} 
Now letting $\gamma=1/2ma^2$, the kinetic energy operator becomes
\begin{equation}
T=-\gamma L.
\end{equation}
This defines a CTQW, i.e. the propagation 
of a quantum particle with kinetic energy when confined to a lattice. 
Unlike the case of the Markov process, now the parameter 
$\gamma\in\mathbb{R}^+$ corresponds to the amplitude rate 
of the walk. A higher rate corresponds to a 
particle with smaller mass, since a less massive particle scatters
 more readily \cite{wong2016laplacian}.
\subsection{Planar lattice graph}
\label{subsec:plg}
\begin{figure}[!h]
	\centering
	\includegraphics[width=0.45\textwidth]{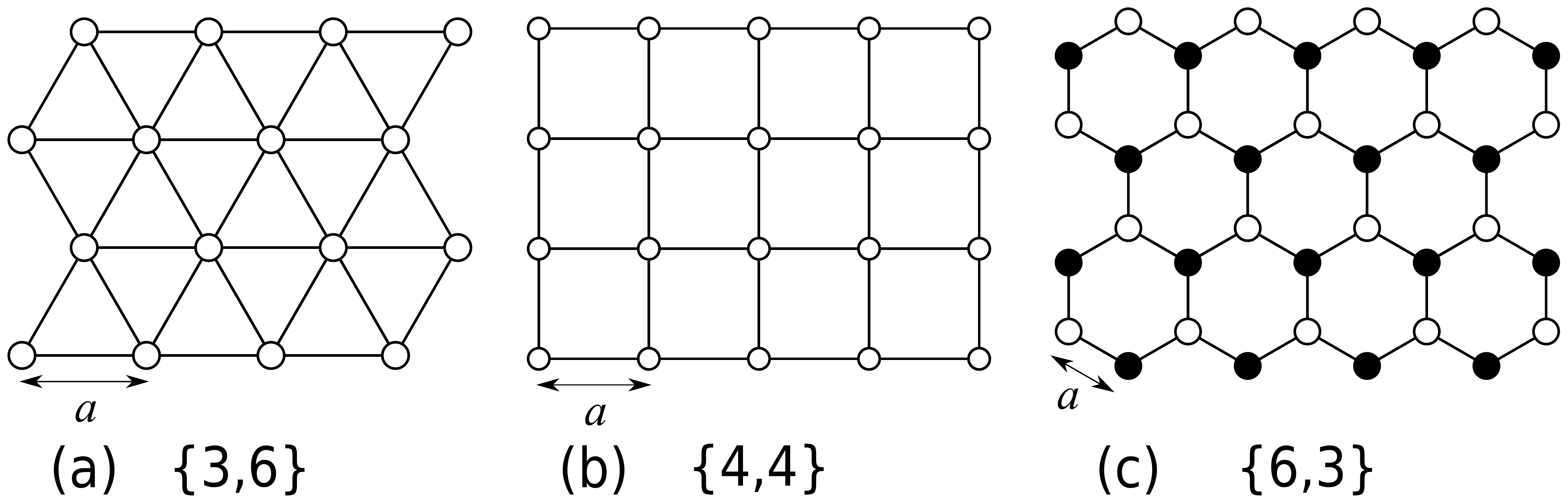}
	\caption{The three regular tessellations of the Euclidean plane: (a) equilateral triangles, (b) squares, and (c) regular hexagons. Below each tessellation the corresponding Schl\"{a}fli symbol is reported. These tessellations lead, respectively, to triangular, square, and honeycomb lattice graph. Equivalent vertices are represented with same circles and $a$ denotes the lattice parameter.}
\label{fig:tessel_plg}
\end{figure}

In graph theory, a graph $G$ is said to be \textit{planar} if it can 
be drawn in the plane in such a way that pairs of edges intersect 
only at vertices, if at all. Such a drawing is a \textit{planar 
embedding} of $G$ \cite{harris2008combinatorics,west2001introduction}. 
A \textit{lattice graph} is a graph possessing a drawing whose embedding 
in a Euclidean space $\mathbb{R}^n$ forms a \textit{regular tiling} \cite{WolframLatticeGraph,harary1967graph,dale2017applying}. It is a 
simple graph with a distance measurement (called metric) of a geometric 
object, it is a regular graph and each edge has the same weight or 
represents the same distance in Euclidean space as in other spaces \cite{chen2014digital}. A \textit{tiling} of regular polygons (in 
two dimensions), polyhedra (in three dimensions), or polytopes (in $n$ 
dimensions) is called a \textit{tessellation}. In other words, we 
may say that a tessellation is \textit{regular} if it has regular 
faces and a regular vertex figure at each vertex. There are exactly 
three regular tessellations composed of regular polygons symmetrically 
tiling the plane: equilateral triangles, squares and regular hexagons 
(Fig. \ref{fig:tessel_plg}) \cite{coxeter1969introduction,WolframTessellation,wells1991penguin}. Tessellations can be specified using a \textit{Schl\"{a}fli symbol}, 
which is a symbol of the form $\left\lbrace p,q,r,\dots\right\rbrace$ 
used to describe regular polygons, polyhedra, and their 
higher-dimensional counterparts. The symbol $\left\lbrace p,q 
\right\rbrace$ denotes a tessellation of regular $p$-gons, $q$ 
surrounding each vertex \cite{coxeter1969introduction,WolframSchlafi}.
In view of these preliminary definitions, we call \textit{planar 
lattice graph} (PLG) a graph possessing a drawing whose embedding 
in a Euclidean plane forms a regular tiling, i.e. a regular 
tessellation. This leads only to triangular, square, and honeycomb 
lattice graphs.

In a Bravais lattice both the arrangement and orientation of the array of vertices must appear the same from every vertex in the lattice. Unlike in the triangular and square lattice graph, which are clearly Bravais lattices and all their vertices are equivalent, in the honeycomb one vertices are not all equivalent. Structural relations are identical, but not orientational relations, so the vertices of a honeycomb do not form a Bravais lattice \cite{ashcroft1976solid}.

We introduce here below the notation adopted in the following. When considering a lattice, a generic vertex (site) $V$ is identified by a couple of discrete indices $(j_V,k_V)\in \mathbb{Z}^2$. We denote the lattice parameter by $a$, the coordinates of the vertex $V$ by $(x_V,y_V)$, and a generic scalar function of the position by $f(x_V,y_V)$. In the following, since the explicit use of discrete indices or coordinates might be misleading and confusing (see Appendix \ref{subapp:index_coords} for details), we will refer to a generic vertex $V$ and its nearest-neighbors (NNs) as shown in Table \ref{tab:plg_nn_coords}, and we simplify the notation according to $f_V :=f(x_V,y_V)$. Moreover, the honeycomb lattice graph is characterized by two classes of non-equivalent vertices, $\{\circ,\bullet\}$ (see Fig. \ref{fig:tessel_plg}(c)). Thus, for this PLG, we define the variable $\odot \in \{\circ,\bullet\}$, we denote by $\bar{\odot}$ its complement in the same set, i.e. $\bar{\circ}=\bullet$ and $\bar{\bullet}=\circ$, and we define
\begin{equation}
\operatorname{sgn}(\odot)=
\begin{cases}
+1 & \text{if }\odot=\circ\,,\\
-1 & \text{if }\odot=\bullet\,.
\end{cases}
\end{equation}

\begin{table}[!h]
\renewcommand\arraystretch{1.75}
\center
\begin{ruledtabular}
\begin{tabular}{cl}
\textbf{PLG} &\textbf{NNs of $V=(x_V,y_V)$}\\\hline
\textit{Square} & {\small $\operatorname{deg}(V)=4$}\\
\multirow{3}{0.13\textwidth}{\parbox[t]{1em}{\includegraphics[width=0.13\textwidth]{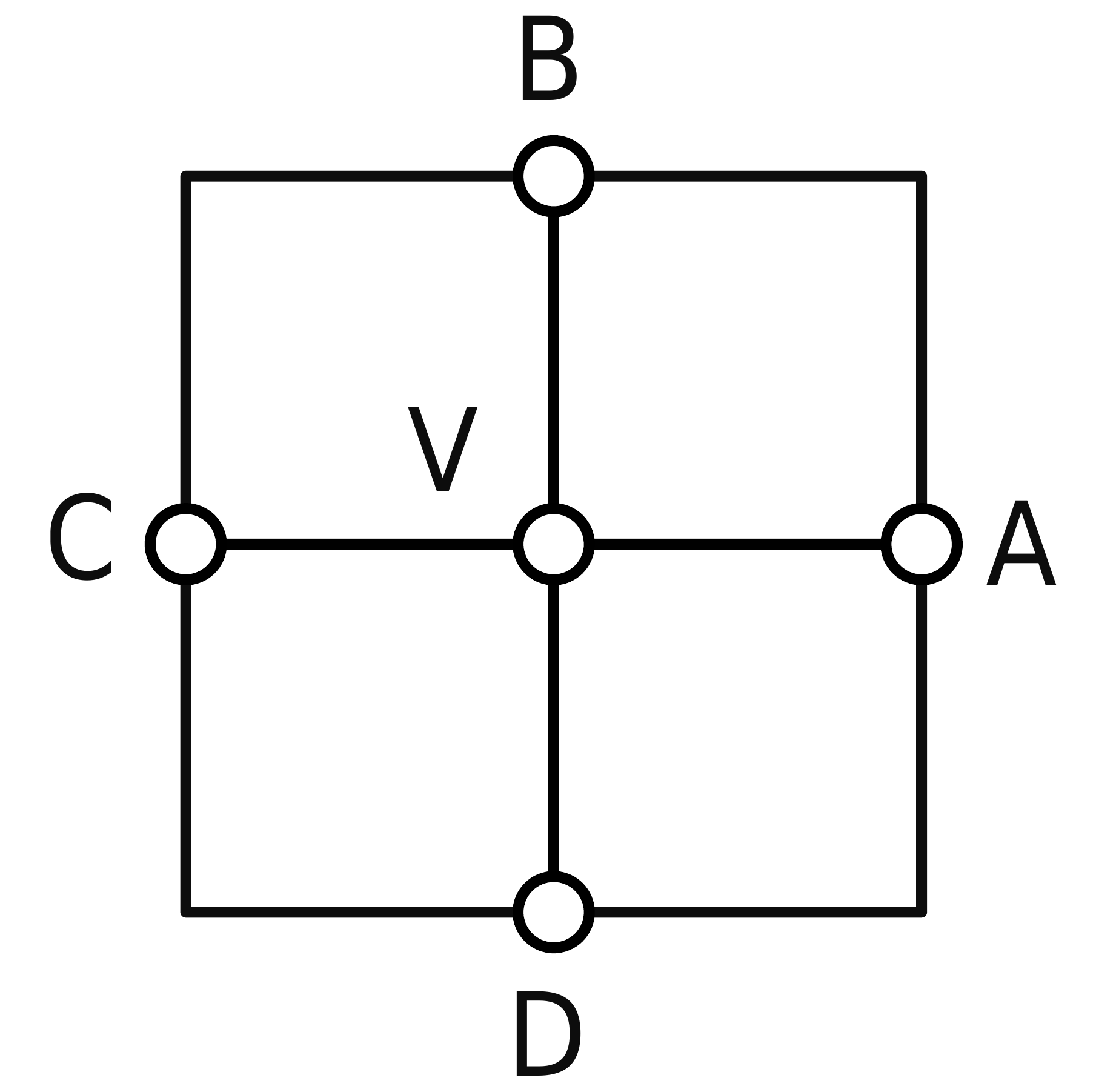}}}
&{\small $A=(x_V+a,y_V)$}\\
&{\small $B=(x_V,y_V + a)$}\\
&{\small $C=(x_V-a,y_V)$}\\
&{\small $D=(x_V,y_V-a)$}\\
\textit{Triangular} & {\small $\operatorname{deg}(V)=6$}\\
\multirow{3}{0.13\textwidth}{\parbox[t]{1em}{\includegraphics[width=0.13\textwidth]{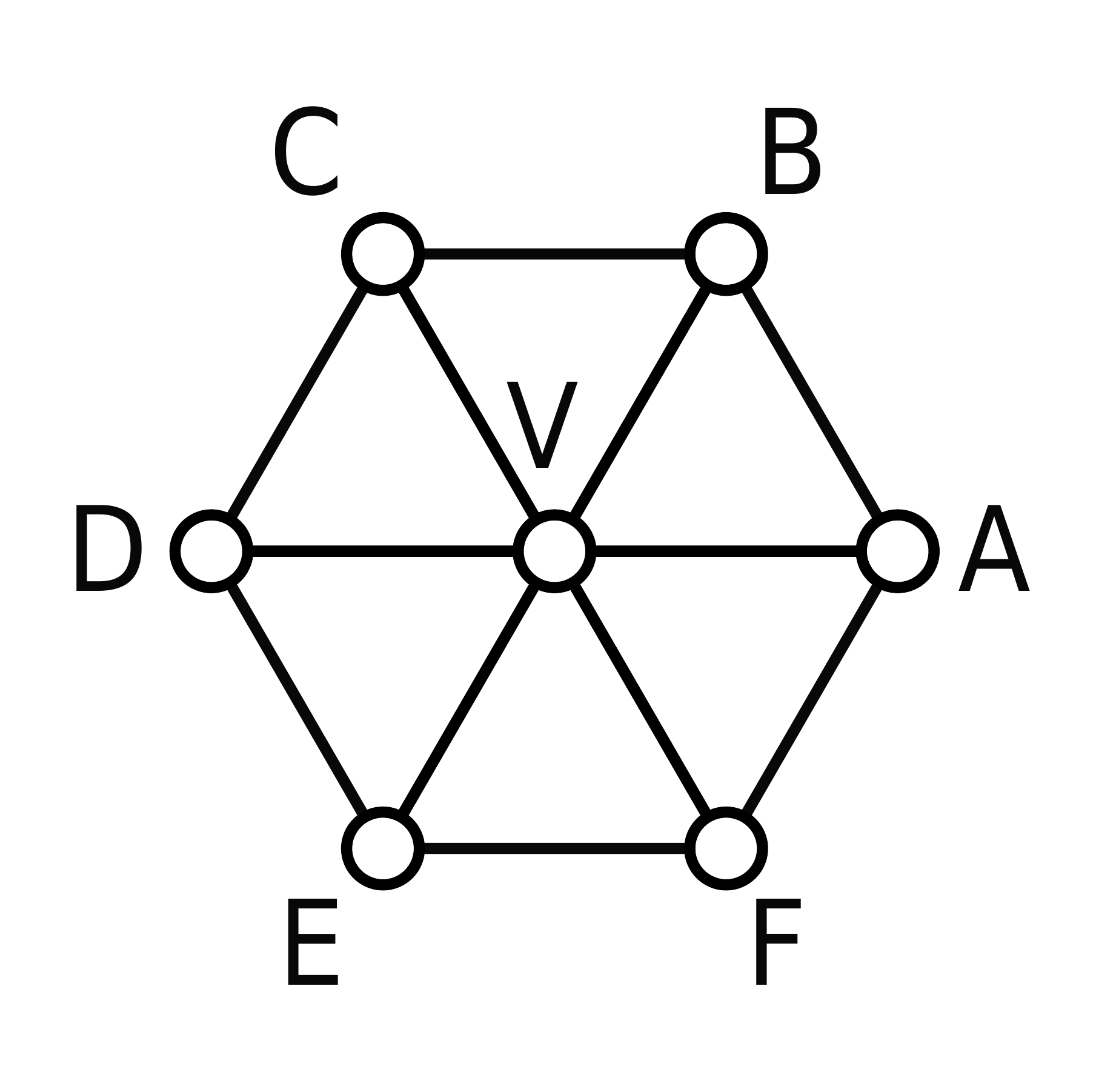}}}
&{\small $A=(x_V+a,y_V)$}\\
&{\small $B=(x_V+a/2,y_V + \sqrt{3}a/2)$}\\
&{\small $C=(x_V-a/2,y_V + \sqrt{3}a/2)$}\\
&{\small $D=(x_V-a,y_V)$}\\
&{\small $E=(x_V-a/2,y_V - \sqrt{3}a/2)$}\\
&{\small $F=(x_V+a/2,y_V - \sqrt{3}a/2)$} \\
\textit{Honeycomb, $(V,\circ)$} & {\small $\operatorname{deg}(V,\circ)=3$}\\
\multirow{3}{0.13\textwidth}{\parbox[t]{1em}{\includegraphics[width=0.13\textwidth]{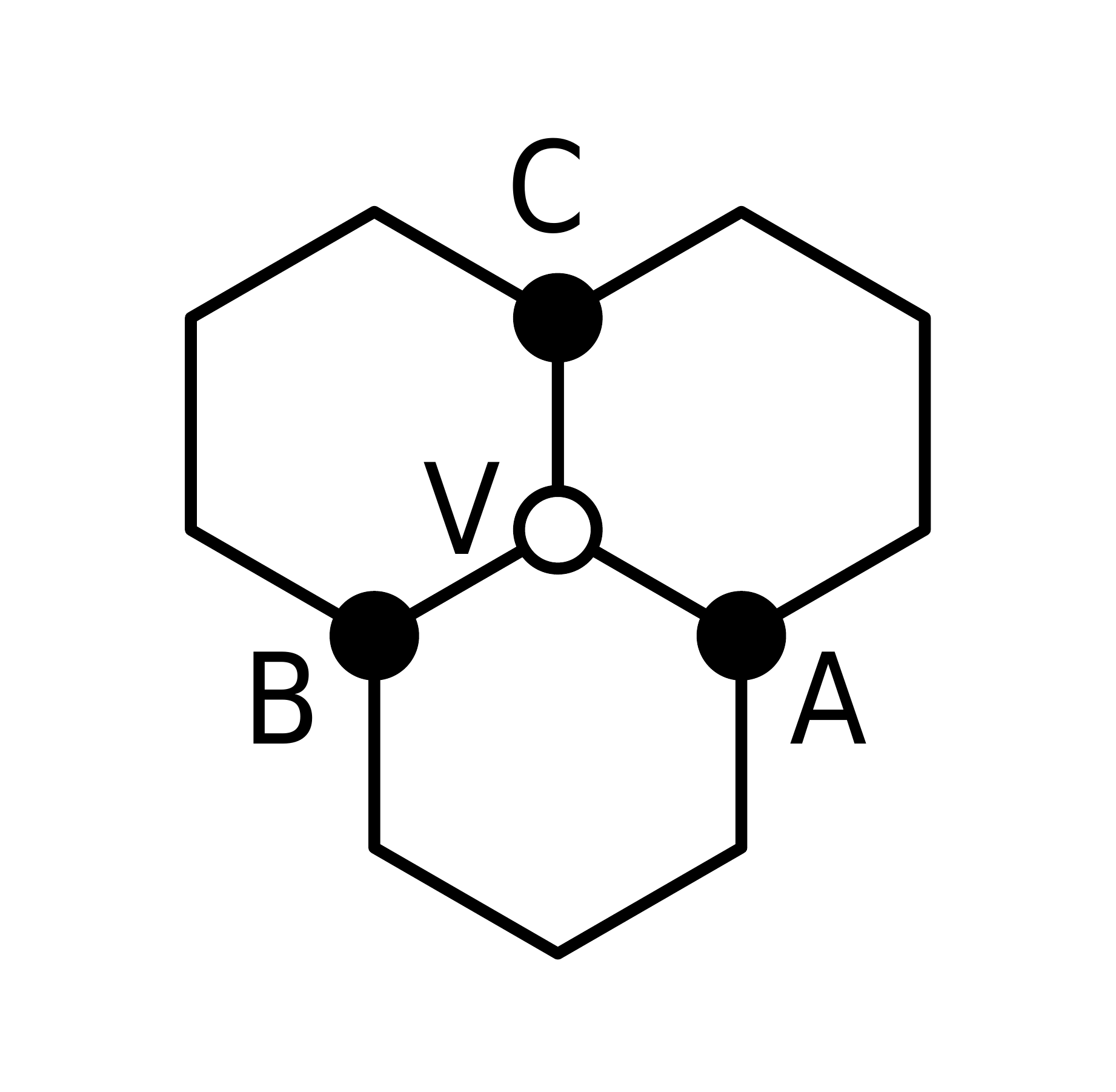}}}
&{\small $(A,\bullet)=(x_V+\sqrt{3}a/2,y_V-a/2)$}\\
&{\small $(B,\bullet)=(x_V-\sqrt{3}a/2,y_V-a/2)$}\\
&{\small $(C,\bullet)=(x_V,y_V + a)$}\\
\textit{Honeycomb, $(V,\bullet)$} & {\small$\operatorname{deg}(V,\bullet)=3$}\\
\multirow{3}{0.13\textwidth}{\parbox[t]{1em}{\includegraphics[width=0.13\textwidth]{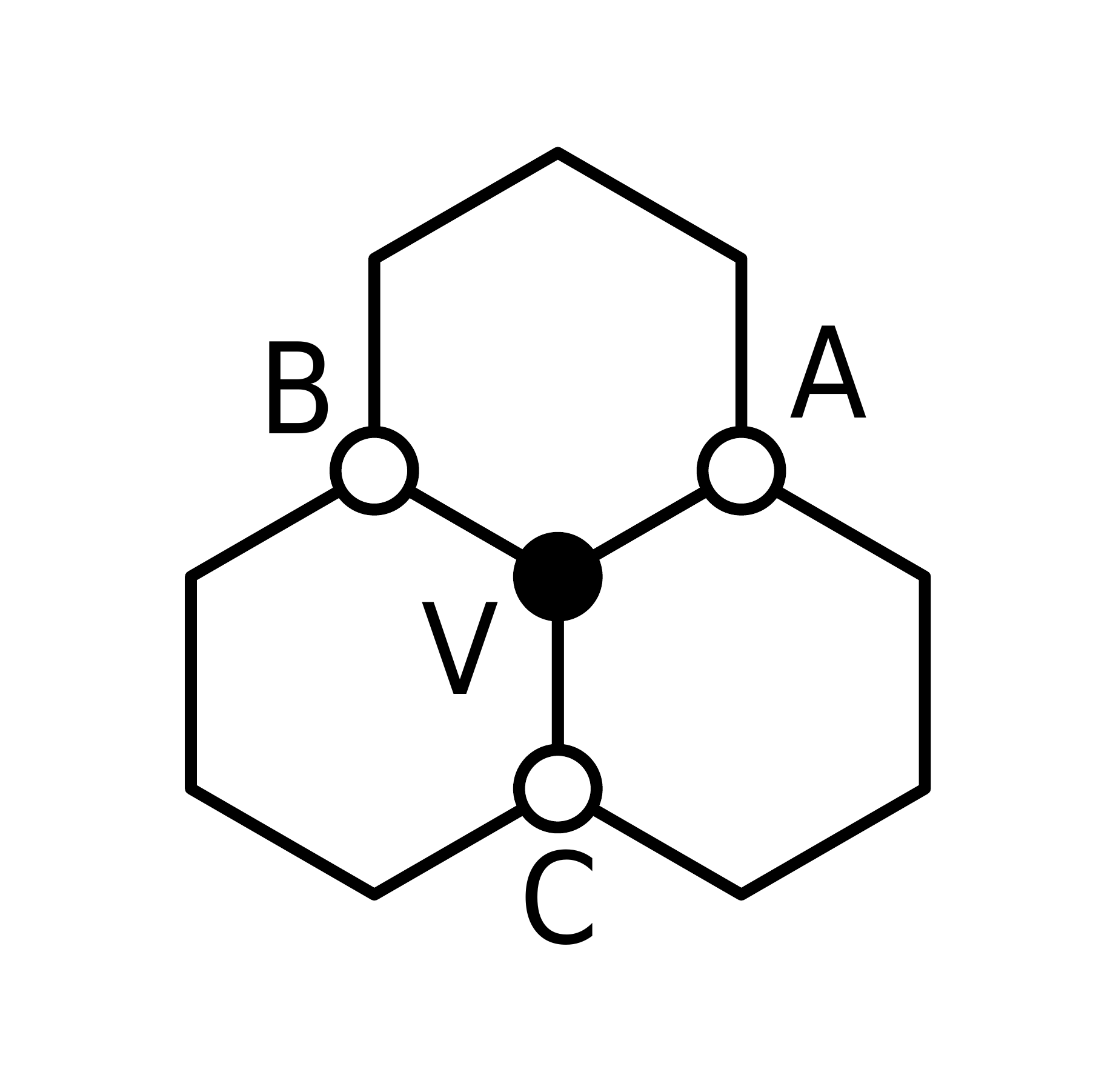}}}
&{\small $(A,\circ)=(x_V+\sqrt{3}a/2,y_V+a/2)$}\\
&{\small $(B,\circ)=(x_V-\sqrt{3}a/2,y_V+a/2)$}\\
&{\small $(C,\circ)=(x_V,y_V - a)$}
\end{tabular}
\end{ruledtabular}
\caption{NNs of a vertex $V$ in a square, triangular, and honeycomb lattice graph. The two classes of non-equivalent vertices in a honeycomb lattice graph are denoted as $\{\circ,\bullet\}$. The number of NNs is given by the degree of the vertex.}
\label{tab:plg_nn_coords}
\end{table}

\subsection{The CTQW Hamiltonian on PLGs}
\label{subsec:CTQW_H_free}
We first consider the CTQW of the free particle, whose Hamiltonian merely consists of the kinetic term. We have therefore to spatially discretize Eq. \eqref{eq:kin_term_free} according to the different PLGs. In doing so, we follow the same idea underlying the heuristic proof of the origin of the graph Laplacian in Sec. \ref{subsec:CTQW_graph}, i.e. Taylor expanding a scalar function $f$ evaluated in the NNs about the given vertex and combining the resulting expansions in such a way that the discrete version of the Laplacian $\nabla^2f=(\partial_x^2+\partial_y^2)f$ is found in terms of finite differences. For further details, please refer to Appendix \ref{app:math_discretization}.
 
We anticipate that in each PLG the discrete Laplacian turns 
out to be of the form 
\begin{equation}
\nabla^2f_V \sim \sum_{W\in N\!N(V)}f_W-\deg(V)f_V\,,
\end{equation}
with $N\!N(V)$ the set of NNs of $V$, consistently with Eq. \eqref{eq:graphL}. The reason why we compute the Laplacian by means of Taylor expansion, even though it is analogous to the graph Laplacian, whose definition is much more manageable, is that this approach allows us to actually take into account the underlying geometry of the PLG. Indeed the graph Laplacian is a `ready-made' operator and there is no computation telling us how the hopping amplitude of the resulting CTQW Hamiltonian changes in the different PLGs. Instead, using Taylor expansion is a `constructive' way to determine the Laplacian and the resulting Hamiltonian has a different hopping amplitude depending on the graph. This is a valuable feature, because by changing the degree of a vertex we expect the hopping amplitude to change accordingly.

\subsubsection{Square lattice graph}
\label{subsubsec:fd_taylor_squ}
A vertex $V$ has four NNs, namely $A$, $B$, $C$ and $D$ (see Table \ref{tab:plg_nn_coords}). We evaluate the following Taylor expansions about $V$ up to the second order:
\begin{align}
f_A &\approx f_V +a\partial_x f_V+\frac{a^2}{2}\partial_x^2 f_V\,,\\
f_B &\approx f_V +a\partial_y f_V+\frac{a^2}{2}\partial_y^2 f_V\,,\\
f_C &\approx f_V-a\partial_x f_V+\frac{a^2}{2}\partial_x^2 f_V\,,\\
f_D &\approx f_V -a\partial_y f_V+\frac{a^2}{2}\partial_y^2 f_V\,.
\end{align}
Now we consider a linear combination of the expressions above, understood as equalities:
\begin{align}
\alpha f_A &+\beta f_B +\gamma f_C +\delta f_D\nonumber\\
&=(\alpha + \beta +\gamma +\delta)f_V\nonumber\\
&\quad+a(\alpha - \gamma)\partial_x f_V + a(\beta - \delta)\partial_y f_V \nonumber\\
&\quad+ \frac{a^2}{2}(\alpha + \gamma)\partial_x^2 f_V + \frac{a^2}{2}(\beta + \delta)\partial_y^2 f_V\,.
\label{eq:lc_taylor_squ}
\end{align}
If we set $\alpha=\beta=\gamma=\delta=1$, we get
\begin{equation}
\nabla^2 f_V=\frac{1}{a^2}\left( f_A +f_B +f_C +f_D-4f_V\right)\,.
\label{eq:squ_lap_fd}
\end{equation}
The resulting finite-difference formula is the same used in numerical analysis \cite{abramowitz1970handbook}. According to this graph Laplacian, the Hamiltonian reads then as follows:
\begin{align}
\mathcal{\hat{H}}=&-J_S\sum_V \left(\dyad{A}{V} +\dyad{B}{V}+\dyad{C}{V}\right.\nonumber\\
&\left.+\dyad{D}{V}-4\dyad{V}{V}\right)\,,
\label{eq:H_SQU_plg_free}
\end{align}
where the hopping amplitude is
\begin{equation}
J_S:=\frac{\hbar^2}{2ma^2}\,.
\label{eq:J_SQU_plg}
\end{equation}
\subsubsection{Triangular lattice graph}
\label{subsubsec:fd_taylor_tri}
\begin{figure}[!h]
	\centering
	\includegraphics[width=0.45\textwidth]{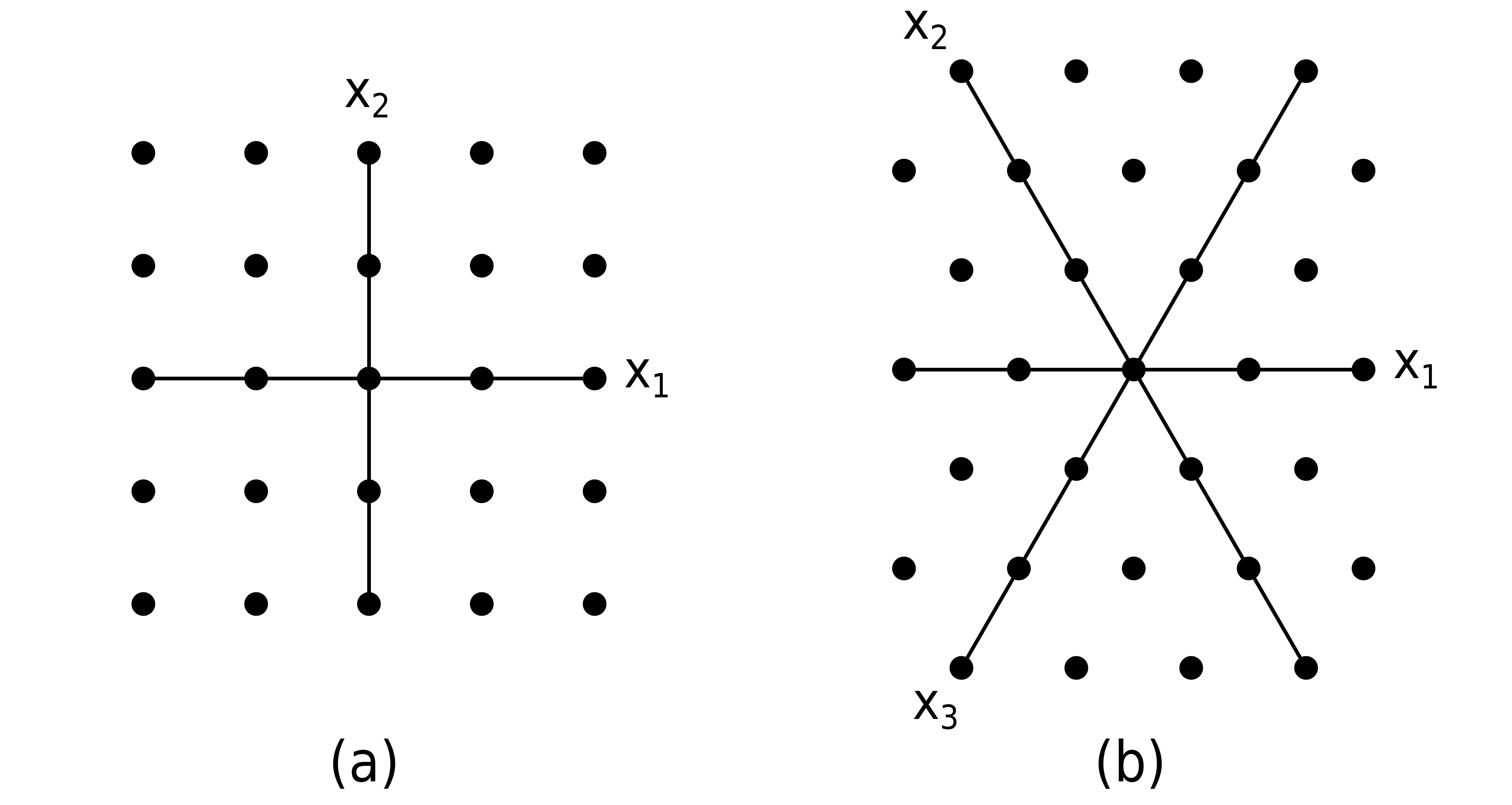}
	\caption{$n$ symmetrically distributed directions in a (a) square ($n=2$) and (b) triangular ($n=3$) lattice graph.}
	\label{fig:tri_squ_symmetric_dir}
\end{figure}

A vertex $V$ has six NNs, namely $A$, $B$, $C$, $D$, $E$, and $F$ (see Table \ref{tab:plg_nn_coords}). We evaluate the following Taylor expansions 
about $V$ up to the second order:
\begin{align}
f_A &\approx f_V +a\partial_x f_V+\frac{a^2}{2}\partial_x^2 f_V\,,\\
f_B &\approx f_V+\frac{a}{2}\partial_x f_V +\frac{\sqrt{3}a}{2}\partial_y f_V+\frac{a^2}{8}\partial_x^2 f_V+\frac{3a^2}{8}\partial_y^2 f_V\,,\\
f_C &\approx f_V-\frac{a}{2}\partial_x f_V +\frac{\sqrt{3}a}{2}\partial_y f_V+\frac{a^2}{8}\partial_x^2 f_V+\frac{3a^2}{8}\partial_y^2 f_V\,,\\
f_D &\approx f_V -a\partial_x f_V+\frac{a^2}{2}\partial_x^2 f_V\,,\\
f_E &\approx f_V-\frac{a}{2}\partial_x f_V -\frac{\sqrt{3}a}{2}\partial_y f_V+\frac{a^2}{8}\partial_x^2 f_V+\frac{3a^2}{8}\partial_y^2 f_V\,,\\
f_F &\approx f_V+\frac{a}{2}\partial_x f_V -\frac{\sqrt{3}a}{2}\partial_y f_V+\frac{a^2}{8}\partial_x^2 f_V+\frac{3a^2}{8}\partial_y^2 f_V\,.
\end{align}
Now we consider a linear combination of the expressions above, understood as equalities:
\begin{align}
\alpha f_A &+\beta f_B +\gamma f_C +\delta f_D +\varepsilon f_E +\phi f_F\nonumber\\
&=(\alpha + \beta +\gamma +\delta +\varepsilon +\phi)f_V \nonumber\\
&\quad+\frac{a}{2}(2\alpha + \beta - \gamma -2\delta - \varepsilon + \phi)\partial_x f_V\nonumber\\
&\quad+ \frac{\sqrt{3}a}{2}(\beta + \gamma-\varepsilon - \phi)\partial_y f_V \nonumber\\
&\quad+ \frac{a^2}{8}(4\alpha + \beta +\gamma +4\delta +\varepsilon +\phi)\partial_x^2 f_V\nonumber\\
&\quad+ \frac{3a^2}{8}(\beta +\gamma +\varepsilon +\phi)\partial_y^2 f_V\,.
\label{eq:lc_taylor_tri}
\end{align}
If we set $\alpha=\beta=\gamma=\delta=\varepsilon=\phi=1$, we get
\begin{equation}
\nabla^2 f_V=\frac{2}{3a^2}\left( f_A +f_B +f_C +f_D +f_E+ f_F-6f_V\right),
\label{eq:tri_lap_fd}
\end{equation}
which has the same structure of the Laplacian of Eq. \eqref{eq:squ_lap_fd} and is consistent with those
reported in Refs. \cite{woodward1984hexagonal,hamilton2013hexagonal}. In particular, in Ref. \cite{woodward1984hexagonal}, it is also shown that, while the 2D Laplacian is usually represented as a sum of 1D second derivatives in two orthogonal directions $\nabla^2=\partial_{x}^2+\partial_{y}^2$, it may more generally be represented as a summation of 1D second derivatives in any $n\geq 2$ symmetrically distributed directions (Fig. \ref{fig:tri_squ_symmetric_dir})
\begin{equation}
\nabla^2=\frac{2}{n}\sum_{i=1}^n \partial_{x_i}^2\,,
\label{eq:nabla_symm_dir}
\end{equation}
which, for $n=3$ and replacing each $\partial_{x_i}^2$ with its discrete form (see Appendix \ref{subapp:disc_diff_op}), is consistent with Eq. \eqref{eq:tri_lap_fd}. In this case the axes $x_1$, $x_2$, $x_3$ are represented by the unit vectors in $\mathbb{R}^2$:
\begin{equation}
\renewcommand\arraystretch{1.5}
\begin{array}{ccc}
\mathbf{x}_1 =
\left(
\begin{array}{c}
1\\
0
\end{array}
\right),
&
\mathbf{x}_2 =
\left(
\begin{array}{c}
-\frac{1}{2}\\
\frac{\sqrt{3}}{2}
\end{array}
\right),
& 
\mathbf{x}_3 =
\left(
\begin{array}{c}
-\frac{1}{2}\\
-\frac{\sqrt{3}}{2}
\end{array}
\right).
\end{array}
\end{equation}
According to this graph Laplacian, the Hamiltonian reads then as follows:
\begin{align}
\mathcal{\hat{H}}=&-J_T\sum_V\left( \dyad{A}{V} +\dyad{B}{V}+\dyad{C}{V}+\dyad{D}{V}\right.\nonumber\\
&\left. +\dyad{E}{V} +\dyad{F}{V} -6\dyad{V}{V}\right)\,,
\label{eq:H_TRI_plg_free}
\end{align}
where the hopping amplitude is
\begin{equation}
J_T:=\frac{\hbar^2}{3ma^2}=\frac{2}{3}J_S\,.
\label{eq:J_TRI_plg}
\end{equation}

\subsubsection{Honeycomb lattice graph}
\label{subsubsec:fd_taylor_hon}
A vertex $(V,\odot)$, with $\odot \in \{\circ,\bullet\}$, has three NNs, namely $(A,\bar{\odot})$, $(B,\bar{\odot})$, and $(C,\bar{\odot})$ (see Table \ref{tab:plg_nn_coords}). We evaluate the following Taylor expansions about $(V,\odot)$ up to the second order:
\begin{align}
f_{(A,\bar{\odot})} \approx  & f_{(V,\odot)} +\frac{\sqrt{3}a}{2}\partial_x f_{(V,\odot)}-\operatorname{sgn}(\odot)\frac{a}{2}\partial_y f_{(V,\odot)}\nonumber\\
&+\frac{3a^2}{8}\partial_x^2 f_{(V,\odot)}+\frac{a^2}{8}\partial_y^2 f_{(V,\odot)}\,,\\
f_{(B,\bar{\odot})} \approx & f_{(V,\odot)} -\frac{\sqrt{3}a}{2}\partial_x f_{(V,\odot)}-\operatorname{sgn}(\odot)\frac{a}{2}\partial_y f_{(V,\odot)}\nonumber\\
&+\frac{3a^2}{8}\partial_x^2 f_{(V,\odot)}+\frac{a^2}{8}\partial_y^2 f_{(V,\odot)}\,,\\
f_{(C,\bar{\odot})} \approx & f_{(V,\odot)}+\operatorname{sgn}(\odot)a\partial_y f_{(V,\odot)} +\frac{a^2}{2}\partial_y^2 f_{(V,\odot)}\,.
\end{align}
Now we consider a linear combination of the expressions above, understood as equalities:
\begin{align}
\alpha &f_{(A,\bar{\odot})} +\beta f_{(B,\bar{\odot})} +\gamma f_{(C,\bar{\odot})}\nonumber\\
&=(\alpha + \beta +\gamma )f_{(V,\odot)}+\frac{\sqrt{3}a}{2}(\alpha - \beta)\partial_x f_{(V,\odot)}\nonumber\\
&\quad -\operatorname{sgn}(\odot) \frac{a}{2}(\alpha+\beta -2 \gamma)\partial_y f_{(V,\odot)} \nonumber\\
&\quad+ \frac{3a^2}{8}(\alpha + \beta)\partial_x^2 f_{(V,\odot)} + \frac{a^2}{8}(\alpha+\beta + 4\gamma)\partial_y^2 f_{(V,\odot)}\,.
\label{eq:lc_taylor_hon}
\end{align}
If we set $\alpha=\beta=\gamma=1$, we get
\begin{equation}
\nabla^2 f_{(V,\odot)}=\frac{4}{3a^2}\left( f_{(A,\bar{\odot})} + f_{(B,\bar{\odot})} + f_{(C,\bar{\odot})}-3f_{(V,\odot)}\right),
\label{eq:hon_h_lap_fd}
\end{equation}
which has the same structure of the Laplacian of Eq. \eqref{eq:squ_lap_fd}. Notice also that, being the honeycomb lattice a non-Bravais lattice, we cannot obtain the Laplacian from Eq. \eqref{eq:nabla_symm_dir}. According to this graph Laplacian, the Hamiltonian reads then as follows:
\begin{align}
\hat{\mathcal{H}}=& -J_H\sum_{\odot\in\{\circ,\bullet\}}\sum_{(V,\odot)} \left(\dyad{A,\bar{\odot}}{V,\odot} + \dyad{B,\bar{\odot}}{V,\odot}\right.\nonumber\\
&\left.+\dyad{C,\bar{\odot}}{V,\odot}-3\dyad{V,\odot}{V,\odot} \right) \,,
\label{eq:H_HON_plg_free}
\end{align}
where the hopping amplitude is
\begin{equation}
J_H:=\frac{2\hbar^2}{3ma^2}=\frac{4}{3}J_S\,.
\label{eq:J_HON_plg}
\end{equation}

\subsection{Numerical simulation}
\label{subsec:num_sim_free}

\subsubsection{Parameter setting}
\label{subsubsec:par_set_free}
\paragraph*{Units.} For the computational implementation we set $a = \hbar =1$, where $a$ is the lattice parameter, and $\hbar$ the reduced Planck's constant. According to this choice, the dimensions and the units of the fundamental quantities are examined in Appendix \ref{app:units}.

\paragraph*{Time evolution.} In the \textit{Schr\"{o}dinger picture} the time evolution of the state of a quantum system is ruled by the unitary \textit{time-evolution operator}
\begin{equation}
\hat{\mathcal{U}}(t,t_0)=e^{-i\hat{\mathcal{H}}(t-t_0)}\,,
\label{eq:time_evol_op_schro}
\end{equation}
where $t_0$ and $t$ denote the initial and final time, respectively. Because of the previous units choice, the mass is left as the only dimensional quantity and it is controlled through the hopping amplitude $J$, as shown, e.g., in Eq. \eqref{eq:J_SQU_plg}. Such parameter enters the Hamiltonian as a global multiplicative factor, thus, if we focus on the time-evolution operator in Eq. \eqref{eq:time_evol_op_schro}, we can appreciate its role as a time-scaling factor in $i J\sum_V [\ldots] (t-t_0)$: the greater $J$, the lighter $m$, the faster the time evolution, whereas the lower $J$, the heavier $m$, the slower the time evolution. The quantum system, therefore, has a characteristic time given by $\tau=1/J$. We set $J_S=1$ and $J_T$ and $J_H$ follow according to Eqs. \eqref{eq:J_TRI_plg} and \eqref{eq:J_HON_plg}, respectively. This is equivalent to fixing the mass of the walker and comparing its CTQW on the different PLGs. Because of the aforementioned role of the hopping amplitude, in order to have a proper comparison of the results, these will be expressed as a function of the adimensional time $Jt$ (where $J$ takes the proper value in the different PLGs).

\paragraph*{Lattice size.} Unlike the square lattice graph, for which we can define the size as $N_x \times N_y$, where $N_x$ ($N_y$) is the number of vertices along the $x$ ($y$) direction, for the triangular and honeycomb ones the definition of the size is not straightforward: the `directions' to be considered might be polylines (see Appendix \ref{subsubapp:index}). We refer to $N_j$ ($N_k$) as the number of vertices along the $j$ ($k$) polyline, which plays the role of the $x$ ($y$) direction, and the resulting size of the graph is therefore $N_j \times N_k=\dim(\mathscr{H})$, where $\mathscr{H}$ denotes the Hilbert space of the system. We consider a \textit{finite} $(2n+1)\times(2m+1)$ PLG (see Appendix \ref{subapp:bc}), with $n,m \in \mathbb{N}$, since it has a properly defined center in $(n+1,m+1)$, of coordinates $(x_c,y_c)$ (in the following we partially restore the two-indices notation for labeling sites, see Sec. \ref{subsec:plg}). We set $N_j=N_k=41$ for the triangular lattice graph, $N_j=N_k=31$ for the square one, and $N_j=31$, $N_k=21$ for the honeycomb one. The size chosen for these graphs allows to make the system evolve for a long enough time, at a reasonable computational cost, to observe interesting effects before the wavefunction reaches the boundaries. The choice of setting $N_k < N_j$ for the honeycomb lattice graph is due to the following reason: two adjacent vertices $(j,k)$ and $(j+1,k)$ differs  by $\sqrt{3}a/2$ along the $x$ direction, whereas $(j,k)$ and $(j,k+1)$ by $a$ or $2a$ along the $y$ direction (see Appendix \ref{subapp:index_coords}, Fig. \ref{fig:labeling_plg}(c)). A honeycomb lattice graph with $N_j=N_k$ would be strongly unbalanced and the wavefunction would reach the $j$-boundary much earlier than the $k$-one. 

\paragraph*{Quantities of interest.} We study the time evolution of an initial state $\ket{\psi(0)}$ localized in the central vertex $(x_c,y_c)$ of the PLG (hence it is an eigenstate of $\hat{x}$ and $\hat{y}$). We look at the probability distribution of the walker and at the variance of the space coordinates as a function of time. We therefore introduce the \textit{probability density} $\rho_{j,k}(t)=\abs{\psi_{j,k}(t)}^2$ of finding the walker in the site $(j,k)$ at the time $t$. Maps of the probability density are to be understood according to Appendix \ref{subapp:maps}, and axis ticks according to Appendix \ref{subsubapp:index} (the indexing of vertices runs along polylines). The variance of the space coordinates is computed after recovering the spatial coordinates $(x_j,y_k)$ of vertices (see Appendix \ref{subsubapp:coords}) according to $\sigma_x^2=\langle \hat{x}^2 \rangle-\langle \hat{x} \rangle^2$, where $\langle \hat{x} \rangle=\sum_{j,k=1}^{N_j,N_k} \rho_{j,k}\, x_{j,k}$, since, in general, the $x$ coordinate of a vertex depends on both the indices (e.g. in the honeycomb and triangular lattice graphs). Analogously for $\sigma_y^2$.

\subsubsection{Results}
\begin{figure}[!h]
	\centering	
	\includegraphics[width=0.45\textwidth]{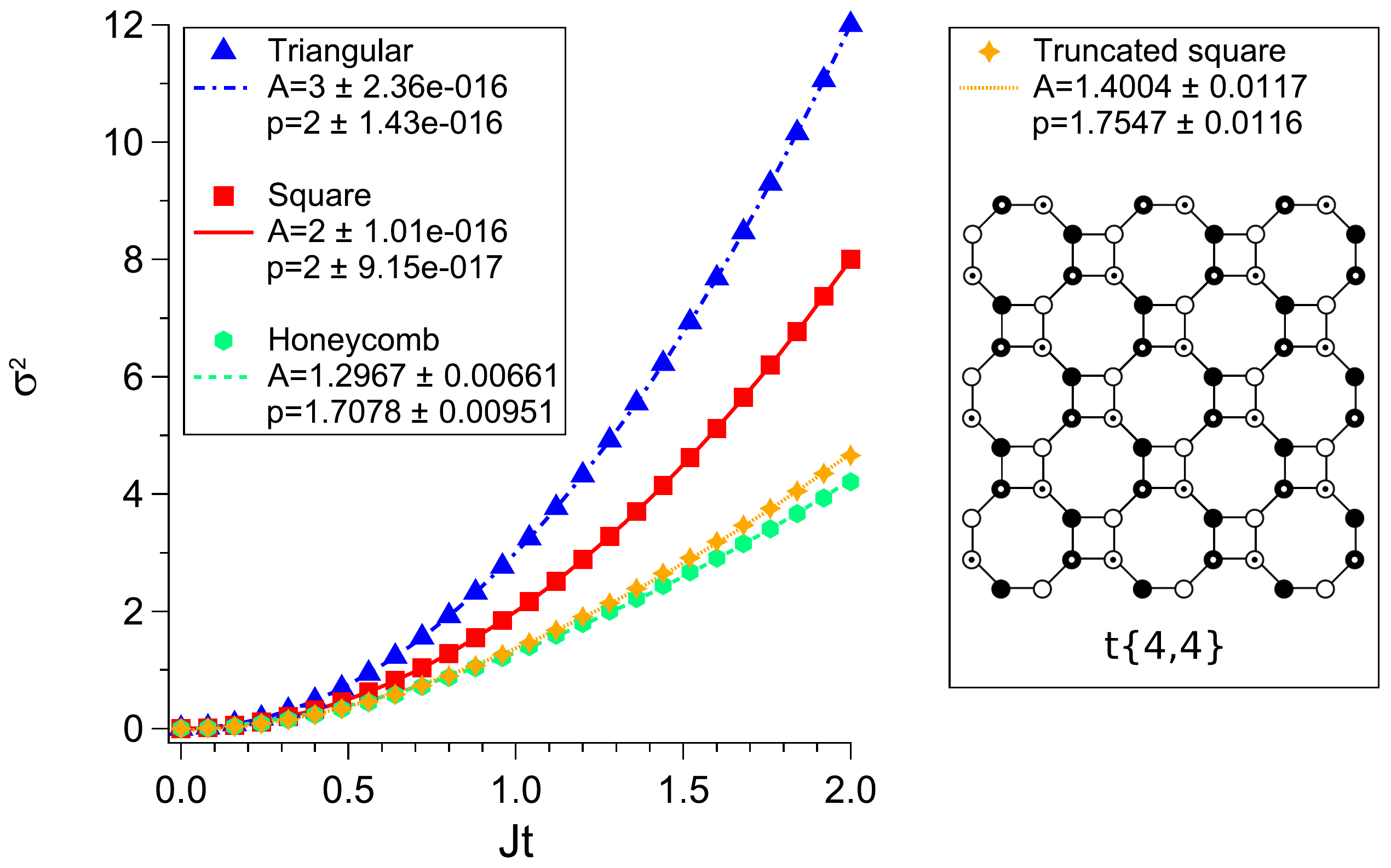}
	\caption{Variance of the space coordinates obtained in a CTQW of a free particle on a square (red squares), triangular (blue triangles), honeycomb (green hexagons), and truncated square (yellow four-pointed star) lattice graph. The latter, shown in the legend, consists of squares and octagons and it is characterized by four classes of non-equivalent vertices $\{\whitecirc,\blackcirc,\whitecircdot,\blackcircdot\}$. The variance of the two spatial coordinates is equal, $\sigma_x^2(Jt)=\sigma_y^2(Jt)=\sigma^2(Jt)$. Lines denote the fitting curves in Eq. \eqref{eq:fit}.}
	\label{fig:variance_plg_fit_free}
\end{figure}

The first study concerns the CTQW of a free particle on the different PLGs. For such a CTQW we expect a ballistic spreading of the wavefunction, i.e. $\sigma^2(Jt)\propto (Jt)^2$. Therefore we analyze the resulting variance of the space coordinates (Fig. \ref{fig:variance_plg_fit_free}) according to the fitting curve \footnote{No offset is introduced since the walker is initially localized in a single vertex, thus, as we checked, $\sigma_{x}^2(0)=\sigma_{y}^2(0)=0$.}
\begin{equation}
f(Jt)=A(Jt)^p.
\label{eq:fit}
\end{equation}
CTQWs on a square or triangular lattice graph show the same ballistic behavior for both the spatial coordinates, i.e. $\sigma_{x}^2(Jt)=\sigma_{y}^2(Jt)\propto (Jt)^2$. On the other hand, for the CTQW on a honeycomb lattice graph we observe $\sigma_{x}^2(Jt)=\sigma_{y}^2(Jt)\propto (Jt)^p$, with $1<p<2$, i.e. a behavior which is neither ballistic ($p=2$) nor diffusive ($p=1$), but \textit{sub-ballistic}. It is important to note that this numerical result puts limits to the universal ballistic spreading for both 1D and 2D QWs, as instead suggested in Ref. \cite{tang2018experimental}. The reason is believed to reside in the fact that, unlike the triangular and square lattice graphs, which are Bravais lattices, the honeycomb lattice graph is a non-Bravais lattice. Whereas in the former ones we can always go further along the same direction, in the latter one when we move one step from a vertex to an adjacent one, we change class of vertex and the NNs of the final vertex are arranged and oriented differently from those of the initial one (see Sec. \ref{subsec:plg}). This difference turns out to slow down the spreading of the quantum walker. We also notice that $\sigma_H^2 (Jt)\leq \sigma_S^2 (Jt)\leq \sigma_T^2 (Jt)$, i.e. the largest variance is obtained in the triangular lattice graph, while the lowest one in the honeycomb lattice graph. This behavior can be related to the different degree of a vertex in each PLG: $6$ in the triangular, $4$ in the square, and only $3$ in the honeycomb lattice graph.

A further clue that CTQWs behave differently on Bravais and non-Bravais lattice is provided by the CTQW on another 2D non-Bravais lattice: the \textit{truncated square tiling} (or \textit{truncated quadrille} \cite{conway2008symmetries}), whose Schl\"{a}fli symbol is $t\left\lbrace 4,4\right\rbrace$ (legend of Fig. \ref{fig:variance_plg_fit_free}). It is a \textit{semiregular} or \textit{Archimedean} tessellation, all of whose tiles are regular polygons, with one square and two octagonal tiles about each vertex, and the tiling pattern around each vertex being the same \cite{wells1991penguin}. Weakening the definition of lattice graph in order to include also semiregular tilings, in the following we will refer to this non-Bravais lattice as the truncated square lattice graph. In such graph a generic vertex has $\operatorname{deg}(V)=3$, as in the honeycomb lattice graph, but unlike the latter, here there are four classes of non-equivalent vertices $\{\whitecirc,\blackcirc,\whitecircdot,\blackcircdot\}$. The CTQW Hamiltonian matrix $\mathcal{H}=-J L$ has been here defined according to Eqs. \eqref{eq:adj_matr}--\eqref{eq:deg_matr}. Indeed, since for a given vertex the hopping directions are not symmetrically distributed, defining the Laplacian by means of finite-difference formulae from Taylor expansion is ill-defined, since it provides different hopping terms depending on the direction. Even in this case we observe $\sigma_{x}^2(Jt)=\sigma_{y}^2(Jt)\propto (Jt)^p$, with $1<p<2$, i.e. a sub-ballistic spreading.

\begin{figure}[!h]
	\centering	
	\includegraphics[width=0.45\textwidth]{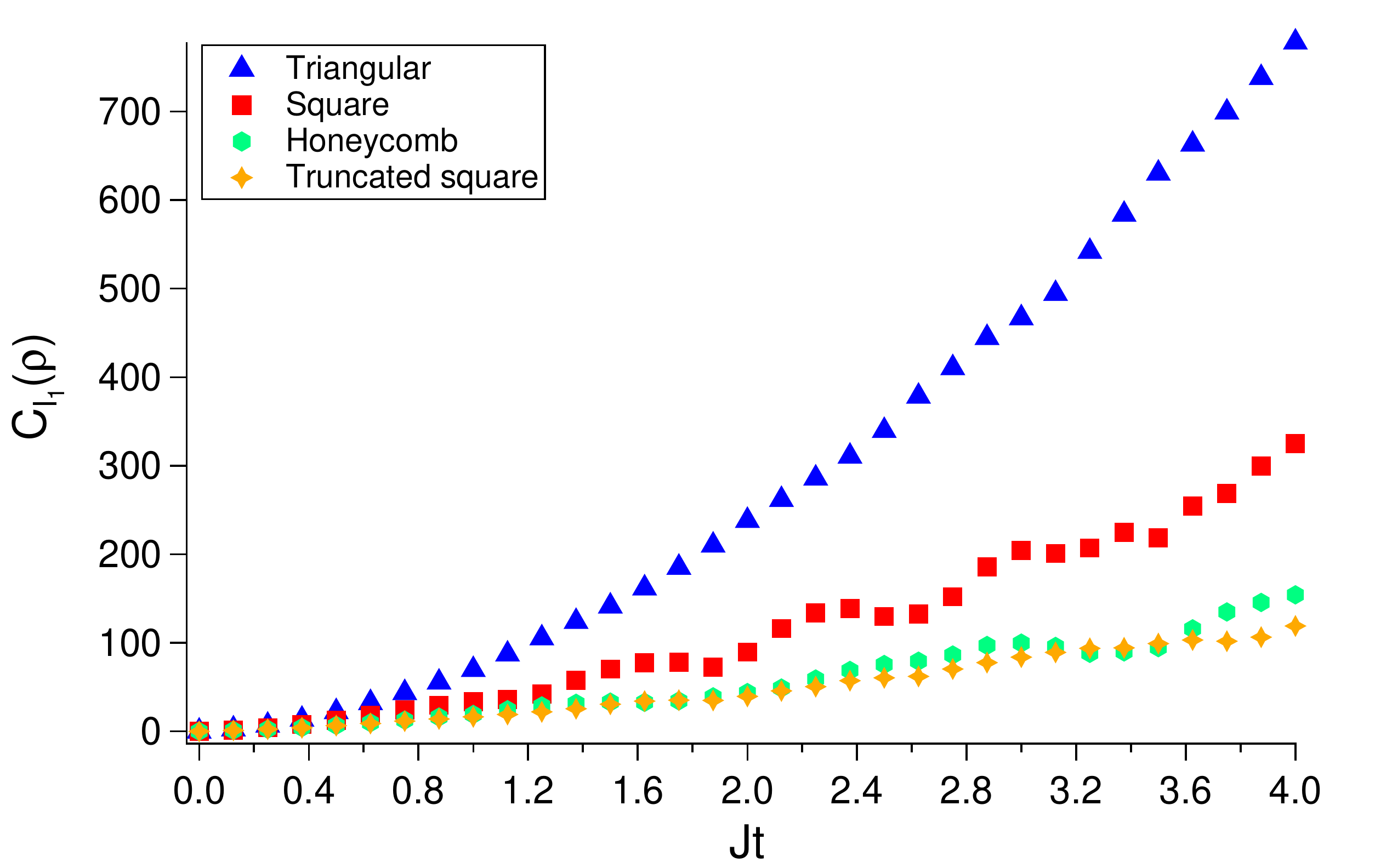}
	\caption{Quantum coherence of a CTQW of a free particle on a square (red squares), triangular (blue triangles), honeycomb (green hexagons), and truncated square (yellow four-pointed star) lattice graph. In this computation, the same lattice size has been adopted for all the PLGs, in order to have Hilbert spaces of the same dimension and so a proper comparison.}
	\label{fig:qcoher_free}
\end{figure}

A proper measure of quantum coherence is provided by the $l_1$ norm of coherence \cite{baumgratz2014quantifying,zhu2018axiomatic}
\begin{equation}
C_{l_1}(\rho)=\sum_{m\neq n} \abs{\rho_{mn}}=\sum_{m,n} \abs{\rho_{mn}}-1\,,
\end{equation}
i.e. the sum of the absolute values of the off-diagonal elements of the density matrix. According to this definition, after writing the density matrix in the vertex states basis, we observe that the PLGs causing more coherence are those in which the  CTQW is properly ballistic, \textit{vice versa} CTQWs on the non-Bravais PLGs are characterized by a lower coherence and a sub-ballistic spreading (Fig. \ref{fig:qcoher_free}). This is in agreement with the idea that the ballistic spreading is due to interference phenomena.

\begin{figure*}[t]
	\centering	
	\includegraphics[width=0.75\textwidth]{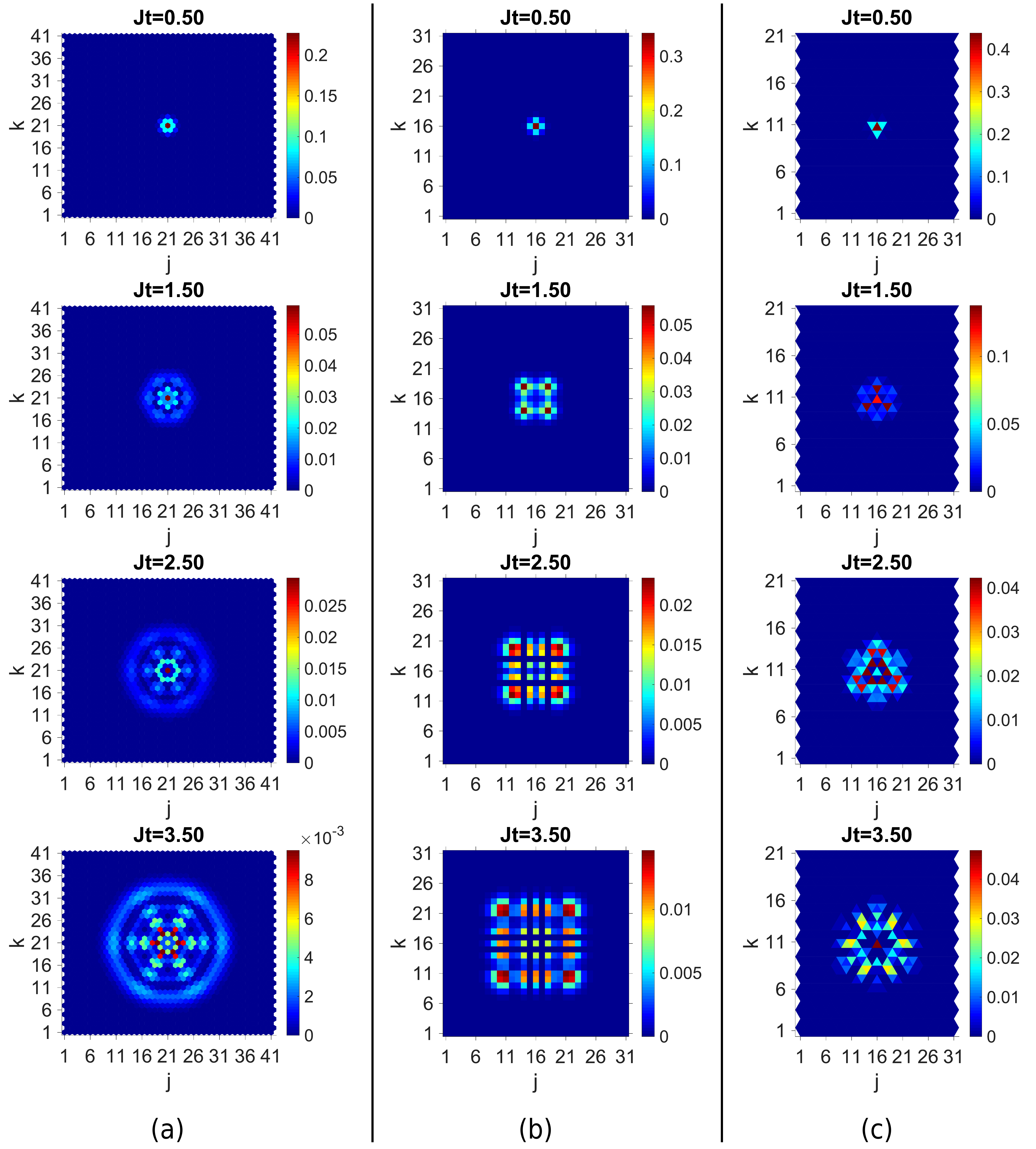}
	\caption{Maps of the time evolution of the probability density according to the CTQW of a free particle on a (a) $41 \times 41$ triangular, (b) $31 \times 31$ square, and $31 \times 21$ (c) honeycomb lattice graph.}
	\label{fig:free_QW_maps}
\end{figure*}

Maps of the time-evolving probability density are shown in Fig. \ref{fig:free_QW_maps}. In each case the spread path is characterized by the symmetry of the underlying lattice.

\section{A charged quantum walker in a magnetic field}
\label{sec:charge_ptc_B}

\subsection{The Hamiltonian of the system: the issue of the spatial discretization}
\label{subsec:charge_ptc_B_Ham}
The Hamiltonian of a particle of mass $m$ and charge $q$ in a plane in the presence of a electromagnetic field is obtained on the basis of the Hamiltonian of the free particle through the minimal substitution $\hat{\mathbf{p}}\rightarrow\hat{\mathbf{p}}-q\mathbf{A}$ and by inserting the electric potential, namely
\begin{equation}
\hat{\mathcal{H}}=\frac{1}{2m}\left( \hat{\mathbf{p}}-q\mathbf{A}\right)^2+q\phi,
\label{eq:Ham_mq_em}
\end{equation}
where $\phi$ and $\mathbf{A}$ are, respectively, the scalar and vector potential of the electric field $\mathbf{E}=-\nabla\phi-\partial_t \mathbf{A}$ and magnetic field $\mathbf{B}=\nabla\times\mathbf{A}$. In order to study a charged particle in the presence of the perpendicular magnetic field only, we set $\phi=0$ and choose a time-independent vector potential $\mathbf{A}=A^x(x,y)\vu{i}+A^y(x,y)\vu{j}$. The Hamiltonian is then
\begin{equation}
\hat{\mathcal{H}}=\frac{1}{2m}\hat{\mathbf{p}}^2-\frac{q}{2m}\left ( \hat{\mathbf{p}}\cdot\mathbf{A}+\mathbf{A}\cdot\hat{\mathbf{p}}\right )+\frac{q^2}{2m}\mathbf{A}^2\,,
\label{eq:Ham_charged_ptc_cont}
\end{equation}
where $\hat{\mathbf{p}}\cdot\mathbf{A}$ acts on the wavefunction as $\hat{\mathbf{p}}\cdot(\mathbf{A}\psi(\vb{r}))$. 

In the light of the strict connection between the generator of the evolution of the CTQW and the Hamiltonian (see Sec. \ref{subsec:CTQW_graph}), the straightforward approach to get a CTQW Hamiltonian is to spatially discretize the Hamiltonian of the corresponding system in the continuum. Several works addressed the presence of potentials \cite{wong2016quantum},  defects or disorder \cite{izaac2013continuous} which depend on the vertices and interactions between the walkers when in the same vertex or in NNs \cite{preiss2015strongly}. However, this spatial dependence has been usually considered for 1D systems or graphs, the latter intended as mathematical objects for algorithmic purposes \cite{wong2016qwsearch}. When inserting the magnetic field, the vector potential has an actual spatial dependence, we can not prescind from the spatial coordinates of vertices. Moreover, unlike the aforementioned cases, the Hamiltonian in Eq. \eqref{eq:Ham_charged_ptc_cont} includes a cross-term $\sim(\vb{A}\cdot \hat{\vb{p}}+\hat{\vb{p}}\cdot \vb{A})$, the orbital paramagnetic term, which couples the field to the particle's orbital motion, so that the linear momentum is present both at first and second order. This means that not only the Laplacian, but also the first-order differential operator is required.  We expect the Laplacian to behave like the graph one, for which we already have a general definition in Eq. \eqref{eq:graphL}, but we expect also the first-order differential operator, for which we do not have an analogous general definition, to be sensitive to the geometry of the lattice and to return not only non-negative results for NNs, as instead the kinetic term (the graph Laplacian) does ($L_{jk}=1$ if $j \neq k$ and connected). Here lies the crux of the present work and its peculiarity, i.e. the hard task of spatially discretizing Eq. \eqref{eq:Ham_charged_ptc_cont} - and so the differential operators - in a 2D space according to the different geometries characterizing the PLGs. Indeed, to the best of our knowledge, such issue has not been addressed for CTQWs yet. However, there are works considering DTQWs under artificial magnetic fields on square lattices involving Peierls phase-factors \cite{yalccinkaya2015two,boada2017quantum}, and this is a first hint to treat our CTQW in the presence of a magnetic field without explicitly involving the spatial discretization of differential operators. On the other hand, we are also interested in finding a way to spatially discretize the Hamiltonian in Eq. \eqref{eq:Ham_charged_ptc_cont} according to the different geometries of the PLGs (Fig. \ref{fig:tessel_plg}).

In the free-particle Hamiltonian we know the hopping must be equiprobable along the allowed directions, i.e. the walker must have the same jumping rate forward or backward, along a direction or another  (see Sec. \ref{subsec:CTQW_graph}). Such requirement is usually satisfied computing $\nabla^2\psi_V$ in a given vertex $V$ by means of \textit{central} finite difference formulae \footnote{The use of \textit{backward} or \textit{forward} finite difference formulae, instead, would produce a bias in the hopping of the walker, a non-hermitian Hamiltonian, and so the system would not describe a CTQW.}, which involve all the NNs of $V$. Moreover, this ensures the hermiticity of the Hamiltonian (as regards the terms in $\hat{\vb{p}}$). The latter, we recall, is the ultimate condition for having a CTQW, since any hermitian operator abiding the topology of the graph can describe a CTQW. Let us consider the 1D case for the free particle: the central difference formula to compute the $\nabla^2\psi (x_n)$, with $x_n=n \in \mathbb{Z}$, involves the vertex $x_n$ itself and its NNs $x_{n\pm 1}$. Considering the hopping terms, the central difference formula allows the walker to jump from $x_n$ to $x_{n\pm1}$, with the same jumping rate (see Eq. \eqref{eq:Lapl_central_1D}). Since it holds $\forall n$, this Hamiltonian matrix is symmetric (hermiticity for a real-valued matrix), meaning that the hopping term from $x_n$ to $x_{n\pm1}$ is the same as the one from $x_{n\pm1}$ to $x_n$. This reasoning also applies to a complex-valued Hamiltonian matrix and hermiticity, where the hopping terms between two NNs are one the hermitian conjugate of the other.

We assume a hopping to NNs which takes into account the contribution of the magnetic field. We therefore explore the two approaches: (i) the introduction of the Peierls phase-factors (Sec. \ref{sec:peierls_model}), according to which the tunneling matrix element of the free particle becomes complex, accompanied by the Peierls phase due to the vector potential; (ii) the spatial discretization of the Hamiltonian of Eq. \eqref{eq:Ham_charged_ptc_cont} in terms of finite-difference formulae (Sec. \ref{sec:spatial_disc_H_fd}). In the first case the assumption on the hopping to NNs is fulfilled for free, since the model is based on the CTQW Hamiltonian of the free particle (ultimately on the graph Laplacian); in the second case the differential operators must be discretized according to the NNs of a given vertex by means of `central difference'-like formulae.

\subsection{Numerical simulation: parameter setting}
\label{subsec:par_set_magn}
In addition to what stated in Sec. \ref{subsubsec:par_set_free}, we set the following:

\paragraph*{Units.} For the computational implementation we set the electric charge $q=1$ (see Appendix \ref{app:units} for units and dimensions).

\paragraph*{Gauge and magnetic field.} A uniform magnetic field $\mathbf{B}=B\hat{\mathbf{k}}$ is introduced by means of the vector potential (symmetric gauge) $\mathbf{A}=\frac{B}{2}\left (-(y-y_c),(x-x_c),0\right)$. In this gauge, as known, the Hamiltonian in Eq. \eqref{eq:Ham_charged_ptc_cont} turns out to be the Hamiltonian of a 1D harmonic oscillator, whose degenerate energy levels are the so-called \textit{Landau levels} \cite{shankar1994principles} and which is characterized by the \textit{cyclotron frequency} $\omega_0 = qB/m$, which depends on the magnetic field. This choice of gauge breaks translational symmetry in both the $x$ and the $y$ directions, but it does preserve rotational symmetry about the center $(x_c,y_c)$ of the PLG. This means that the angular momentum together with the Landau level are good quantum numbers \cite{tong2016lectures} to label states. The angular momentum is classically defined as $\mathbf{L}=\mathbf{r}\times\mathbf{p}$, but since our charged particle lies in the $xy$ plane, the angular momentum is $\vb{L}=L_z\vu{k}$ and the corresponding operator is
\begin{equation}
\hat{L}_z =(\hat{x}\hat{p}_y-\hat{y}\hat{p}_x)\,.
\end{equation}
Since the symmetric gauge belongs to the Coulomb gauge, where $\nabla \cdot \mathbf{A}=0$ so $[\hat{\mathbf{p}},\hat{\mathbf{A}}]=0$, it can be proved that $[\hat{\mathcal{H}},\hat{L}_z]=0$, i.e. $\hat{\mathcal{H}}$ and $\hat{L}_z$ represent a complete set of compatible observables. The states belonging to the lowest Landau level, i.e. the ground state, are characterized by a ring-shaped probability density. So, if we allow the PLG to better follow the rotational symmetry of the Hamiltonian, we expect more circular structures in the probability density of the walker. In particular we expect the triangular lattice graph, because of its six NNs per vertex, to provide the best discrete approximation of a circle among the PLGs; instead, because of the only three NNs per vertex, we expect the honeycomb lattice graph to provide the worst one. Moreover, due to structure of non-equivalent vertices of the latter, symmetries may struggle to emerge. It is important to keep in mind this premise about the rotational symmetry and the harmonic oscillator because it will be of help in the interpretation of the results in the following.

The \textit{magnetic length} is the fundamental characteristic length scale for any quantum phenomena in the presence of a magnetic field \cite{tong2016lectures} and it imposes an upper bound to the interval of fields investigated. Indeed, it is defined as 
\begin{equation}
l_B := \sqrt{\frac{\hbar}{qB}} = B^{-\frac{1}{2}}\,,
\label{eq:magnetic_length}
\end{equation}
where the last equality holds because of our units ($\hbar=q=a=1$), so, since for $B>1$ the magnetic length becomes smaller than the lattice constant $a$, we consider $B\in[0,1]$.

\section{CTQW under magnetic field: the Peierls model}
\label{sec:peierls_model}

\subsection{The Peierls phase-factors}
\label{subsec:peierls_ph_factor}
The motion of a charged particle in a magnetic field is accompanied by a geometric phase, the Aharonov-Bohm phase \cite{aharonov1959significance}. On a lattice these phases are introduced in the form of the so-called \textit{Peierls phases} that a particle picks up when hopping in the lattice. Such phases allow to rewrite the tight-binding Hamiltonian of a charged particle in a magnetic field as the tight-binding Hamiltonian of a free particle where tunneling matrix elements are complex and hopping in the lattice is accompanied by the Peierls phase \cite{aidelsburger2015artificial,bernevig2013topological}. The spectrum of the Hamiltonian so obtained is the famous Hofstadter butterfly \cite{hofstadter1976energy}. According to Feynman, the Hamiltonian having such Peierls phase-factors can be traced, in some limits, to the well-known Hamiltonian in Eq. \eqref{eq:Ham_mq_em} \cite{feynman2011feynmanII,feynman2011feynmanIII}.

\begin{figure}[!h]
	\centering
	\includegraphics[width=0.2\textwidth]{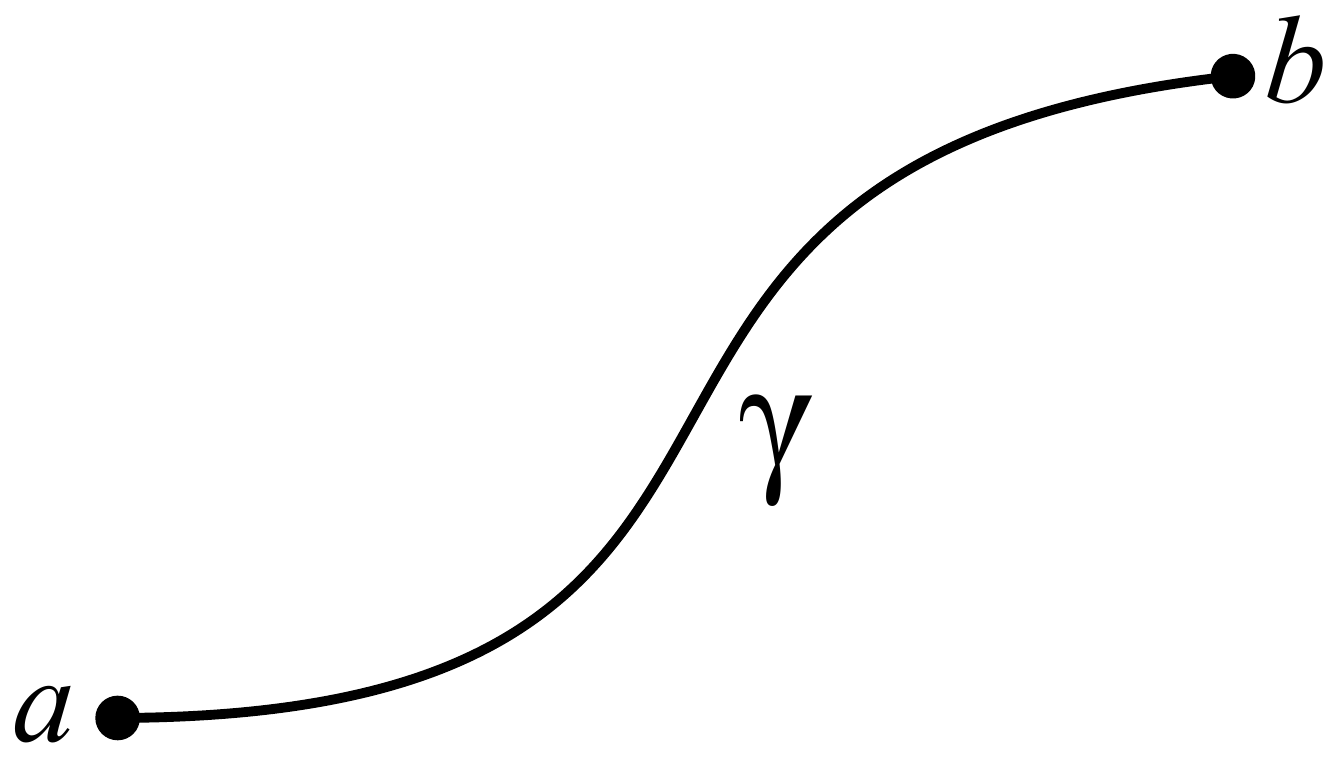}
	\caption{The probability amplitude to go from $a$ to $b$ along the path $\gamma$, in the presence of a vector potential $\mathbf{A}$, is proportional to $\operatorname{exp}\left[\frac{iq}{\hbar}\int_a^b\mathbf{A}\cdot \diff \mathbf{s}\right]$.}
	\label{fig:Feynman_Peierls_path}
\end{figure}

\begin{proof}
The Feynman's argument develops as follows. An external magnetic field is described by a vector potential. The probability amplitude that a particle goes from one place to another, along a certain path when there is a field present (Fig. \ref{fig:Feynman_Peierls_path}), is
\begin{equation}
\langle b \vert a \rangle_{\text{in }\mathbf{A}} = \langle b \vert a \rangle_{A=0}\cdot \operatorname{exp}\left[\frac{iq}{\hbar}\int_a^b\mathbf{A}\cdot \diff\mathbf{s}\right]\,,
\label{eq:feynman_phase}
\end{equation}
i.e. it is the same as that of the particle going along the same path when there is no field, multiplied by a phase factor which depends on the line integral of the vector potential.

Feynman considers then a simple example in which instead of having a continuous situation there is a line of atoms along the $x$ axis with the spacing $a$, an electron has a probability amplitude $-K$ to jump from one atom to another when there is no field, and there is a vector potential in the $x$ direction $A_x(x, t)$. The rate of change of the probability amplitude $C(x)$ to find the electron at the atom ``$n$'' located at $x$ is given by the following equation:
\begin{align}
i\hbar \partial_t C(x)=&E_0C(x)-Ke^{-iaf(x+a/2)}C(x+a)\nonumber\\
&-Ke^{+iaf(x-a/2)}C(x-a)\,,
\label{eq:feynman_Cx_prob}
\end{align}
where $E_0$ is the energy of the electron if located at $x$, $f(x):=(q/\hbar)A_x$, and $-KC(x\pm a)$ is the probability amplitude for the electron to have jumped backward or forward, respectively, one step from
atom ``$n \pm 1$'', located at $x \pm a$. If $A_x$ is not changing appreciably in one atomic spacing, the integral can be written as just the value of $A_x$ at the midpoint times the spacing $a$, resulting in a phase factor $\exp{\pm iaf (x \pm a/2)}$. The sign of the phase shift reflects the direction of the hopping: backward ($-$) or forward ($+$).

If the function $C(x)$ is smooth enough (long wavelength limit), and if we let the atoms get closer together ($a\to 0$), Eq. \eqref{eq:feynman_Cx_prob} will approach the behavior of an electron in free space. So the next step is to Taylor expand the right-hand side of Eq. \eqref{eq:feynman_Cx_prob} ($C(x)$, $f(x)$, and the exponentials) in powers of $a$, to collect the terms up to $\mathcal{O}(a^2)$ and to recast all into
\begin{equation}
i\hbar\partial_t C(x) = (E_0 -2K) C(x)-Ka^2\left[ \partial_x-if(x)\right]^2C(x)\,.
\label{eq:feynman_Cx_dx}
\end{equation}
The solutions for zero magnetic field represent a particle with an effective mass $m_\textup{eff}$ given by
\begin{equation}
Ka^2 = \frac{\hbar^2}{2m_\textup{eff}}\,.
\end{equation}
After setting $E_0 = 2K$ and restoring $f(x) = (q/\hbar)A_x$, we can easily check that Eq. \eqref{eq:feynman_Cx_dx} is the same as the first part of Eq. \eqref{eq:Ham_mq_em}. Hence, the proposition of Eq. \eqref{eq:feynman_phase} that the vector potential changes all the probability amplitudes by the exponential factor is the same as the rule that the momentum operator $-i\hbar\nabla$ gets replaced by
\begin{equation}
-i\hbar\nabla - q\mathbf{A}\,,
\end{equation}
as we see in the Schr\"{o}dinger equation of Eq. \eqref{eq:Ham_mq_em}.
\end{proof}

Since resorting to Peierls phase-factors is equivalent to the minimal substitution in the Hamiltonian, they can be used to study the CTQW in the presence of a magnetic field, with no need of discrete differential operators but the graph Laplacian. In other words, we may simply correct the free-particle Hamiltonian according to the Peierls substitution \cite{peierls1933theorie}, i.e. by making the tunneling matrix element complex:
\begin{equation}
J \longrightarrow J \operatorname{exp}\left[ \frac{iq}{\hbar}\int_{\mathbf{r}_a}^{\mathbf{r}_b}\mathbf{A}\cdot \diff\mathbf{r}\right]\,,
\label{eq:J_Jpeierls}
\end{equation}
where $J$ is the NN hopping amplitude and the integral is evaluated along the edge connecting $\mathbf{r}_a$ and $\mathbf{r}_b$, i.e. the initial and final positions (vertices) of the particle, respectively. However, the Peierls phase-factors are equivalent to the minimal substitution in the continuum limit. This means that the quadratic term in $\vb{A}$ is recovered only in such limit, and so it is not present in this Hamiltonian. Indeed, such term would affect the diagonal elements, the on-site energies, but in this model they are left as the degree of the vertex (or set equal to zero, being the lattice graph regular - $\operatorname{deg}(V)=const$ - and so providing an irrelevant global phase to the wavefunction).

The Peierls phase-factor in Eq. \eqref{eq:J_Jpeierls} involves a line integral which does depend on the chosen path (Fig. \ref{fig:Feynman_Peierls_path}), and it is calculated as follows:
\begin{align}
\int_\gamma \vb{A}(\vb{r})\cdot \diff\vb{r}&=\int_{a}^{b}\vb{A}(\vb{r}(t))\cdot \vb{r}'(t)\diff t\nonumber\\
&=\int_{a}^{b}\left [ A^x(\vb{r}(t))x'(t)+A^y(\vb{r}(t))y'(t)\right]\diff t \,,
\end{align}
where $\vb{r}(t):\left [ a,b\right ]\rightarrow \gamma$ is a bijective parametrization of the curve $\gamma$ such that $\vb{r}_a:=\vb{r}(a)$ and $\vb{r}_b:=\vb{r}(b)$ give the endpoints of $\gamma$. In particular, $\vb{r}(t)=x(t)\vu{i}+y(t)\vu{j}$ and $\vb{r}'(t)=\dv{\vb{r}}{t}$. Such line integral has to be evaluated along the edges of the PLG, i.e. pieces of straight lines that we parametrize as follows
\begin{equation}
\begin{cases}
x(t)=x_0+t(x_1-x_0)\\
y(t)=y_0+t(y_1-y_0)
\end{cases},\quad t\in[0,1]\,,
\end{equation}
from which $x'=x_1-x_0$ and $y'=y_1-y_0$ are constants. Then the integral is approximated according to the \textit{trapezoidal rule} \cite{burden2010numerical,butenko2014numerical}:
\begin{align}
\int_{0}^{1}\vb{A}(\vb{r}(t))\cdot \vb{r}'(t)\diff t\approx &\frac{1}{2}\left[ (x_1-x_0)\left  (A^x(\vb{r}_0)+A^x(\vb{r}_1)\right )\right.\nonumber\\
&\left.\quad+(y_1-y_0)\left (A^y(\vb{r}_0)+A^y(\vb{r}_1)\right )\right ]\,,
\label{eq:approx_trapez_line_int}
\end{align}
where $\vb{r}_0=(x_0,y_0)$ and $\vb{r}_1=(x_1,y_1)$ are the coordinates of the initial and final vertex, respectively.
Moreover, if the vector potential components depend linearly on the $x$ and $y$ coordinates, e.g. in the Landau and in the symmetric gauge (providing a uniform magnetic field $\vb{B}=B\vu{k}$),  Eq. \eqref{eq:approx_trapez_line_int} is exact and it holds as equality. Indeed, let $f(x)=mx+q$, then
\begin{align}
\int_a^b f(x)\diff x &=\frac{b-a}{2}(ma+mb+2q)\nonumber\\
&=\frac{b-a}{2}\left (f(a)+f(b)\right )\,.
\end{align}

\subsection{The CTQW Hamiltonian}

\subsubsection{Square lattice graph}
With reference to Table \ref{tab:plg_nn_coords}, the Hamiltonian describing the CTQW according to the Peierls model is:
\begin{align}
\mathcal{\hat{H}}=&-J_S\sum_V\left[\exp{\frac{iqa}{2\hbar}(A^x_V+A^x_A)}\dyad{A}{V}\right.\nonumber\\
&+\exp{\frac{iqa}{2\hbar}(A^y_V+A^y_B)}\dyad{B}{V}\nonumber\\
&+\exp{-\frac{iqa}{2\hbar}(A^x_V+A^x_C)}\dyad{C}{V}\nonumber\\
&\left.\vphantom{\exp{-\frac{iqa}{2\hbar}}}+\exp{-\frac{iqa}{2\hbar}(A^y_V+A^y_D)}\dyad{D}{V}-4\dyad{V}{V}\right]\,,
\label{eq:H_SQU_plg_peierls}
\end{align}
where $J_S$ is defined in Eq. \eqref{eq:J_SQU_plg}.

\subsubsection{Triangular lattice graph}
With reference to Table \ref{tab:plg_nn_coords}, the Hamiltonian describing the CTQW according to the Peierls model is:
\begin{align}
\mathcal{\hat{H}}=&-J_T\sum_V\left[\exp{\frac{iqa}{2\hbar}(A^x_V+A^x_A)}\dyad{A}{V}\right.\nonumber\\
&+\exp{\frac{iqa}{4\hbar}\left[A^x_V+A^x_B+\sqrt{3}(A^y_V+A^y_B)\right]}\dyad{B}{V}\nonumber\\
&+\exp{-\frac{iqa}{4\hbar}\left[A^x_V+A^x_C-\sqrt{3}(A^y_V+A^y_C)\right]}\dyad{C}{V}\nonumber\\
&+\exp{-\frac{iqa}{2\hbar}(A^x_V+A^x_D)}\dyad{D}{V}\nonumber\\ 
&+\exp{-\frac{iqa}{4\hbar}\left[A^x_V+A^x_E+\sqrt{3}(A^y_V+A^y_E)\right]}\dyad{E}{V}\nonumber\\ 
&+\exp{\frac{iqa}{4\hbar}\left[A^x_V+A^x_F-\sqrt{3}(A^y_V+A^y_F)\right]}\dyad{F}{V}\nonumber\\
&\left.\vphantom{\exp{\frac{iqa}{2\hbar}}}-6\dyad{V}{V}\right]\,,
\label{eq:H_TRI_plg_peierls}
\end{align}
where $J_T$ is defined in Eq. \eqref{eq:J_TRI_plg}.

\subsubsection{Honeycomb lattice graph}
With reference to Table \ref{tab:plg_nn_coords}, the Hamiltonian describing the CTQW according to the Peierls model is:
\begin{align}
\hat{\mathcal{H}}= &-J_H\sum_{\odot\in\{\circ,\bullet\}}\sum_{(V,\odot)} \left[\vphantom{\sum}
e^{i\theta_{AV}} \dyad{A,\bar{\odot}}{V,\odot} \right.\nonumber\\
&+e^{i\theta_{BV}} \dyad{B,\bar{\odot}}{V,\odot}+e^{i\theta_{CV}}\dyad{C,\bar{\odot}}{V,\odot}\nonumber\\
&\left.\vphantom{\sum}-3 \dyad{V,\odot}{V,\odot} \right] \,,
\label{eq:H_HON_plg_peierls}
\end{align}
where $J_H$ is defined in Eq. \eqref{eq:J_HON_plg}, and we have defined:
\begin{align}
\theta_{AV} :=& \frac{qa}{4\hbar}\left [ \sqrt{3}\left (A^x_{(V,\odot)}+A^x_{(A,\bar{\odot})}\right )\right.\nonumber\\
&\left.-\operatorname{sgn}(\odot)\left (A^y_{(V,\odot)}+A^y_{(A,\bar{\odot})}\right )\right ]\,,\\
\theta_{BV} :=& -\frac{qa}{4\hbar}\left[ \sqrt{3}\left (A^x_{(V,\odot)}+A^x_{(B,\bar{\odot})}\right)\right.\nonumber\\
&\left.+\operatorname{sgn}(\odot)\left (A^y_{(V,\odot)}+A^y_{(B,\bar{\odot})}\right)\right ]\,,\\
\theta_{CV} :=& \operatorname{sgn}(\odot)\frac{qa}{2\hbar}\left (A^y_{(V,\odot)}+A^y_{(C,\bar{\odot})}\right )\,.
\end{align}

\subsection{Numerical simulation: results}
\label{subsec:results_magn_peierls}
\begin{figure}[!h]
	\centering
	\includegraphics[width=0.45\textwidth]{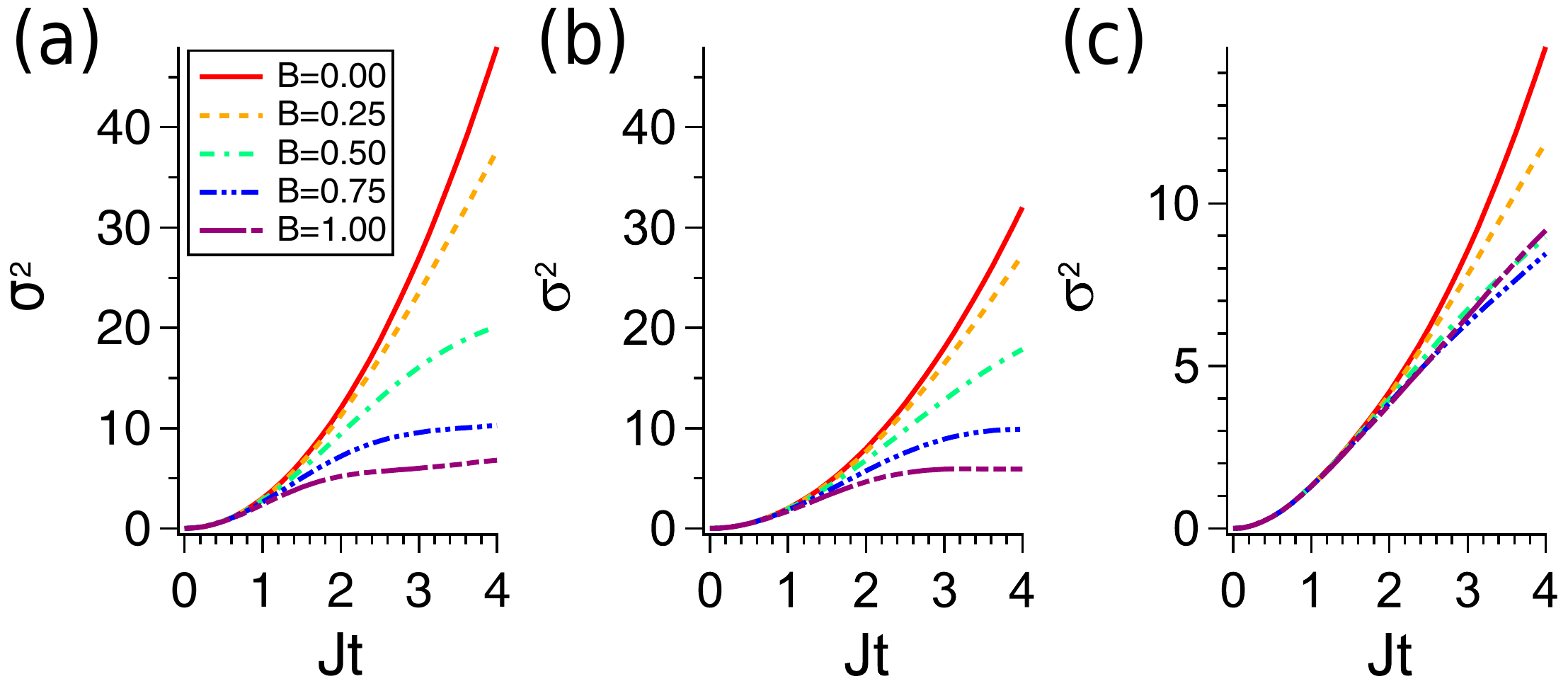}
	\caption{Variance of the space coordinates obtained in a CTQW of a charged particle in the (a) triangular, (b) square, and (c) honeycomb lattice graph for increasing values of the modulus $B$ of the perpendicular uniform magnetic field. As the latter increases, the variance deviates from the curve of the free particle. The variance of the two spatial coordinates is equal, $\sigma_x^2(t)=\sigma_y^2(t)$. The stronger the magnetic field, the smaller the variance. The CTQW Hamiltonian is obtained from the Peierls model.}
	\label{fig:var_B_2_zero_peierls}
\end{figure}

The behavior of the variance of the space coordinates is shown in Fig. \ref{fig:var_B_2_zero_peierls}, and it is the same for both the $x$ and $y$ coordinate, i.e. $\sigma_{x}^2(Jt)=\sigma_{y}^2(Jt)$. We observe that, as the modulus of the magnetic field increases, the curve of the variance of the space coordinates deviates from that of the free particle, decreasing.

\begin{figure}[!h]
	\centering
	\includegraphics[width=0.45\textwidth]{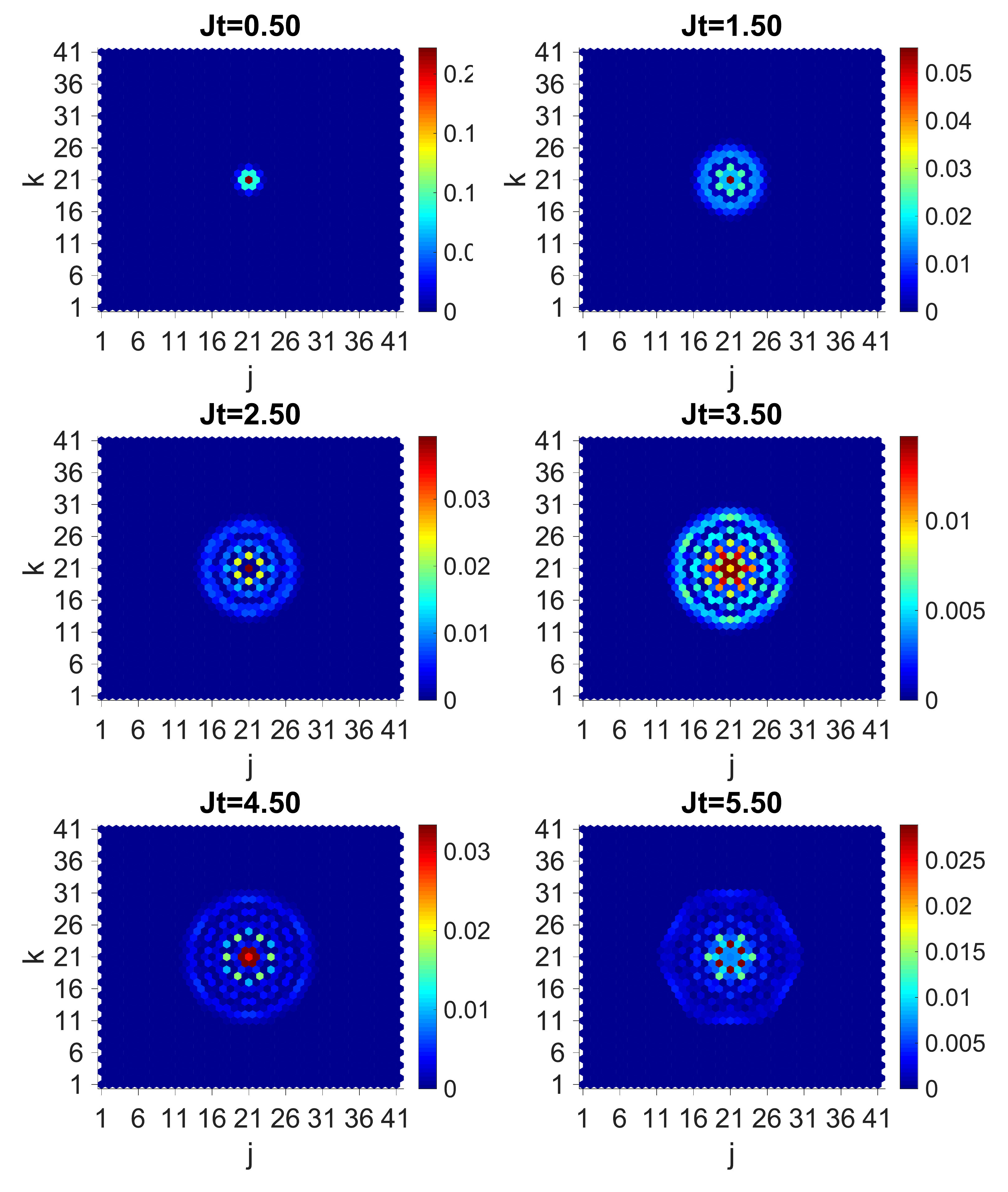}
	\caption{Map of the time evolution of the probability density according to the CTQW of a charged particle on a triangular lattice graph in the presence of a perpendicular uniform magnetic field ($B=0.6$). The CTQW Hamiltonian is obtained from the Peierls model.}
	\label{fig:rho_TRI_B_peierls}
\end{figure}

\begin{figure}[!h]
	\centering
	\includegraphics[width=0.45\textwidth]{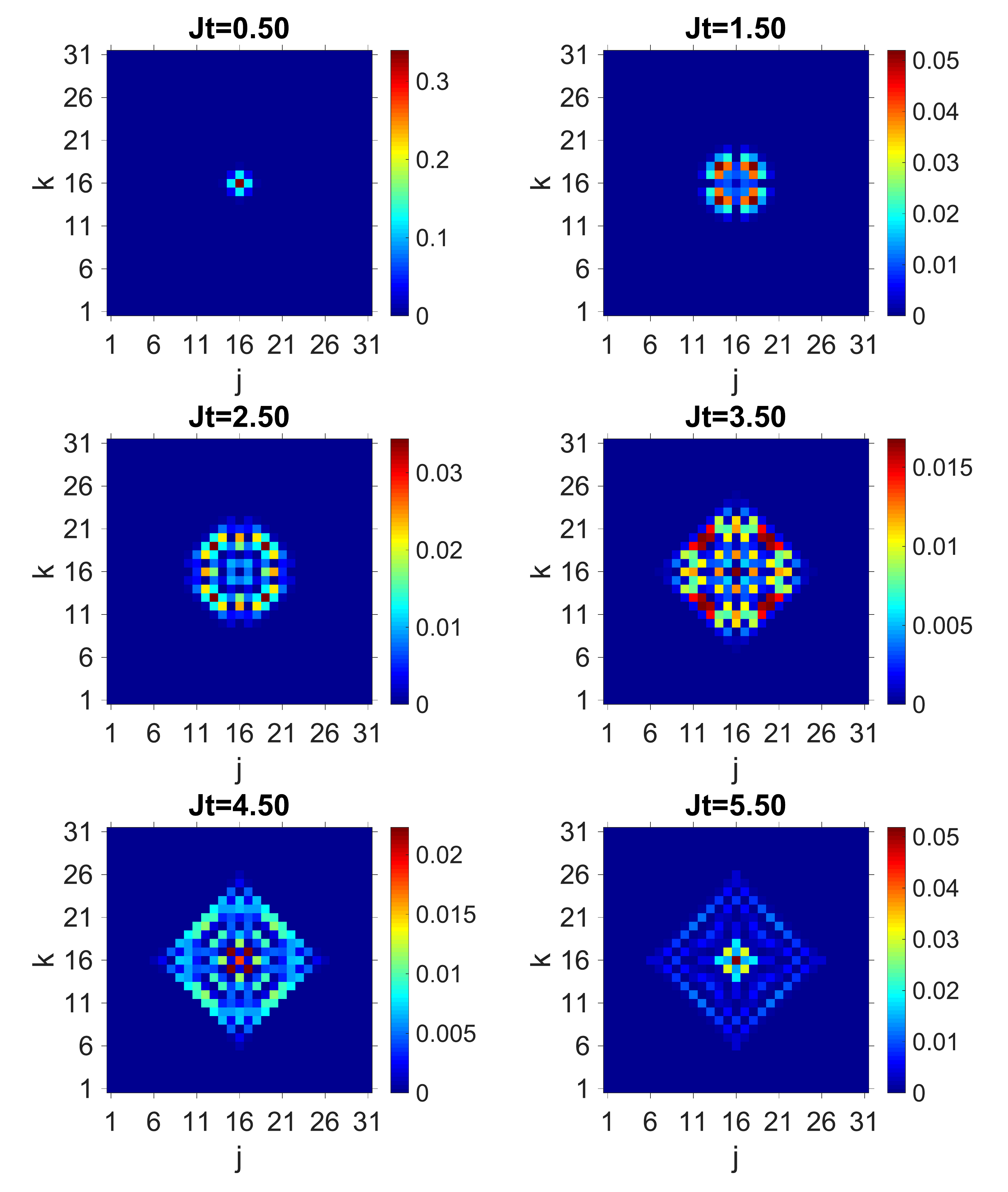}
	\caption{Map of the time evolution of the probability density according to the CTQW of a charged particle on a square lattice graph in the presence of a perpendicular uniform magnetic field ($B=0.6$). The CTQW Hamiltonian is obtained from the Peierls model.}
	\label{fig:rho_SQU_B_peierls}
\end{figure}

\begin{figure}[!h]
	\centering
	\includegraphics[width=0.45\textwidth]{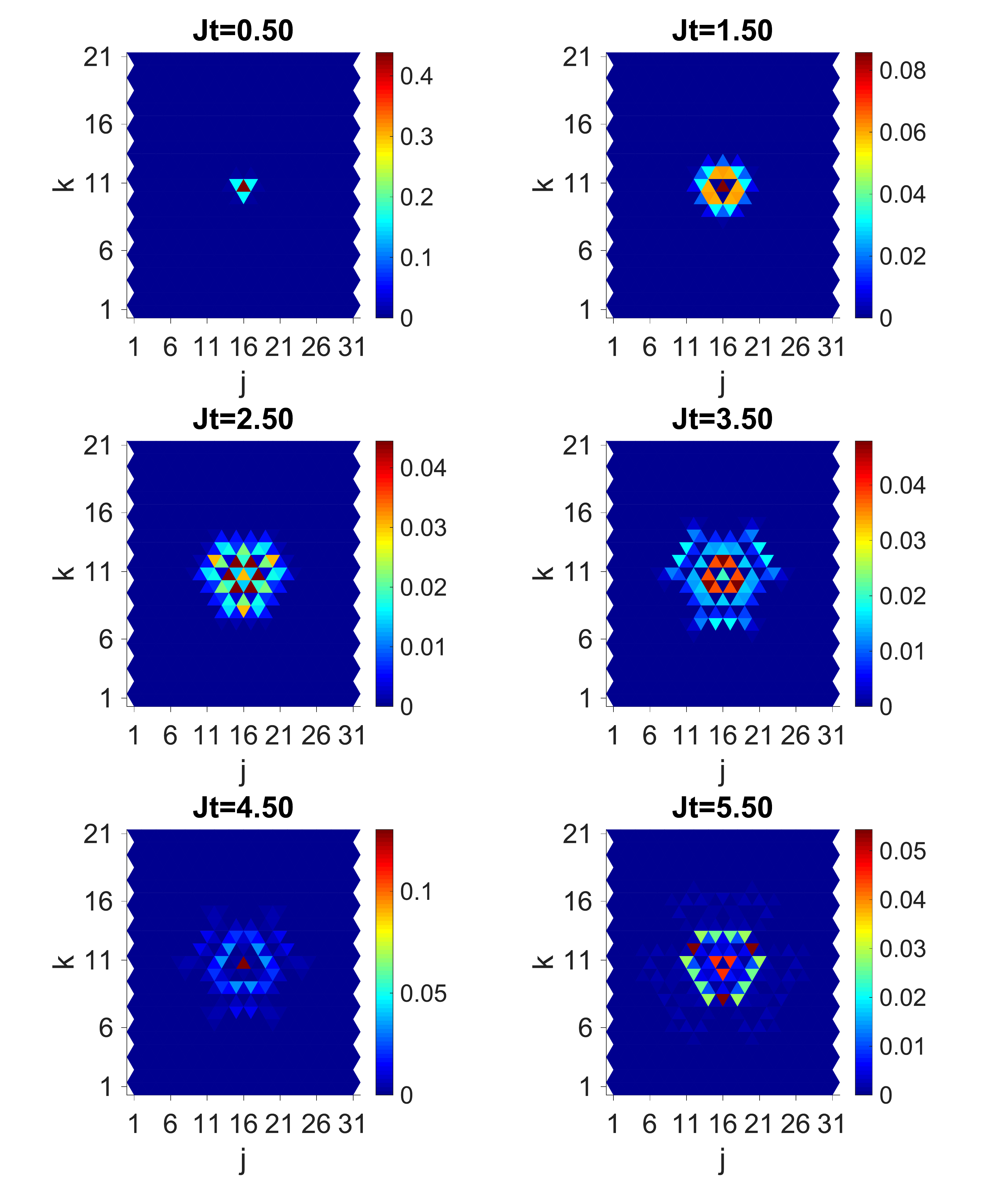}
	\caption{Map of the time evolution of the probability density according to the CTQW of a charged particle on a honeycomb lattice graph in the presence of a perpendicular uniform magnetic field ($B=0.6$). The CTQW Hamiltonian is obtained from the Peierls model.}
	\label{fig:rho_HON_B_peierls_peierls}
\end{figure}

Maps of the time evolution of the probability density are shown in Figs. \ref{fig:rho_TRI_B_peierls}--\ref{fig:rho_HON_B_peierls_peierls} and are characterized by a trade-off between the circular symmetry due to the gauge and the symmetry of the underlying lattice. In general, we observe a distribution of probability which initially spreads over the lattice, then the maxima come back towards the initial vertex and eventually move away from it. However, during the time evolution, the tails of the wavefunction continue to get away from the center of the lattice graph. Indeed, in the Peierls model, being its Hamiltonian based on the graph Laplacian, there is no term confining or limiting the spreading of the walker, since the quadratic term in $\vb{A}$ is not explicitly present but only recovered, in the continuum limit, from the Peierls phase-factors of the hopping terms.

\section{CTQW under magnetic field: the spatial discretization of the Hamiltonian}
\label{sec:spatial_disc_H_fd}

\subsection{Finite-difference formulae with Taylor expansion}
\label{subsec:fd_taylor}
Here we show how far we can go in using the Taylor expansion in order to get finite-difference formulae for differential operators, as usually done for a square lattice. The same approach has been already used in Sec. \ref{subsec:CTQW_H_free} to obtain the Laplacian in the different PLGs, so all we are left to do is to determine the discrete version of the first partial derivatives of a scalar function $f$. The idea is, again, to Taylor expand $f$ evaluated in the NN vertices about the given one $V$ and combining the resulting expansions to obtain $\partial_x f_V$ and $\partial_y f_V$ in terms of finite differences. We point out that after combining such Taylor expansions we have then to solve a system of linear equations specific for each term we are interested in: the corresponding coefficient of the linear combination will be set to 1, whereas all the others to 0. In particular, such systems consist of five equations (one condition on the coefficient of $f_V$, two on the first partial derivatives, and two on the second partial derivatives forming the Laplacian) in $\operatorname{deg}(V)$ unknowns.

\subsubsection{Square lattice graph}
To find the first partial derivatives of $f$ we recall the linear combination in Eq. \eqref{eq:lc_taylor_squ} and we impose the following systems of equations:
\begin{equation}
\renewcommand\arraystretch{1.5}
\begin{array}{lccc}
	& \text{(i) } \partial_x f_V 	&& \text{(ii) } \partial_y f_V\\\cline{2-4}
\left\lbrace
\begin{array}{l}
\alpha + \beta +\gamma +\delta\\
a(\alpha - \gamma)\\
a(\beta - \delta)\\
\frac{a^2}{2}(\alpha + \gamma)\\
\frac{a^2}{2}(\beta + \delta)
\end{array}
\right.
&
\begin{array}{ll}
=&0\\
=&1\\
=&0\\
=&0\\
=&0
\end{array}
&\text{ and }&
\begin{array}{ll}
=&0\\
=&0\\
=&1\\
=&0\\
=&0\,.
\end{array}
\end{array}
\end{equation}
\begin{enumerate}[(i)]
\item $\partial_x f_V$ is obtained from the solution of a system of five equations in four unknowns: if we consider only the last four equations, the resulting system of four equation is definite, i.e. it admits the unique solution $(\alpha, \beta, \gamma, \delta)=\left(\frac{1}{2a},0,-\frac{1}{2a},0\right)$, which also satisfies the first equation. This leads to
\begin{equation}
\partial_x f_V = \frac{1}{2a}\left(f_A-f_C \right).
\label{eq:squ_dx_taylor}
\end{equation}
\item $\partial_y f_V$ is obtained from the solution of a system of five equations in four unknowns: if we consider only the last four equations, the resulting system of four equation is definite and admits the unique solution $(\alpha, \beta, \gamma, \delta)=\left(0,\frac{1}{2a},0,-\frac{1}{2a}\right)$, which also satisfies the first equation. This leads to
\begin{equation}
\partial_y f_V = \frac{1}{2a}\left(f_B-f_D \right).
\label{eq:squ_dy_taylor}
\end{equation}
\end{enumerate}
This approach, on a square lattice graph, provides the finite-difference formulae both for the first partial derivatives, Eqs. \eqref{eq:squ_dx_taylor}--\eqref{eq:squ_dy_taylor}, and for the Laplacian, Eq. \eqref{eq:squ_lap_fd}, and these are consistent with those used in numerical analysis \cite{abramowitz1970handbook}. A point we want to stress is that the systems of equations returning the first partial derivatives are characterized by a coefficient matrix whose rank is the same as that of the augmented matrix and equal to the number of unknowns. This is the reason why we can state that solutions are unique.

\subsubsection{Triangular lattice graph}
To find the first partial derivatives of $f$ we recall the linear combination in Eq. \eqref{eq:lc_taylor_tri}. The resulting systems consist of five equations in six unknowns, hence we can not have a (unique) solution. Even if we increase the order of the Taylor expansion (in order to have systems of more equations than unknowns), the rank of the coefficient matrix turns out to be less than the number of unknowns. So, there is no way of finding a unique solution, if any. This approach, on a triangular lattice graph, can only provide the finite-difference formula of the Laplacian, Eq. \eqref{eq:tri_lap_fd}. 

\subsubsection{Honeycomb lattice graph}
To find the first partial derivatives of $f$ we recall the linear combination in Eq. \eqref{eq:lc_taylor_hon}. The resulting systems consist of five equations in three unknowns. The rank of the coefficient matrix is equal to the number of unknowns, the rank of the augmented matrix of the system for $\partial_x f_{(V,\odot)}$ is equal to the rank of the coefficient matrix, but the rank of the augmented matrix of the system for $\partial_y f_{(V,\odot)}$ is greater than the rank of the coefficient matrix. This approach, on a honeycomb lattice graph, does not provide finite-difference formulae of both first partial derivatives (only $\partial_x f_{(V,\odot)}$ is returned), whereas it does for the Laplacian, Eq. \eqref{eq:hon_h_lap_fd}.

\subsection{Conservative finite-difference methods}
\label{subsec:cons_fd_methods}
\begin{figure}[!h]
	\centering
	\includegraphics[width=0.4\textwidth]{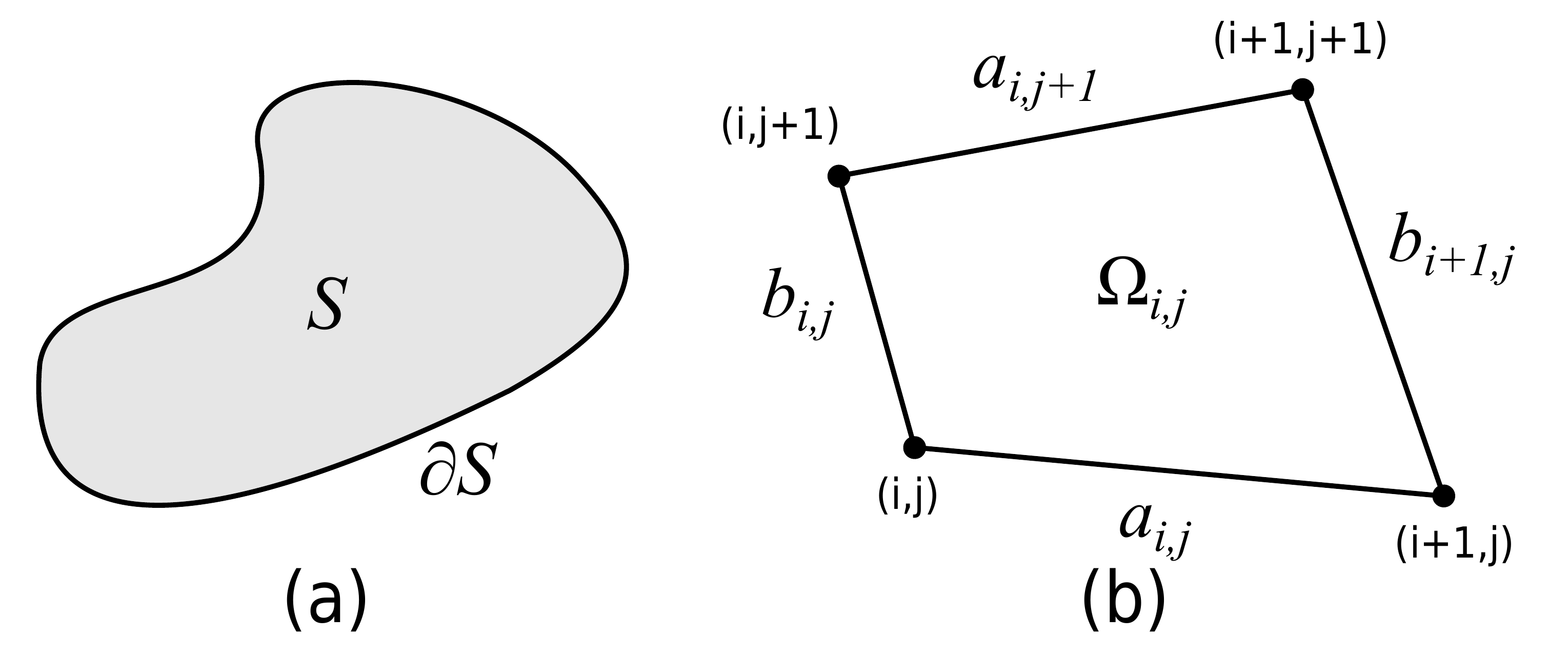}
	\caption{Continuous and discrete version of the region needed to compute the first partial derivatives of a scalar function $f$ according to Green's formulae, Eqs. \eqref{eq:green_form_dx}--\eqref{eq:green_form_dy} . (a) A region $S$ in $\mathbb{R}^2$ with boundary $\partial S$ (line); (b) a cell grid $\Omega_{i,j}$ whose boundary is the union $a_{i,j} \cup b_{i+1,j} \cup a_{i,j+1} \cup b_{i,j}$.}
	\label{fig:surface_boundary}
\end{figure}

Numerically solving problems has shown that the best results are usually obtained by using discrete models that reproduce fundamental properties of the original continuum model of the underlying physical problem, such as conservation, symmetries of the solution, etc. The development of the discrete algorithms that capture all the important characteristics of the physical problem becomes more and more difficult with the increasing complexity of the latter (number of involved physical processes, shape of the physical domain, etc.). Hence the need of having a discretization method that is sufficiently general to be applied to a wide range of physical systems. In Ref. \cite{shashkov1996conservative} it is shown how to construct, by using the \textit{support-operators method} \cite{samarskii1981operational,samarskii1982employment}, high-quality finite-difference schemes such that the resulting discrete difference operators mimic the crucial properties of the continuum differential operators, e.g. symmetry, conservation, stability, and the integral identities between the gradient, curl, and divergence. Moreover, many of the standard finite difference methods, e.g. the finite-volume methods, are special cases of the support-operators method. Unlike the former ones, the latter one can be used to construct finite-difference schemes on grids of arbitrary structure and, because invariant operators are used, the method can be easily used in any coordinate system. However, there are some points of such method differing from our constraints and purposes (see Sec. \ref{subsec:charge_ptc_B_Ham}), so that it can not directly apply to the present work:
\begin{enumerate}[(i)]
\item there are two main types of scalar functions of a discrete argument depending on the discretization adopted: \textit{nodal discretization}, where the values of the function correspond to the nodes, or \textit{cell-valued} (or \textit{cell-centered}) \textit{discretization}, where the value of a function does not correspond to a specific point in the cell but corresponds to the cell as a whole geometrical object (Fig. \ref{fig:nodal_cell_disc}). It is shown in Ref. \cite{shashkov1996conservative} that if we choose the nodal discretization for the scalar function (since we know the wavefunction on the nodes of the graph), then the difference analog, e.g., of the derivative $\partial_x$, which is the discrete operator $D_x$, acts as follows:
\begin{equation}
D_x : HN \longrightarrow HC\,,
\end{equation}
where $HN$ and $HC$ denote the spaces of discrete scalar functions according to nodal and cell-valued discretization, respectively. In other words, if we know the scalar function $f$ on the nodes of the graph, then its first partial derivatives are assigned to the cell used for the discretization as a whole, but we want them to be assigned to a node;

\begin{figure}[!h]
	\centering
	\includegraphics[width=0.45\textwidth]{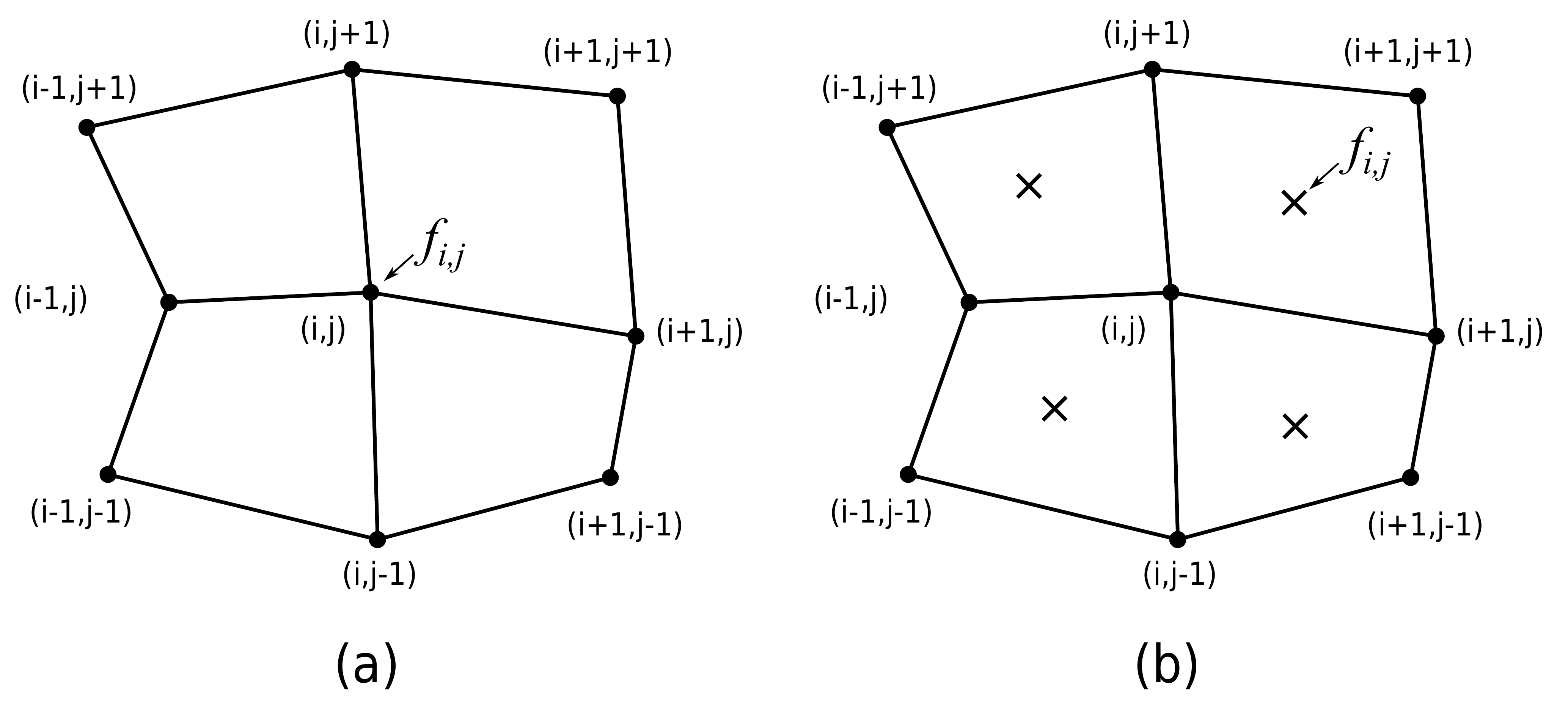}
	\caption{(a) Nodal and (b) cell-valued discretization of a scalar function $f$ in 2D.}
	\label{fig:nodal_cell_disc}
\end{figure}

\item this method involves quadrangular cells, because related to 2D logically rectangular grids (very suitable for algorithmic implementation), and it is inspired to the \textit{forward} difference method. In our case, then, we can not remap our PLGs into rectangular grids, because otherwise the resulting discrete operators would not involve all and only the NNs of a given node;

\item the Laplacian is rightly seen as the divergence of the gradient, but in terms of finite differences this means that the Laplacian is computed as difference of differences, so involving further nodes. A first cell is needed to compute the gradient of a scalar function $f\in HN$, then computing the divergence of $\nabla f$ requires the differences of the components of the latter, so the adjacent cells are involved (Fig. \ref{fig:lap_grad_nodes}). Therefore, in order to approximate the second derivative, we must construct another difference analog for the first derivative
\begin{equation}
\mathcal{D}_x : HC \longrightarrow HN\,,
\end{equation}
so that the discrete analog of the second derivative is
\begin{equation}
\mathcal{D}_x D_x : HN \longrightarrow HN\,.
\end{equation}

\begin{figure}[!h]
	\centering
	\includegraphics[width=0.4\textwidth]{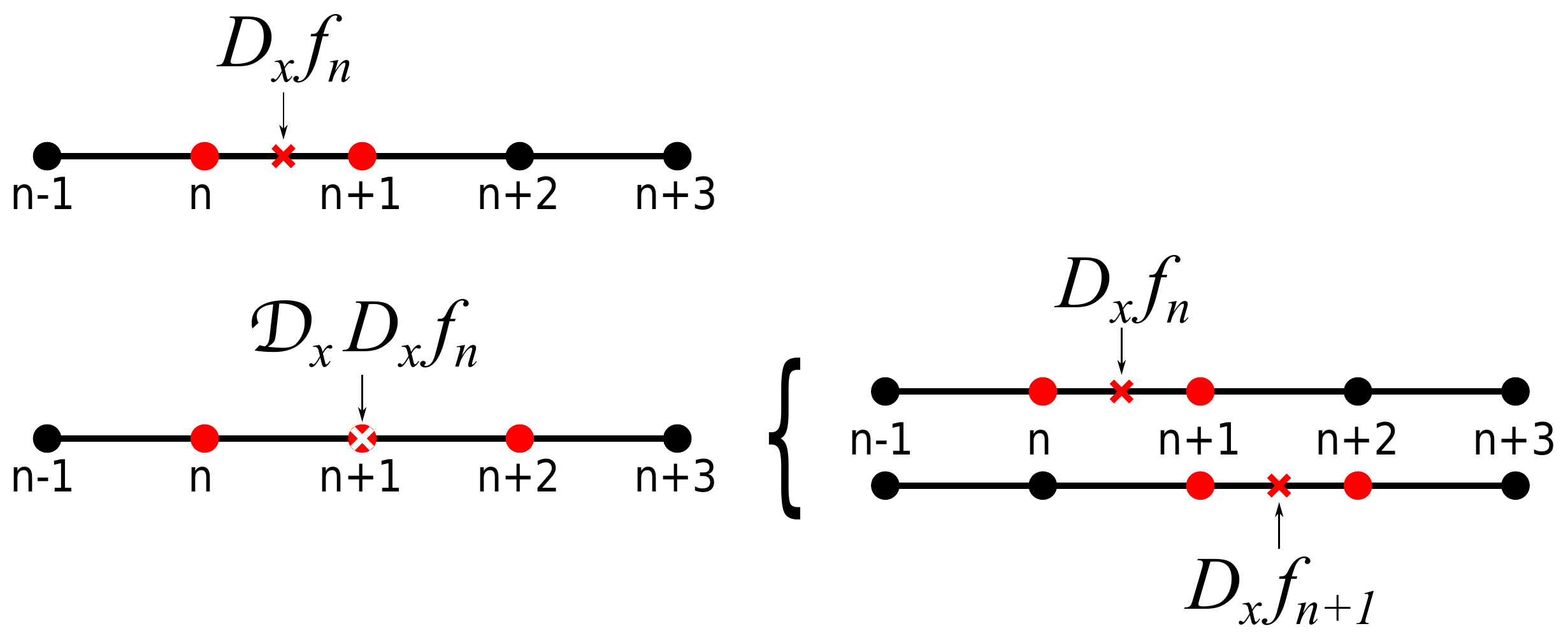}
	\caption{1D example of the different nodes involved in the computation of the discrete analog of the gradient and the Laplacian of a discrete scalar function $f\in HN$. The discrete operator $D: HN \rightarrow HC$, whereas $\mathcal{D}: HC \rightarrow HN$, so that the Laplacian $\nabla^2=\nabla\cdot\nabla$ reads $\mathcal{D}\cdot D: HN \rightarrow HN$.}
	\label{fig:lap_grad_nodes}
\end{figure}

\end{enumerate}
Despite these issues, this method provides an effective tool to compute the first partial derivatives. Green's formulae \cite{aris1962vectors}, which are the key to determine the discrete version $D=(D_x,D_y)$ of the differential operator $\nabla=(\partial_x, \partial_y)$, descend from the proof of the Green's theorem in a plane (Appendix \ref{subapp:green_th}) and read as follows:
\begin{align}
\partial_x f &=\lim_{S\rightarrow0}\frac{\oint_{\partial S}f\, \diff y}{S} \,, \label{eq:green_form_dx}\\
\partial_y f &=-\lim_{S\rightarrow0}\frac{\oint_{\partial S}f\, \diff x}{S} \,, \label{eq:green_form_dy}
\end{align}
where $S$ is some area and $\partial S$ its boundary (Fig. \ref{fig:surface_boundary}(a)).

In a discrete case the role of $S$ is played by the grid cell $\Omega_{ij}$ and therefore the boundary $\partial S$ is the union of sides $a_{i,j}$, $b_{i+1,j}$, $a_{i,j+1}$, and $b_{i,j}$ (Fig. \ref{fig:surface_boundary}(b)). For approximation of the contour integral in the RHS of Eqs. \eqref{eq:green_form_dx}--\eqref{eq:green_form_dy} we divide the contour integral into four integrals each over the corresponding side of quadrangle $\Omega_{ij}$ and for the approximate evaluation of each integral we use the trapezoidal rule. According to this, as a result, we get the following expression for the difference analog of the derivative $\partial_x f$:
\begin{align}
\left( D_x f \right)_{i,j}=&\frac{1}{\Omega_{i,j}}\left[ \frac{f_{i+1,j}+f_{i,j}}{2}(y_{i+1,j}-y_{i,j})\right.\nonumber\\
&+\frac{f_{i+1,j+1}+f_{i+1,j}}{2}(y_{i+1,j+1}-y_{i+1,j})\nonumber\\
&+\frac{f_{i,j+1}+f_{i+1,j+1}}{2}(y_{i,j+1}-y_{i+1,j+1})\nonumber\\
&\left.+\frac{f_{i,j}+f_{i,j+1}}{2}(y_{i,j}-y_{i,j+1})\right]\,,
\end{align}
where $y_{i,j}$ denotes the $y$ coordinate of the node $(i,j)$ and $\Omega_{i,j}$ is also the area of the grid cell. Notice that this area is the area of the region bounded by the contour of integration. In the same way the difference analog of the derivative $\partial_y f$ can be found.

\begin{figure}[!h]
	\centering
	\includegraphics[width=0.45\textwidth]{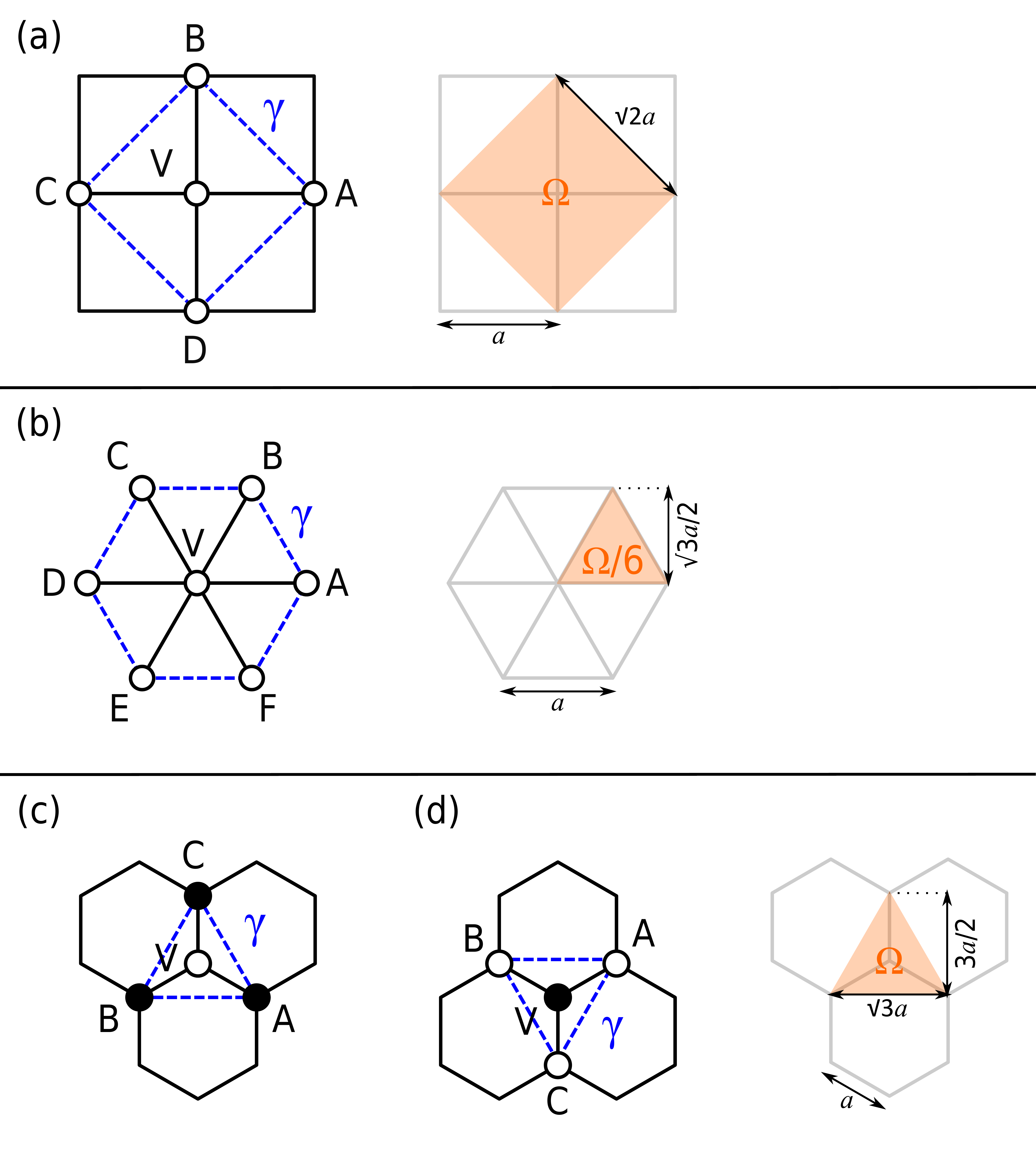}
	\caption{NNs of a vertex $V$ and closed path $\gamma$ (blue dashed line) involved in the computation of the discrete analogs of $\partial_x f_V$ and $\partial_y f_V$ in the different PLGs. The geometry of the region of area $\Omega$ (orange shade) bounded by the curve $\gamma$ is also reported. (a) Square ($\Omega=2a^2$), (b) triangular ($\Omega=\frac{3\sqrt{3}}{2}a^2$), and (c)--(d) honeycomb ($\Omega=\frac{3\sqrt{3}}{4}a^2$) lattice graph, with lattice parameter $a$.}
\label{fig:plg_discrete_area}
\end{figure}

In the present work we analogously apply the Green's formulae to our purposes, i.e. by defining a suitable closed path crossing the nodes of interest (Fig. \ref{fig:plg_discrete_area}) and then performing a discrete evaluation of the contour integral according to the trapezoidal rule. Indeed, as previously said, the first partial derivatives of a scalar function $f\in HN$ are cell-valued, hence the need of designing our approach in such a way that all the NNs of a node are involved, so that the result can be reasonably intended as node-valued. In the following, for sake of simplicity, we will denote by $\partial_x$, $\partial_y$, and $\nabla^2$ also their discrete version, and with $\Omega$ the area of the region bounded by the closed curve $\gamma$. Notice that for a PLG such area is constant, $\Omega_{i,j}=\Omega \, \forall \, (i,j)\in \mathbb{Z}^2$.

\subsubsection{Square lattice graph}
\label{subsubsec:cons_fd_squ}
A vertex $V$ has four NNs, i.e. $A$, $B$, $C$, and $D$ (see Fig. \ref{fig:plg_discrete_area}(a) and Table \ref{tab:plg_nn_coords}). We denote by 
$\gamma$ the closed path crossing such adjacent vertices and bounding a region of area $\Omega=2a^2$. According to Green's formulae, the discrete analogs of the first partial derivatives read:
\begin{align}
\partial_x f_V &= \frac{1}{2a}(f_A-f_C)\,, \label{eq:SQU_grad_x}\\
\partial_y f_V &= \frac{1}{2a}(f_B-f_D)\,,
\label{eq:SQU_grad_y}
\end{align}
since
\begin{align}
\oint_\gamma f \, \diff y &\approx a(f_A-f_C)\,,\\
\oint_\gamma f \, \diff x &\approx -a(f_B-f_D)\,.
\end{align}
The finite-difference formulae so computed are the same used in numerical analysis \cite{abramowitz1970handbook} and already seen in Eqs. \eqref{eq:squ_dx_taylor}--\eqref{eq:squ_dy_taylor}.

\subsubsection{Triangular lattice graph}
\label{subsubsec:cons_fd_tri}
A vertex $V$ has six NNs, i.e. $A$, $B$, $C$, $D$, $E$, and $F$ (see Fig. \ref{fig:plg_discrete_area}(b) and Table \ref{tab:plg_nn_coords}). We denote by 
$\gamma$ the closed path crossing such adjacent vertices and bounding a region of area $\Omega=\frac{3\sqrt{3}}{2}a^2$. According to Green's formulae, the discrete analogs of the first partial derivatives read:
\begin{align}
\partial_x f_V &= \frac{1}{6a}(2f_A+f_B-f_C-2f_D-f_E+f_F)\,,\label{eq:TRI_grad_x}\\
\partial_y f_V &= \frac{1}{2\sqrt{3}a}(f_B+f_C-f_E-f_F)\,, \label{eq:TRI_grad_y}
\end{align}
since
\begin{align}
\oint_\gamma f \, \diff y &\approx \frac{\sqrt{3}}{4}a(2f_A+f_B-f_C-2f_D-f_E+f_F)\,,\\
\oint_\gamma f \, \diff x &\approx -\frac{3}{4}a(f_B+f_C-f_E-f_F)\,.
\end{align}

\subsubsection{Honeycomb lattice graph}
\label{subsubsec:cons_fd_hon}
A vertex $V$ has three NNs, i.e. $A$, $B$, and $C$ (see Fig. \ref{fig:plg_discrete_area}(c)--(d) and Table \ref{tab:plg_nn_coords}). We denote by $\gamma$ the closed path crossing such adjacent vertices and bounding a region of area $\Omega=\frac{3\sqrt{3}}{4}a^2$. According to Green's formulae, the discrete analogs of the first partial derivatives read:
\begin{align}
\partial_x f_{(V,\odot)} &= \frac{1}{\sqrt{3}a}\left(f_{(A,\bar{\odot})}-f_{(B,\bar{\odot})}\right)\,,\label{eq:HON_grad_x}\\
\partial_y f_{(V,\odot)} &= \frac{\operatorname{sgn}(\odot)}{3a}\left(2f_{(C,\bar{\odot})}-f_{(A,\bar{\odot})}-f_{(B,\bar{\odot})}\right)\,,\label{eq:HON_grad_y}
\end{align}
since
\begin{align}
\oint_\gamma f \, \diff y &\approx \frac{3}{4}a\left(f_{(A,\bar{\odot})}-f_{(B,\bar{\odot})}\right)\,,\\
\oint_\gamma f \, \diff x &\approx -\operatorname{sgn}(\odot)\frac{\sqrt{3}}{4}a\left(2f_{(C,\bar{\odot})}-f_{(A,\bar{\odot})}-f_{(B,\bar{\odot})}\right)\,.
\end{align}

\subsection{The CTQW Hamiltonian}
Whereas the Peierls phase-factors are a suitable solution to our problem, the issue about the spatial discretization of the Hamiltonian satisfying our assumptions is still open, in particular in the triangular and honeycomb lattice graph. If on the one hand the finite difference formulae from Taylor expansion are well-behaved only in the square lattices and ill-defined in the other PLGs, on the other hand this approach returns, for all the PLGs, a discrete Laplacian which is analogous to the graph one. The discrete first partial derivatives, instead, are provided by the discretization of the Green's formulae. Following the latter approach, the discrete Laplacian is given by the divergence of the gradient, i.e. as a finite difference of finite differences, thus involving next NNs of a given vertex. This point is at odds with our assumption of hopping only to NNs (Sec. \ref{subsec:charge_ptc_B_Ham}), because the kinetic term of the free particle would be accountable for the hopping up to next NNs, whereas the orbital paramagnetic term for the hopping only to NNs. In view of these results, we therefore suggest a hybrid method which combines the above mentioned results: the discrete first partial derivatives are provided by conservative finite-difference methods (Sec. \ref{subsec:cons_fd_methods}), whereas the Laplacian by finite difference formulae from Taylor expansion (Sec. \ref{subsec:fd_taylor}). According to this approach, we spatially discretize Eq. \eqref{eq:Ham_charged_ptc_cont} in order to obtain the Hamiltonian describing the CTQW of a charged particle on the different PLGs in the presence of a perpendicular magnetic field. As we are going to see below, what we obtain is reminiscent of the Peierls model (Sec. \ref{sec:peierls_model}). Indeed, the hopping terms can be regarded as the first-order Taylor expansion of the Peierls phase-factors, but now the diagonal elements of the Hamiltonian matrix, i.e. the on-site terms, also include the quadratic term in $\vb{A}$.

\subsubsection{Square lattice graph}
\label{subsubsec:hyb_squ}
With reference to Table \ref{tab:plg_nn_coords}, according to the Laplacian in Eq. \eqref{eq:squ_lap_fd} and the first partial derivatives in Eqs. \eqref{eq:SQU_grad_x}--\eqref{eq:SQU_grad_y}, the resulting CTQW Hamiltonian reads:
\begin{align}
\mathcal{\hat{H}}=&-J_S\sum_V\left\lbrace\vphantom{\frac{q^2}{\hbar^2}}\left[1+i\frac{qa}{2\hbar}\left(A_V^x+A_A^x\right)\right]\dyad{A}{V}\right.\nonumber\\
&+\left[1+i\frac{qa}{2\hbar}\left(A_V^y+A_B^y\right)\right]\dyad{B}{V}\nonumber\\
&+\left[1-i\frac{qa}{2\hbar}\left(A_V^x+A_C^x\right)\right]\dyad{C}{V}\nonumber\\
&+\left[1-i\frac{qa}{2\hbar}\left(A_V^y+A_D^y\right)\right]\dyad{D}{V}\nonumber\\
&\left. -\left[4+ \frac{q^2a^2}{\hbar^2}\left({A_V^x}^2+{A_V^y}^2\right)\right]\dyad{V}{V}\right\rbrace\,,
\label{eq:H_SQU_plg}
\end{align}
where $J_S$ is defined in Eq. \eqref{eq:J_SQU_plg}.

\subsubsection{Triangular lattice graph}
\label{subsubsec:hyb_tri}
With reference to Table \ref{tab:plg_nn_coords}, according to the Laplacian in Eq. \eqref{eq:tri_lap_fd} and the first partial derivatives in Eqs. \eqref{eq:TRI_grad_x}--\eqref{eq:TRI_grad_y}, the resulting CTQW Hamiltonian reads:
\begin{align}
\mathcal{\hat{H}}=&-J_T\sum_V\left\lbrace\vphantom{\frac{q^2}{\hbar^2}}\left[1+i\frac{qa}{2\hbar}\left(A_V^x+A_A^x\right)\right]\dyad{A}{V}\right.\nonumber\\
&+\left[1+i\frac{qa}{4\hbar}\left(A_V^x+A_B^x+\sqrt{3}\left(A_V^y+A_B^y\right)\right )\right]\dyad{B}{V}\nonumber\\
&+\left[1-i\frac{qa}{4\hbar}\left(A_V^x+A_C^x-\sqrt{3}\left (A_V^y+A_C^y\right)\right )\right]\dyad{C}{V}\nonumber\\
&+\left[1-i\frac{qa}{2\hbar}\left(A_V^x+A_D^x\right)\right]\dyad{D}{V}\nonumber\\ 
&+\left[1-i\frac{qa}{4\hbar}\left(A_V^x+A_E^x+\sqrt{3}\left(A_V^y+A_E^y\right)\right )\right]\dyad{E}{V}\nonumber\\ 
&+\left[1+i\frac{qa}{4\hbar}\left(A_V^x+A_F^x-\sqrt{3}\left (A_V^y+A_F^y\right)\right )\right]\dyad{F}{V}\nonumber\\
&\left.-\left[6+\frac{3q^2a^2}{2\hbar^2}\left({A_V^x}^2+{A_V^y}^2\right)\right]\dyad{V}{V}\right\rbrace\,,
\label{eq:H_TRI_plg}
\end{align}
where $J_T$ is defined in Eq. \eqref{eq:J_TRI_plg}.

\subsubsection{Honeycomb lattice graph}
\label{subsubsec:hyb_hon}
With reference to Table \ref{tab:plg_nn_coords}, according to the Laplacian in Eq. \eqref{eq:hon_h_lap_fd} and the first partial derivatives in Eqs. \eqref{eq:HON_grad_x}--\eqref{eq:HON_grad_y}, the resulting CTQW Hamiltonian reads:
\begin{align}
\hat{\mathcal{H}}= &-J_H\sum_{\odot\in\{\circ,\bullet\}}\sum_{(V,\odot)} \left[\vphantom{\sum}h_{AV} \dyad{A,\bar{\odot}}{V,\odot}  \right.\nonumber\\
&+ h_{BV} \dyad{B,\bar{\odot}}{ V,\odot}+ h_{CV} \dyad{C,\bar{\odot}}{V,\odot}\nonumber\\
&\left.\vphantom{\sum} +h_{VV} \dyad{V,\odot}{V,\odot} \right] \,,
\label{eq:H_HON_plg}
\end{align}
where $J_H$ is defined in Eq. \eqref{eq:J_HON_plg}, and we have defined:
\begin{align}
h_{AV}:=&1+i\frac{qa}{4\hbar}\left[ \sqrt{3}\left(A_{(V,\odot)}^x+A_{(A,\bar{\odot})}^x\right)\right.\nonumber\\
&\left. - \operatorname{sgn}(\odot)\left(A_{(V,\odot)}^y+A_{(A,\bar{\odot})}^y\right)\right]\,,\label{eq:AV_hon_fd}\\
h_{BV}:=&1-i\frac{qa}{4\hbar}\left[ \sqrt{3}\left(A_{(V,\odot)}^x+A_{(B,\bar{\odot})}^x\right)\right.\nonumber\\
&\left.+\operatorname{sgn}(\odot)\left(A_{(V,\odot)}^y+A_{(B,\bar{\odot})}^y\right)\right]\,,\\
h_{CV}:=&1+ \operatorname{sgn}(\odot)i\frac{qa}{2\hbar}\left( A_{(V,\odot)}^y+A_{(C,\bar{\odot})}^y\right)\,,\\
h_{VV}:=&-\left[3+ \frac{3q^2a^2}{4\hbar^2}\left({A^x}_{(V,\odot)}^2+{A^y}_{(V,\odot)}^2\right)\right]\,.\label{eq:VV_hon_fd}
\end{align}

\subsection{Numerical simulation: results}
\label{subsec:results_magn_fd}
\begin{figure}[!h]
	\centering
	\includegraphics[width=0.45\textwidth]{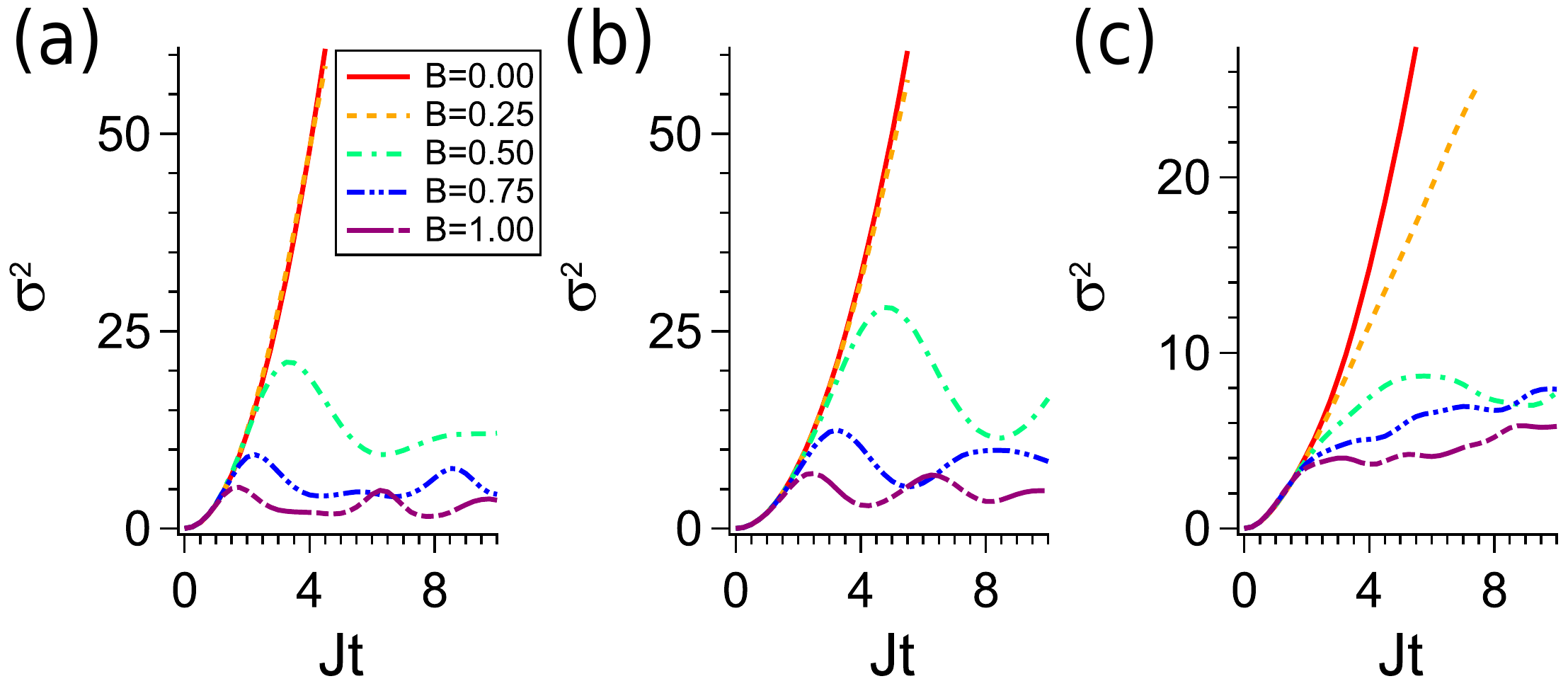}
	\caption{Variance of the space coordinates obtained in a CTQW of a charged particle in the (a) triangular, (b) square, and (c) honeycomb lattice graph for increasing values of the modulus $B$ of the perpendicular uniform magnetic field. As the latter increases, the variance deviates from the curve of the free particle. The stronger the magnetic field, the smaller the variance and with a higher frequency the \textit{pseudo}-oscillations. The variance of the two spatial coordinates is equal, $\sigma_x^2(t)=\sigma_y^2(t)$. The CTQW Hamiltonian is obtained from the spatial discretization of Eq. \eqref{eq:Ham_charged_ptc_cont}.}
	\label{fig:var_B_2_zero}
\end{figure}

The behavior of the variance of the space coordinates is shown in Fig. \ref{fig:var_B_2_zero}, and it is the same for both the $x$ and $y$ coordinate, i.e. $\sigma_{x}^2(Jt)=\sigma_{y}^2(Jt)$. We observe that, as the modulus of the magnetic field increases, the curve of the variance of the space coordinates deviates from that of the free particle, it shows a maximum which lowers, and a oscillation having increasing frequency. This is more evident in the square and triangular lattice graph than in the honeycomb one.

\begin{figure}[!h]
	\centering
	\includegraphics[width=0.45\textwidth]{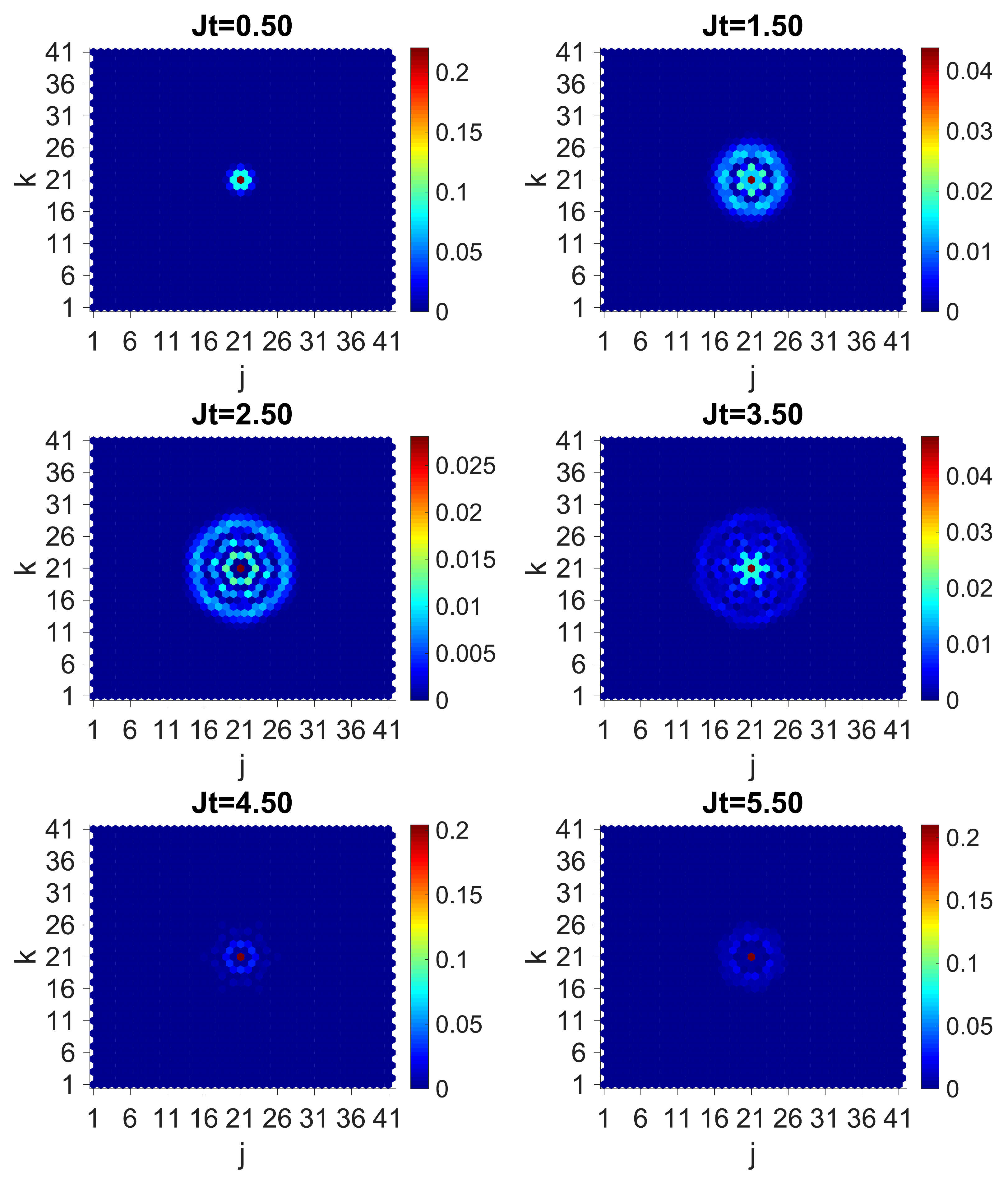}
	\caption{Map of the time evolution of the probability density according to the CTQW of a charged particle on a triangular lattice graph in the presence of a perpendicular uniform magnetic field ($B=0.6$). The CTQW Hamiltonian is obtained from the spatial discretization of Eq. \eqref{eq:Ham_charged_ptc_cont}.}
	\label{fig:rho_TRI_B}
\end{figure}

\begin{figure}[!h]
	\centering
	\includegraphics[width=0.45\textwidth]{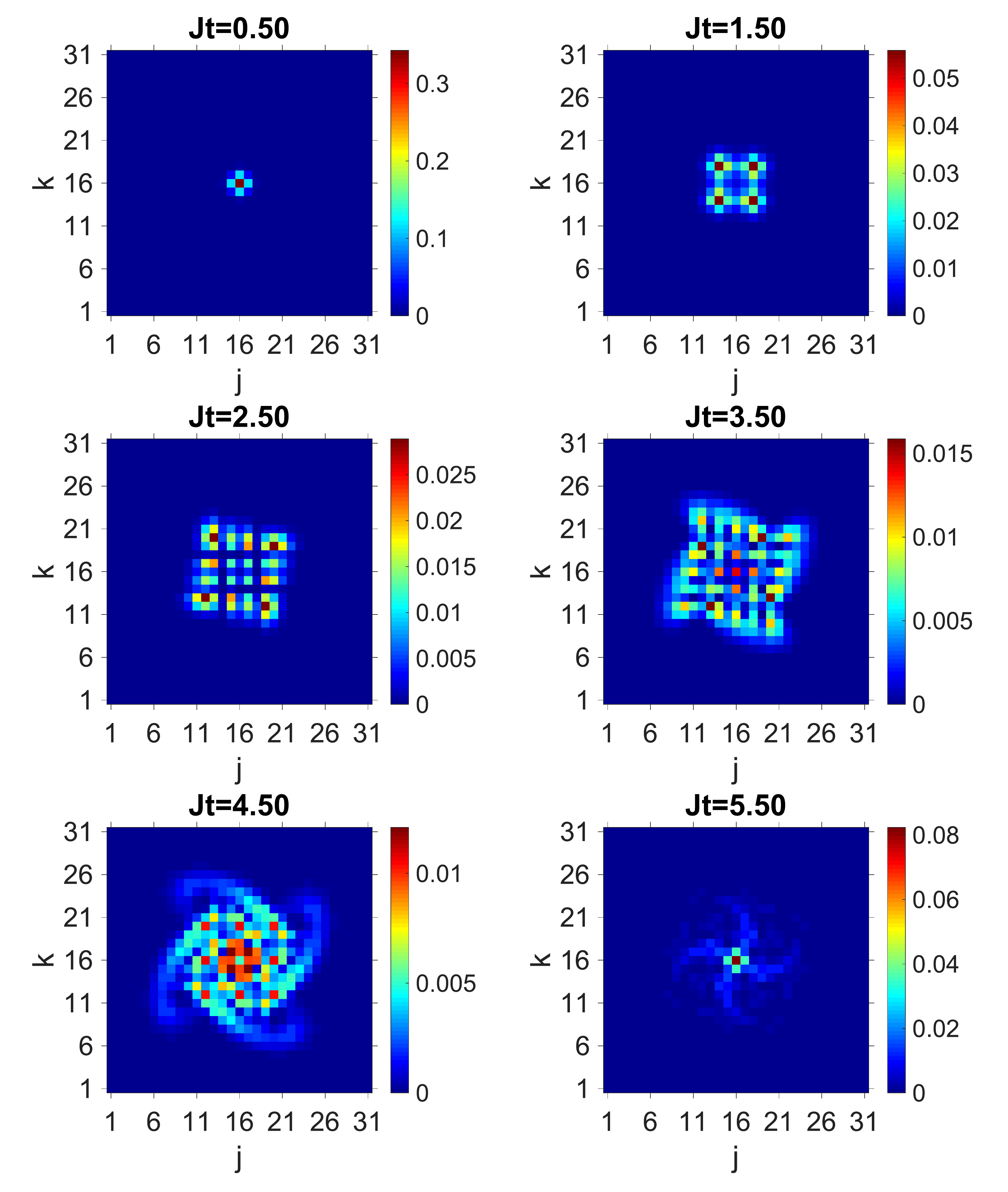}
	\caption{Map of the time evolution of the probability density according to the CTQW of a charged particle on a square lattice graph in the presence of a perpendicular uniform magnetic field ($B=0.6$). The CTQW Hamiltonian is obtained from the spatial discretization of Eq. \eqref{eq:Ham_charged_ptc_cont}.}
	\label{fig:rho_SQU_B}
\end{figure}

\begin{figure}[!h]
	\centering
	\includegraphics[width=0.45\textwidth]{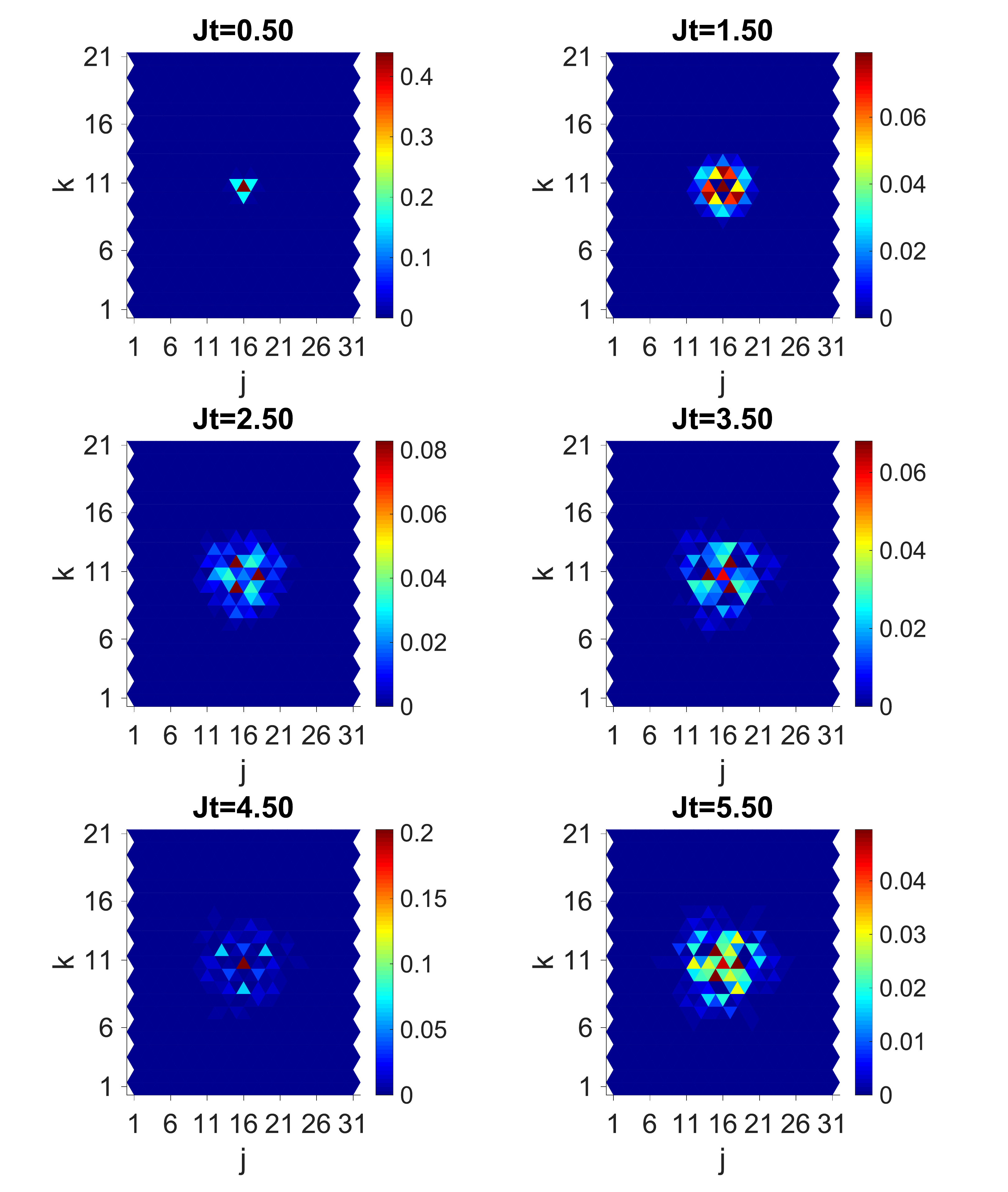}
	\caption{Map of the time evolution of the probability density according to the CTQW of a charged particle on a honeycomb lattice graph in the presence of a perpendicular uniform magnetic field ($B=0.6$). The CTQW Hamiltonian is obtained from the spatial discretization of Eq. \eqref{eq:Ham_charged_ptc_cont}.}
	\label{fig:rho_HON_B}
\end{figure}

Maps of the time evolution of the probability density are shown in Figs. \ref{fig:rho_TRI_B}--\ref{fig:rho_HON_B}. We observe that CTQWs on PLGs are characterized by an oscillating (spiral) probability density which arises from the trade-off among the symmetry of the lattice, the rotational symmetry of the Hamiltonian in the continuum and the harmonic oscillator behind the latter. Moreover the probability density, in time, seems to rotate, mimicking the effects of the Lorentz force (this is particularly evident on the square lattice graph, Fig. \ref{fig:rho_SQU_B}).

\begin{figure}[!h]
	\centering	
	\includegraphics[width=0.45\textwidth]{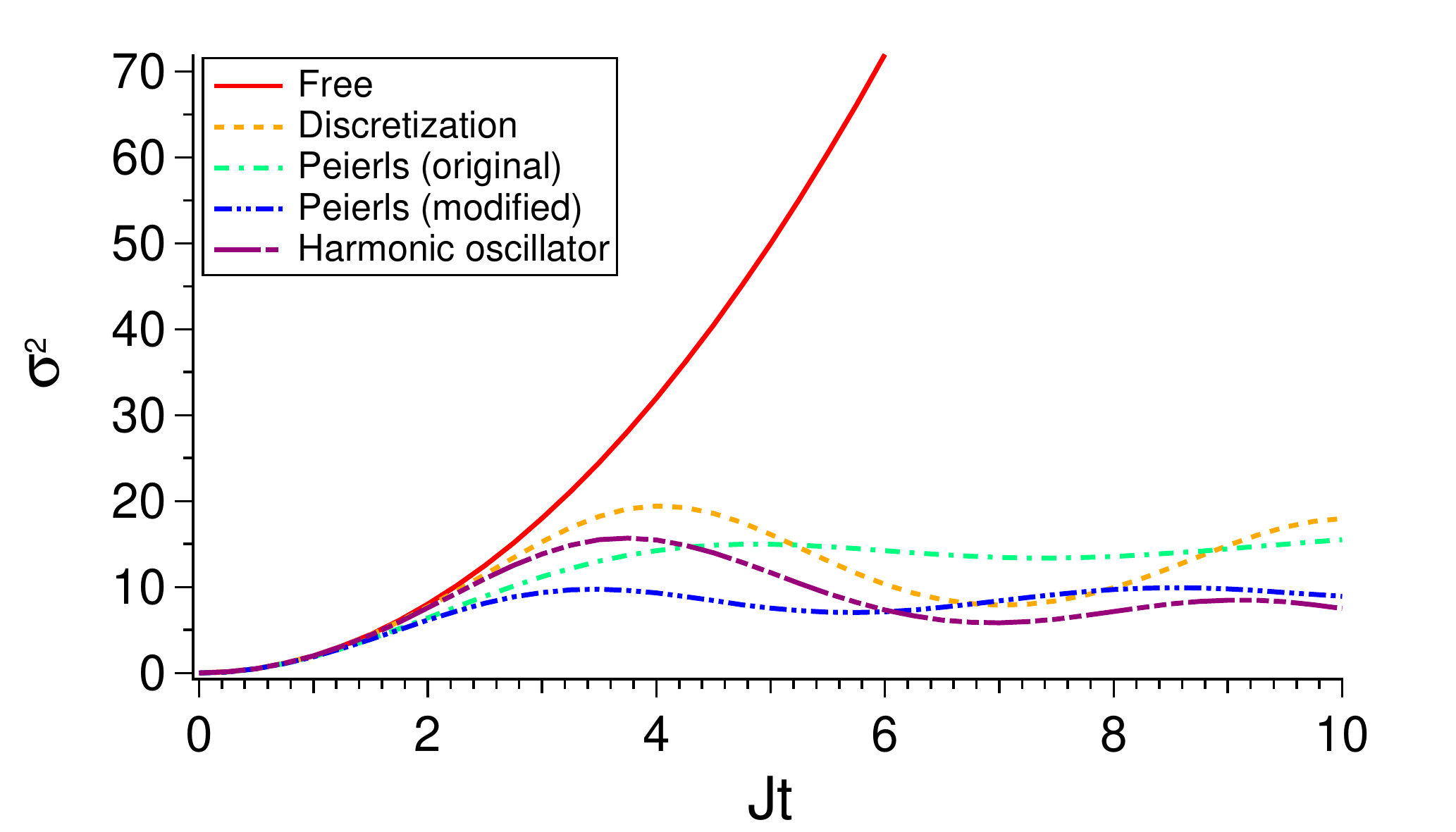}
	\caption{Comparison of the variance of the space coordinates obtained in a CTQW of a charged particle on a square lattice graph in the presence of a perpendicular uniform magnetic field ($B=0.6$) according to different models: `Peirls (original)', the model according to which the presence of the magnetic field is encoded in the Peierls phase-factors describing the hopping terms; `Peirls (modified)', a correction to the Peirls model by including in the Hamiltonian the on-site energies due to the quadratic term $q^2\vb{A}^2/2m$; `Discretization', the spatial discretization of the Hamiltonian of the corresponding system in the continuum; `Harmonic oscillator', the CTQW of a particle in a harmonic potential equivalent to the quadratic term $q^2\vb{A}^2/2m$ ($\vb{A}$ in the symmetric gauge).
	The curve of the free particle is reported as reference. The variance of the two spatial coordinates is equal, $\sigma_x^2(Jt)=\sigma_y^2(Jt)$.}
	\label{fig:var_SQU_models}
\end{figure}

Clues of the harmonic oscillator behind the Hamiltonian in the continuum are revealed by the maps of the time evolution of the probability density, alternating phases of expansion and contraction, and by the variance of the space coordinates, alternating local maxima and minima which become more frequent for increasing magnetic field (a reminiscence of the cyclotron frequency). However, the observed behavior is not exactly oscillating and periodic because of the spatial discretization. Even in the Peirls model we observe something similar, e.g. the probability distribution has an expansion and then a contraction, but this model is based on the Hamiltonian of the free particle, and indeed the tails of the wavefunction continue to move away. Instead, when spatially discretizing Eq. \eqref{eq:Ham_charged_ptc_cont}, the quadratic term in $\vb{A}$ - absent in the Peirls model - is here explicitly present. In the symmetric gauge such term reads as
\begin{equation}
\frac{q^2}{2m}\vb{A}^2=\frac{q^2B^2}{8m}\left [(x-x_c)^2+(y-y_c)^2\right ]\,,
\label{eq:quadratic_A}
\end{equation}
i.e. it is a 2D harmonic potential. The role of this term is clearer in Fig. \ref{fig:var_SQU_models}, where we report the variance of the space coordinate for a CTQW on a square lattice according to different models: the original Peierls model (Sec. \ref{sec:peierls_model}), i.e. the presence of the magnetic field is encoded in the Peierls phase-factors describing the hopping terms; then we correct such model by including in the Hamiltonian the on-site energies due to Eq. \eqref{eq:quadratic_A}; the spatial discretization of the Hamiltonian of the corresponding system in the continuum (Sec. \ref{sec:spatial_disc_H_fd}); in the end, we consider the CTQW of a particle in a harmonic potential equivalent to Eq. \eqref{eq:quadratic_A}.

\section{Conclusions}
\label{sec:conclusions}
In the present work, we have studied the continuous-time quantum walks (CTQWs) of a charged particle in the presence of a perpendicular uniform magnetic field on planar lattice graphs (PLGs), i.e. graphs possessing a drawing whose embedding in a Euclidean plane forms a regular tessellation (triangular, square, and honeycomb lattice graphs). Because of the strict connection between the generator of the evolution of the CTQW and the Hamiltonian, the straightforward approach to get a CTQW Hamiltonian has been to spatially discretize the Hamiltonian of the corresponding system in the continuum. Then we have numerically simulated the CTQWs in order to study the time-evolution of the probability density and variance of the spatial coordinates of a walker initially localized in the center of the PLG.

\paragraph*{CTQW of the free particle.} As a reference, we have first considered the CTQW of the free particle, whose Hamiltonian only consists of the kinetic term. In the vertex states basis, we have spatially discretized the Laplacian by means of finite difference formulae derived from Taylor expansion. The reason for this, even though the discrete Laplacian so obtained has turned out to be analogous to the graph Laplacian, is that this approach allows us to actually take into account the underlying geometry of the PLG. Indeed, by changing the degree of a vertex we expect the hopping amplitude to change accordingly: while for the graph Laplacian, because of its definition in terms of adjacency and diagonal degree matrices, there is no computation telling us how the hopping amplitude changes in the different PLGs, using Taylor expansion is a `constructive' way to determine the discrete Laplacian and the resulting hopping amplitude depends on the PLG. From the numerical simulations of the CTQWs of the free particle, the first result is that the variance of the two spatial coordinates is equal, $\sigma_x^2(t)=\sigma_y^2(t)=:\sigma^2(t)$, and it can be related to the degree of the generic vertex in the different PLGs: $\sigma_H^2(t)\leq \sigma_S^2(t) \leq \sigma_T^2(t)$ and in the honeycomb lattice graph $\operatorname{deg}(V)=3$, in the square lattice graph $\operatorname{deg}(V)=4$, and in the triangular lattice graph $\operatorname{deg}(V)=6$. An analogous behavior has been observed for the quantum coherence.
The second, but more relevant result is that there exist limits to the envisaged universal ballistic spreading for both 1D and 2D QWs: on the square and triangular lattice graphs (Bravais lattices, thus characterized by discrete translation invariance) we have observed a variance of the space coordinates $\sigma^2(t)\propto t^2$ (\textit{ballistic} spreading), whereas on the honeycomb lattice graph (non-Bravais lattice) $\sigma^2(t)\propto t^p$, with $1<p<2$ (\textit{sub-ballistic} spreading, because neither ballistic, $p=2$, nor diffusive, $p=1$). The hypothesis that the underlying reason is to be found in the presence or not of discrete translation invariance is further corroborated by the fact that we have observed an analogous \textit{sub-ballistic} spreading on another 2D non-Bravais lattice, the truncated square tiling, which consists of octagons and squares. After all, the \textit{ballistic} spreading has been usually proved for QWs on a line or on a $n$-D hypercube and in the latter the walker moves one unit in each dimension, thus it clearly reproduces the results on the line, because it is like the QW is taking place on $n$ orthogonal lines (dimensions).

\paragraph*{CTQW under magnetic field.} Then we have turned on the perpendicular uniform magnetic field, considering the vector potential in the symmetric gauge. In such gauge, the Hamiltonian in the continuum is known to be the Hamiltonian of a 1D harmonic oscillator, having degenerate energy levels (the so-called \textit{Landau levels}) and cyclotron frequency $\omega_0 = qB/m$. This choice of gauge breaks translational symmetry in both the $x$ and the $y$ directions, but it does preserve rotational symmetry, i.e. $[\hat{\mathcal{H}},\hat{L}_z]=0$.  Under the assumption of a hopping to nearest neighbors (NN), we have addressed the definition of the CTQW Hamiltonian in the presence of a magnetic field in two ways:
\begin{enumerate}[(i)]
\item Peierls model, i.e. the tunneling matrix elements of the free-particle Hamiltonian are now accompanied by complex Peierls phase-factors due to the vector potential. To our knowledge, this is the way the discrete-time QWs under artificial magnetic field have been studied.
\item Spatial discretization of the Hamiltonian of a spinless charged particle in the presence of a magnetic field. Since the linear momentum is present both at the first and second order, we have faced the non-trivial issue of determining the finite difference formulae approximating the first partial derivatives (from Green's formulae) and the Laplacian (from Taylor expansion, already obtained for the free particle) in the different PLGs.
\end{enumerate}
Again, the first result is that $\sigma_x^2(t)=\sigma_y^2(t)=:\sigma^2(t)$. In both cases we have observed that the variance of the space coordinates lowers as the modulus $B$ of the magnetic field increases and, as expected, we have found more circular structures in the probability density of the walker when allowing the PLG to better follow the rotational symmetry of the Hamiltonian: the triangular lattice graph, $\operatorname{deg}(V)=6$, provides the best discrete approximation of a circle among the PLGs, while the honeycomb lattice graph, $\operatorname{deg}(V)=3$, the worst one. In particular, the maps of the time evolution of the probability density are characterized by a trade-off between the circular symmetry due to the gauge and the symmetry of the underlying lattice. Apart from these qualitatively common features, as soon as we let the CTQW to evolve longer the difference between the two methods shows up and the quadratic term in the vector potential $q^2\vb{A}^2/2m$ plays a crucial role in it.
In the Peierls model the walker initially spreads over the lattice, then the maxima of probability density come back towards the initial vertex and eventually move away from it (as revealed also by the variance). However, during the time evolution, the tails of the wavefunction continue to get away from the center of the lattice graph: being this CTQW Hamiltonian based on the graph Laplacian, there is no term confining or limiting the spreading of the walker, since the quadratic term in $\vb{A}$ is not explicitly present but only recovered, in the continuum limit, from the Peierls phase-factors of the hopping terms. Such term, in the symmetric gauge, plays the role of a 2D harmonic potential. It is explicitly present when spatially discretizing the original Hamiltonian in the continuum and it affects the diagonal elements of the Hamiltonian matrix, i.e. the on-site terms. According to this method, clues of the harmonic oscillator behind the Hamiltonian in the continuum are revealed by the maps of the time evolution of the probability density, alternating phases of expansion and contraction, and by the variance of the space coordinates, alternating local maxima and minima which become more and more frequent for increasing magnetic field (a reminiscence of the cyclotron frequency). However, the observed behavior is not exactly oscillating and periodic because of the spatial discretization. Another difference from the Peierls model is that the probability density, in time, seems to rotate, mimicking the effects of the Lorentz force (this is particularly evident on the square lattice graph).

\acknowledgments
The authors thank Luca Ghirardi for useful discussions. PB and MGAP are members of GNFM-INdAM.

\appendix
\section{Mathematical tools and discretization of the space}
\label{app:math_discretization}

\subsection{Green’s theorem in a plane}
\label{subapp:green_th}
Suppose the functions $P(x,y)$, $Q(x, y)$ and their partial derivatives are single-valued, finite and continuous inside and on the boundary $C$ of some simply connected region $R$ in the $xy$ plane. Green’s theorem in a plane (also known as the `divergence theorem in 2D') then states
\begin{equation}
\oint_C \left( P \diff x + Q \diff y \right) = \iint_R \left( \partial_x Q - \partial_y P\right) \diff x \diff y\,,
\end{equation}
and so relates the line integral around $C$ to a double integral over the enclosed region $R$ \cite{riley2006mathematical}.

\begin{figure}[!h]
	\centering
	\includegraphics[width=0.25\textwidth]{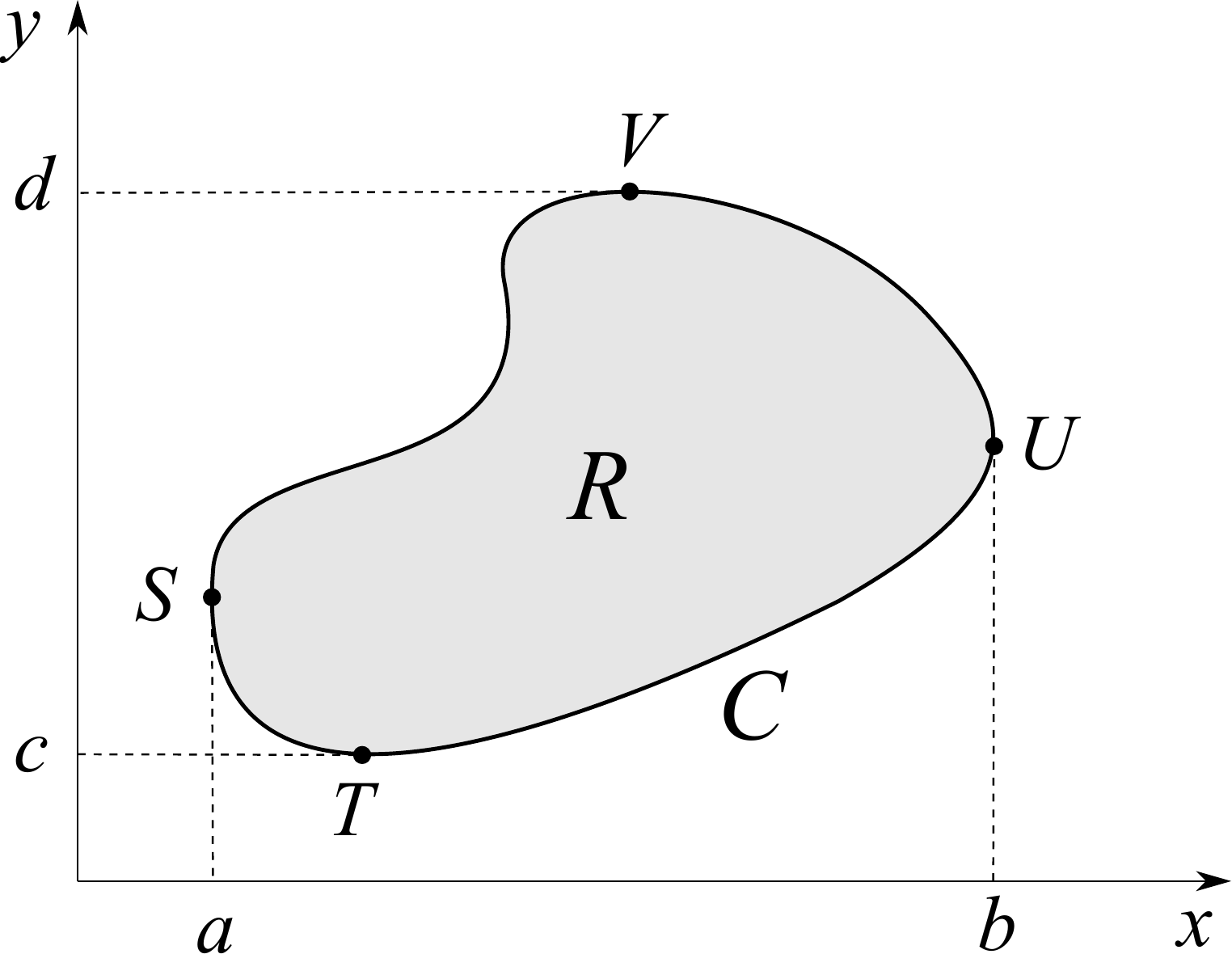}
	\caption{A simply connected region $R$ bounded by the curve $C$.}
\label{fig:green_th}
\end{figure}

\begin{proof}
Consider the simply connected region $R$ in Fig. \ref{fig:green_th}, and let $y = y_1(x)$ and $y = y_2(x)$ be the equations of the curves $STU$ and $SVU$ respectively. We then write
\begin{align*}
\iint_R \partial_y P \diff x \diff y &= \int_a^b \diff x \int_{y_1(x)}^{y_2(x)} \diff y \, \partial_y P \\
&= \int_a^b \diff x \left[ P(x,y) \right]_{y = y_1(x)}^{y = y_2(x)}\\
&=\int_a^b \left[ P(x,y_2(x)) -P(x,y_1(x)) \right]  \diff x\\
& = - \oint_C P \diff x\,,
\end{align*}
where the last equality follows from $\int_a^b P(x,y_2(x)) \diff x \allowbreak = -\int_b^a P(x,y_2(x)) \diff x$. If we now let $x = x_1(y)$ and $x = x_2(y)$ be the equations of the curves $TSV$ and $TUV$ respectively, we can similarly show that
\begin{align*}
\iint_R \partial_x Q \diff x \diff y &= \int_c^d \diff y \int_{x_1(y)}^{x_2(y)} \diff x \, \partial_xQ \\
&= \int_c^d \diff y \left[ Q(x,y) \right]_{x = x_1(y)}^{x = x_2(y)}\\
&=\int_c^d \left[ Q(x_2(y),y) -Q(x_1(y),y) \right] \diff y\\
&= \oint_C Q \diff y\,,
\end{align*}
where the last equality follows from $-\int_c^d Q(x_1(y),y) \diff y \allowbreak =  \int_d^c Q(x_1(y),y) \diff y$. Subtracting these two results gives Green’s theorem in a plane.
\end{proof}

\subsection{Spatial discretization}
\label{subapp:spatial_disc}
Let us consider a particle in a plane: the spatial discretization is accomplished by reducing the Euclidean plane into a square lattice of lattice parameter $a$. Sites of the lattice correspond to positions $(x_j,y_k)=(ja,ka)$, where $j,k\in \mathbb{Z}$, and so each site can be labeled by the couple of indices $(j,k)$. The Hilbert space of such discretized system can be obtained as
\begin{equation}
\mathscr{H}=\mathscr{H}_x \otimes \mathscr{H}_y\,,
\end{equation}
with $\dim (\mathscr{H}) = \dim (\mathscr{H}_x) \times \dim (\mathscr{H}_y)$, i.e. as a tensor product of two Hilbert subspaces - $\mathscr{H}_x$ and $\mathscr{H}_y$ - describing the states of the particle within 1D orthogonal lattices. The basis of each Hilbert subspace is provided, e.g., by the complete set of eigenstates of the position within the corresponding 1D lattice: $\{\vert j \rangle\}_{j=1,\ldots,\dim (\mathscr{H}_x)}$ for  $\mathscr{H}_x$ and $\{\vert k \rangle\}_{k=1,\ldots,\dim (\mathscr{H}_y)}$ for $\mathscr{H}_y$. Then, the basis of the resulting Hilbert space $\mathscr{H}$ is
\begin{equation}
\{\vert j,k \rangle = \vert j \rangle \otimes \vert k \rangle\}_{j,k}\,,
\end{equation}
according to which we outline in Table \ref{tab:spatial_disc_cont} the discrete version of the basic relations involving a generic abstract state and position eigenstates in the position space.

\begin{table*}[!t]
\renewcommand\arraystretch{1.75}
\center
\begin{ruledtabular}
\begin{tabular}{llll}
& \textbf{Continuum} & & \textbf{Lattice}\\\hline
\textit{Position eigenstate} & $\ket{x,y}$ & $\longrightarrow$ & $\ket{j,k}$\\
\textit{Wavefunction} & $\braket{x,y}{\psi}= \psi(x,y)$& $\longrightarrow$ & $\braket{j,k}{\psi} = \psi_{j,k}$\\
\textit{Orthonormality} & $\braket{x',y'}{x,y}= \delta(x'-x)\delta(y'-y)$ & $\longrightarrow$ & $\braket{j',k'}{j,k}= \delta_{j'j}\delta_{k'k}$\\
\textit{Completeness relation} & $\int_{-\infty}^{+\infty}\int_{-\infty}^{+\infty} \dyad{x,y} \diff x \diff y=1$ & $\longrightarrow$ & $\sum_{(j,k)\in \mathbb{Z}^2} \dyad{j,k}=1$\\
\textit{Expansion of a state} &$\ket{\psi} = \int_{-\infty}^{+\infty}\int_{-\infty}^{+\infty} \psi(x,y)\ket{x,y} \diff x \diff y$ & $\longrightarrow$ & $\ket{\psi} =\sum_{(j,k)\in \mathbb{Z}^2} \psi_{j,k}\ket{j,k} $
\end{tabular}
\end{ruledtabular}
\caption{Discrete version of the basic relations involving a generic abstract state $\ket{\psi}$ and position eigenstates in the position space. The lattice is assumed to be infinite and to have lattice parameter $a$, so the discrete positions are $(x_j,y_k)=(ja,ka)\to(j,k)\in\mathbb{Z}^2$. Notice that the discrete version of the Dirac delta is the Kronecker delta.}
\label{tab:spatial_disc_cont}
\end{table*}

\subsection{Discrete differential operators}
\label{subapp:disc_diff_op}
The next step is the discretization of the differential operators (first partial derivatives and Laplacian \footnote{The first partial derivatives are not involved by the Hamiltonian of a free particle but they are when inserting the magnetic field.}) by means of finite difference formulae \cite{abramowitz1970handbook}. If we consider a function $f(x)$ defined on a 1D lattice, whose sites are $x_j=ja$, and assume the lattice parameter $a$ to be small enough, then we can evaluate the following Taylor expansions up to the second order:
\begin{equation}
f(x_j\pm a)\approx f(x_j)\pm f'(x_j)a+\frac{1}{2}f''(x_j)a^2\,.
\end{equation}
Letting $f_j:=f(x_j)$, we have that:
\begin{enumerate}[(i)]
\item the difference of such quantities provides the first derivative of $f$ in the site $j$
\begin{equation}
f'_j \approx \frac{f_{j+1}-f_{j-1}}{2a}\,;
\label{eq:discrete_1st_f}
\end{equation}

\item the sum of such quantities provides the second derivative of $f$ in the site $j$
\begin{equation}
f''_j \approx \frac{f_{j+1}+f_{j-1}- 2 f_j}{a^2}\,.
\label{eq:discrete_2nd_f}
\end{equation}
\end{enumerate}

Notice that the discrete differential operator $\partial_x$ acts on a product of functions $f(x)g(x)$ as on $\left(fg\right)(x)$, i.e.
\begin{align}
\dv{(fg)_j}{x} &= \frac{1}{2a}\left[ (fg)_{j+1}-(fg)_{j-1}\right]\nonumber\\
&=\frac{1}{2a}\left( f_{j+1}g_{j+1}-f_{j-1}g_{j-1}\right)\,,
\label{eq:FD_prod}
\end{align}
since $\left( fg \right)(x)=f(x)g(x)$. 
\begin{proof}
This can be easily proved by considering the Taylor expansion of the product of such functions
\begin{equation}
\left( fg \right)(x\pm a)\approx\left( fg \right)(x)\pm\left( fg \right)'(x)a+\frac{1}{2}\left( fg \right)''(x)a^2\,,
\end{equation}
from which, as before,
\begin{equation}
\left( fg \right)'(x)\approx\frac{1}{2a}\left[\left( fg \right)(x+a)-\left( fg \right)(x-a)\right]\,.
\end{equation}
\end{proof}

Then, in 2D, the discrete versions of the gradient and the Laplacian are obtained by evaluating the partial derivatives according to Eqs. \eqref{eq:discrete_1st_f} and \eqref{eq:discrete_2nd_f}:
\begin{equation}
\nabla f_{j,k}=\frac{1}{2a}\left[\left( f_{j+1,k}-f_{j-1,k}\right)\mathbf{\hat{i}}+\left(f_{j,k+1}-f_{j,k-1}\right)\mathbf{\hat{j}}\right]
\label{eq:grad_2d_QW}
\end{equation}
and
\begin{equation}
\nabla^2 f_{j,k}=\frac{1}{a^2}\left( f_{j+1,k}+f_{j-1,k}+f_{j,k+1}+f_{j,k-1}- 4 f_{j,k}\right)\,,
\label{eq:lapl_2d_QW}
\end{equation}
where $f_{j,k}:=f(x_j,y_k)$ and $\mathbf{\hat{i}}$, $\mathbf{\hat{j}}$ denote the unit vectors of the $x$, $y$ axis, respectively.

\subsection{Linear momentum operator}
\label{subapp:lin_mom_op}
The linear momentum operator in the position space reads as follows:
\begin{equation}
\hat{\mathbf{p}}=-i\hbar \nabla\,.
\end{equation}
For sake of simplicity we consider a 1D space. If we recall the study of the linear momentum as the generator of infinitesimal translations \cite{sakurai2011san}, the action of $\hat{p}$ on a state $
\vert \psi \rangle = \int \diff x'\, \psi(x')\vert x' \rangle$, where $\psi(x')=\langle x'\vert\psi\rangle$, gives
\begin{equation}
\hat{p}\vert \psi \rangle = \int \diff x'\, \vert x' \rangle \left( -i\hbar\partial_{x'} \psi(x')\right)\,,
\end{equation}
or equivalently
\begin{equation}
\langle x'\vert\hat{p}\vert \psi \rangle =-i\hbar\partial_{x'} \psi(x')\,,
\label{eq:xppsi}
\end{equation}
from which, for the matrix element $\hat{p}$ in the $x$-representation, we obtain
\begin{equation}
\langle x' \vert \hat{p} \vert x'' \rangle=-i\hbar\partial_{x'} \delta(x'-x'')\,.
\end{equation}
By repeatedly applying Eq. \eqref{eq:xppsi}, we also have
\begin{equation}
\langle x' \vert \hat{p}^n \vert \psi \rangle=(-i\hbar)^n\partial_{x'}^n \psi(x')\,.
\end{equation}

Now we adapt this result to a discrete 1D space (see also Table \ref{tab:spatial_disc_cont}). The state $\ket{\psi}$ is expanded on the site states basis $\{\vert j \rangle\}_j$, the equivalent of position states, as $\ket{\psi}=\sum_j \psi_j \ket{j}$, where $\psi_j=\braket{j}{\psi}$. Then, by reading $\nabla$ as the \textit{discrete} differential operator acting on the wavefunction according to Eq. \eqref{eq:discrete_1st_f}, we have that
\begin{align}
\langle j \vert \hat{p} \vert \psi \rangle &= \frac{-i\hbar}{2a}\left(\psi_{j+1}-\psi_{j-1}\right)\nonumber\\
&=\frac{-i\hbar}{2a}\left(\langle j+1\vert-\langle j-1\vert\right)\vert \psi \rangle\,,
\end{align}
and so
\begin{align}
\hat{p} \vert \psi \rangle &=\frac{-i\hbar}{2a}\sum_j \vert j \rangle \left(\psi_{j+1}-\psi_{j-1}\right)\nonumber\\
&=\frac{-i\hbar}{2a}\sum_j \psi_j\left(\vert j-1\rangle-\vert j+1 \rangle\right)\nonumber\\
&=\left[\frac{-i\hbar}{2a}\sum_j \left(\vert j-1\rangle\langle j \vert-\vert j+1 \rangle\langle j \vert\right)\right]\vert \psi \rangle\,,
\label{eq:lin_mom_1}
\end{align}
where the second equality follows from rescaling the dummy index of summation, being the latter infinite. Analogously, by reading $\nabla^2$ as the \textit{discrete} Laplacian acting on the wavefunction according to Eq. \eqref{eq:discrete_2nd_f}, we have that
\begin{align}
\hat{p}^2 \vert \psi \rangle &= \frac{(-i\hbar)^2}{a^2}\sum_j\vert j \rangle\left(\psi_{j+1}+\psi_{j-1}-2\psi_j\right)\nonumber\\
&=\frac{(-i\hbar)^2}{a^2}\sum_j \psi_j\left(\vert j-1\rangle+\vert j+1 \rangle-2\vert j \rangle\right)\nonumber\\
&=\left[-\frac{\hbar^2}{a^2}\sum_j \left(\vert j-1\rangle\langle j \vert+\vert j+1 \rangle\langle j \vert-2\vert j \rangle\langle j \vert\right)\right]\vert \psi \rangle\,.
\label{eq:lin_mom_2}
\end{align}

The point we want to stress is that, as well as in the continuum, differential operators act on the wavefunctions, not on the kets. Indices of bras and kets are then accordingly rescaled after the differential operators acted on the wavefunction.

\subsection{Derivation of the discrete Hamiltonian: details}
\label{subapp:disc_H_details}
As seen in Sec. \ref{subapp:lin_mom_op}, the way to obtain the Hamiltonian in terms of projectors all in the form $\dyad{W}{V}$, where $V$ denotes an initial vertex and $W$ one of its NNs, comes through the rescaling of the indices of summation. Another way, more suitable when dealing with PLGs, is to exploit the hermiticity of the Hamiltonian, so that $\mathcal{H}_{WV}=\matrixel{W}{\hat{\mathcal{H}}}{V}=\matrixel{V}{\hat{\mathcal{H}}}{W}^\ast = \left(\mathcal{H}_{VW}\right )^\ast$. This allows us to write the Hamiltonian in terms of projectors in the form $\dyad{W}{V}$ knowing the matrix element describing the opposite hopping $\mathcal{H}_{VW}$. Here below we show further details in the derivation of the CTQW Hamiltonian from the spatial discretization of Eq. \eqref{eq:Ham_charged_ptc_cont} (the free-particle Hamiltonian is recovered for $B=0$). Notice that the term $\hat{\mathbf{p}}\cdot\mathbf{A}\psi_V=-i\hbar\left [\partial_x(A^xf_V)+\partial_y(A^yf_V)\right ]$ is to be computed in the spirit of Eq. \eqref{eq:FD_prod}. The outline to obtain the CTQW Hamiltonian is as follows:
\begin{enumerate}[(i)]
\item we expand the generic state $\vert \psi \rangle$ on the vertex states basis $\{\vert V \rangle\}$ as
\begin{equation}
\vert \psi \rangle = \sum_V \psi_V \vert V \rangle\,,
\end{equation}
where $\psi_V=\langle V \vert \psi \rangle$ is the wavefunction and the index of summation runs over all the vertices in the graph (infinite, in principle);
\item the differential operators act on $\psi_V$;
\item we exploit the hermiticiy of the Hamiltonian in order to write it in terms of projectors in the form $\ketbra{W}{V}$, where $V$ is the starting vertex and $W$ the final one.
\end{enumerate}

\begin{widetext}
\subsubsection{Square lattice graph}
With reference to Eqs. \eqref{eq:squ_lap_fd}, \eqref{eq:J_SQU_plg}, \eqref{eq:SQU_grad_x}, \eqref{eq:SQU_grad_y} and to Table \ref{tab:plg_nn_coords}, the CTQW Hamiltonian acts on a state $\ket{\psi}$ as follows:

\begin{align}
\mathcal{\hat{H}}\ket{\psi} &=-\frac{\hbar^2}{2ma^2}\sum_V\left\lbrace\vphantom{\frac{q^2}{\hbar^2}}\left( \psi_A +\psi_B +\psi_C +\psi_D-4\psi_V\right)-i\frac{qa}{2\hbar}\left[\left(A_V^x+A_A^x\right)\psi_A-\left(A_V^x+A_C^x\right)\psi_C\right]\right.\nonumber\\
&\quad\left.-i\frac{qa}{2\hbar}\left[\left(A_V^y+A_B^y\right)\psi_B-\left(A_V^y+A_D^y\right)\psi_D\right]-\frac{q^2a^2}{\hbar^2}\left({A_V^x}^2+{A_V^y}^2 \right)\psi_V\right\rbrace\ket{V}\nonumber\\
&=-J_S\sum_V\left\lbrace\vphantom{\frac{q^2}{\hbar^2}}\left[1-i\frac{qa}{2\hbar}\left(A_V^x+A_A^x\right)\right]\dyad{V}{A}+\left[1-i\frac{qa}{2\hbar}\left(A_V^y+A_B^y\right)\right]\dyad{V}{B}\right.\nonumber\\
&\quad+\left[1+i\frac{qa}{2\hbar}\left(A_V^x+A_C^x\right)\right]\dyad{V}{C}+\left[1+i\frac{qa}{2\hbar}\left(A_V^y+A_D^y\right)\right]\dyad{V}{D}\nonumber\\
&\left.\quad-\left[4+ \frac{q^2a^2}{\hbar^2}\left({A_V^x}^2+{A_V^y}^2\right)\right]\dyad{V}{V}\right\rbrace\ket{\psi}\nonumber\\
&=\sum_V\left( \mathcal{H}_{V\!A}\dyad{V}{A}+\mathcal{H}_{V\!B}\dyad{V}{B}+\mathcal{H}_{V\!C}\dyad{V}{C}+\mathcal{H}_{V\!D}\dyad{V}{D}+\mathcal{H}_{VV}\dyad{V}{V}\right)\ket{\psi}\,.
\label{eq:H_SQU_plg_psi}
\end{align}
Exploiting $\mathcal{H}^\dagger=\mathcal{H}$, Eq. \eqref{eq:H_SQU_plg_psi} can be recast into Eq. \eqref{eq:H_SQU_plg}.

\subsubsection{Triangular lattice graph}
With reference to Eqs. \eqref{eq:tri_lap_fd}, \eqref{eq:J_TRI_plg}, \eqref{eq:TRI_grad_x}, \eqref{eq:TRI_grad_y} and to Table \ref{tab:plg_nn_coords}, the CTQW Hamiltonian acts on a state $\ket{\psi}$ as follows:
\begin{align}
\mathcal{\hat{H}}\ket{\psi} &=-\frac{\hbar^2}{3ma^2}\sum_V\left\lbrace\vphantom{\frac{q^2}{\hbar^2}}(\psi_A +\psi_B +\psi_C +\psi_D +\psi_E+ \psi_F-6\psi_V)-i\frac{qa}{4\hbar}\left[2\left(A_V^x+A_A^x\right)\psi_A+\left(A_V^x+A_B^x\right)\psi_B\right.\right.\nonumber\\
&\quad\left.-\left(A_V^x+A_C^x\right)\psi_C-2\left(A_V^x+A_D^x\right)\psi_D-\left(A_V^x+A_E^x\right)\psi_E+\left(A_V^x+A_F^x\right)\psi_F\right]-i\frac{\sqrt{3}qa}{4\hbar}\left[\left(A_V^y+A_B^y\right)\psi_B\right.\nonumber\\
&\quad\left.\left.+ \left(A_V^y+A_C^y\right)\psi_C -\left(A_V^y+A_E^y\right)\psi_E-\left(A_V^y+A_F^y\right)\psi_F \right]-\frac{3q^2a^2}{2\hbar^2}\left({A_V^x}^2+{A_V^y}^2 \right)\psi_V\right\rbrace\ket{V}\nonumber\\
&=-J_T\sum_V\left\lbrace\vphantom{\frac{q^2}{\hbar^2}}\left[1-i\frac{qa}{2\hbar}\left(A_V^x+A_A^x\right)\right]\dyad{V}{A}+\left[1-i\frac{qa}{4\hbar}\left(A_V^x+A_B^x+\sqrt{3}(A_V^y+A_B^y)\right)\right]\dyad{V}{B}\right.\nonumber\\ 
&\quad+\left[1+i\frac{qa}{4\hbar}\left(A_V^x+A_C^x-\sqrt{3}(A_V^y+A_C^y)\right)\right]\dyad{V}{C}+\left[1+i\frac{qa}{2\hbar}\left(A_V^x+A_D^x\right)\right]\dyad{V}{D}\nonumber\\ 
&\quad+\left[1+i\frac{qa}{4\hbar}\left(A_V^x+A_E^x+\sqrt{3}(A_V^y+A_E^y)\right)\right]\dyad{V}{E}+\left[1-i\frac{qa}{4\hbar}\left(A_V^x+A_F^x-\sqrt{3}(A_V^y+A_F^y)\right)\right]\dyad{V}{F}\nonumber\\
&\quad\left.-\left[6+\frac{3q^2a^2}{2\hbar^2}\left({A_V^x}^2+{A_V^y}^2\right)\right]\dyad{V}{V}\right\rbrace\ket{\psi}\nonumber\\
&=\sum_V\left( \mathcal{H}_{V\!A}\dyad{V}{A}+\mathcal{H}_{V\!B}\dyad{V}{B}+\mathcal{H}_{V\!C}\dyad{V}{C}+\mathcal{H}_{V\!D}\dyad{V}{D}+\mathcal{H}_{V\!E}\dyad{V}{E}+\mathcal{H}_{V\!F}\dyad{V}{F}\right.\nonumber\\
&\quad\left.+\mathcal{H}_{VV}\dyad{V}{V}\right)\ket{\psi}\,.
\label{eq:H_TRI_plg_psi}
\end{align}
Exploiting $\mathcal{H}^\dagger=\mathcal{H}$, Eq. \eqref{eq:H_TRI_plg_psi} can be recast into Eq. \eqref{eq:H_TRI_plg}.

\subsubsection{Honeycomb lattice graph}
With reference to Eqs. \eqref{eq:hon_h_lap_fd}, \eqref{eq:J_HON_plg}, \eqref{eq:HON_grad_x}, \eqref{eq:HON_grad_y} and to Table \ref{tab:plg_nn_coords}, the CTQW Hamiltonian acts on a state 
\begin{equation}
\ket{\psi}= \sum_{\odot \in \{\circ,\bullet\}}\sum_{(V,\odot)}\psi_{(V,\odot)} \ket{V,\odot}\,, \text{ where } \, \psi_{(V,\odot)}=\braket{V,\odot}{\psi}\,,
\end{equation}
as follows:
\begin{align}
\mathcal{\hat{H}}\ket{\psi} &=-\frac{2\hbar^2}{3ma^2}\sum_{\odot \in \{\circ,\bullet\}}\sum_{(V,\odot)}\left\lbrace\vphantom{\frac{q^2}{\hbar^2}}\left (\psi_{(A,\bar{\odot})}+\psi_{(B,\bar{\odot})}+\psi_{(C,\bar{\odot})}-3\psi_{(V,\odot)}\right )-i\frac{\sqrt{3}qa}{4\hbar}\left[ \left(A_{(V,\odot)}^x+A_{(A,\bar{\odot})}^x\right)\psi_{(A,\bar{\odot})}\right.\right.\nonumber\\
&\quad \left.-\left(A_{(V,\odot)}^x+A_{(B,\bar{\odot})}^x\right)\psi_{(B,\bar{\odot})}\right]-\operatorname{sgn}(\odot)i\frac{qa}{4\hbar}\left[ 2\left(A_{(V,\odot)}^y+A_{(C,\bar{\odot})}^y\right)\psi_{(C,\bar{\odot})}-\left(A_{(V,\odot)}^y+A_{(A,\bar{\odot})}^y\right)\psi_{(A,\bar{\odot})}\right.\nonumber\\
&\quad \left.\left.-\left(A_{(V,\odot)}^y+A_{(B,\bar{\odot})}^y\right)\psi_{(B,\bar{\odot})}\right]-\frac{3q^2a^2}{4\hbar^2}\left ( {A^x}_{(V,\odot)}^2+{A^y}_{(V,\odot)}^2\right )\psi_{(V,\odot)}\right\rbrace\ket{V,\odot}\nonumber\\
&=-J_H\sum_{\odot \in \{\circ,\bullet\}}\sum_{(V,\odot)}\left\lbrace\vphantom{\frac{q^2}{\hbar^2}}\left[ 1-i\frac{qa}{4\hbar}\left( \sqrt{3}\left(A_{(V,\odot)}^x+A_{(A,\bar{\odot})}^x\right) - \operatorname{sgn}(\odot)\left(A_{(V,\odot)}^y+A_{(A,\bar{\odot})}^y\right)\right)\right]\dyad{V,\odot}{A,\bar{\odot}}\right.\nonumber\\
&\quad+\left[ 1+i\frac{qa}{4\hbar}\left( \sqrt{3}\left(A_{(V,\odot)}^x+A_{(B,\bar{\odot})}^x\right)+\operatorname{sgn}(\odot)\left(A_{(V,\odot)}^y+A_{(B,\bar{\odot})}^y\right)\right)\right]\dyad{V,\odot}{B,\bar{\odot}}\nonumber\\
&\quad+\left[ 1- \operatorname{sgn}(\odot)i\frac{qa}{2\hbar}\left( A_{(V,\odot)}^y+A_{(C,\bar{\odot})}^y\right) \right]\dyad{V,\odot}{C,\bar{\odot}}\nonumber\\
&\left.\quad-\left[3+ \frac{3q^2a^2}{4\hbar^2}\left({A^x}_{(V,\odot)}^2+{A^y}_{(V,\odot)}^2\right)\right]\dyad{V,\odot}\right\rbrace \ket{\psi}\nonumber\\
&=\sum_{\odot \in \{\circ,\bullet\}}\sum_{(V,\odot)}\left( \mathcal{H}_{V\!A}\dyad{V,\odot}{A,\bar{\odot}}+\mathcal{H}_{V\!B}\dyad{V,\odot}{B,\bar{\odot}}+\mathcal{H}_{V\!C}\dyad{V,\odot}{C,\bar{\odot}}+\mathcal{H}_{VV}\dyad{V,\odot}{V,\odot}\right)\ket{\psi}\,.
\label{eq:H_HON_plg_psi}
\end{align}
Exploiting $\mathcal{H}^\dagger=\mathcal{H}$, Eq. \eqref{eq:H_HON_plg_psi} can be recast into Eq. \eqref{eq:H_HON_plg}, since $\mathcal{H}_{WV}=-J_H h_{WV}$.\\
\end{widetext}

\section{Computational details}
\label{app:comp_detail}

\subsection{Plotting the maps}
\label{subapp:maps}
\begin{figure}[!h]
	\centering
	\includegraphics[width=0.3\textwidth]{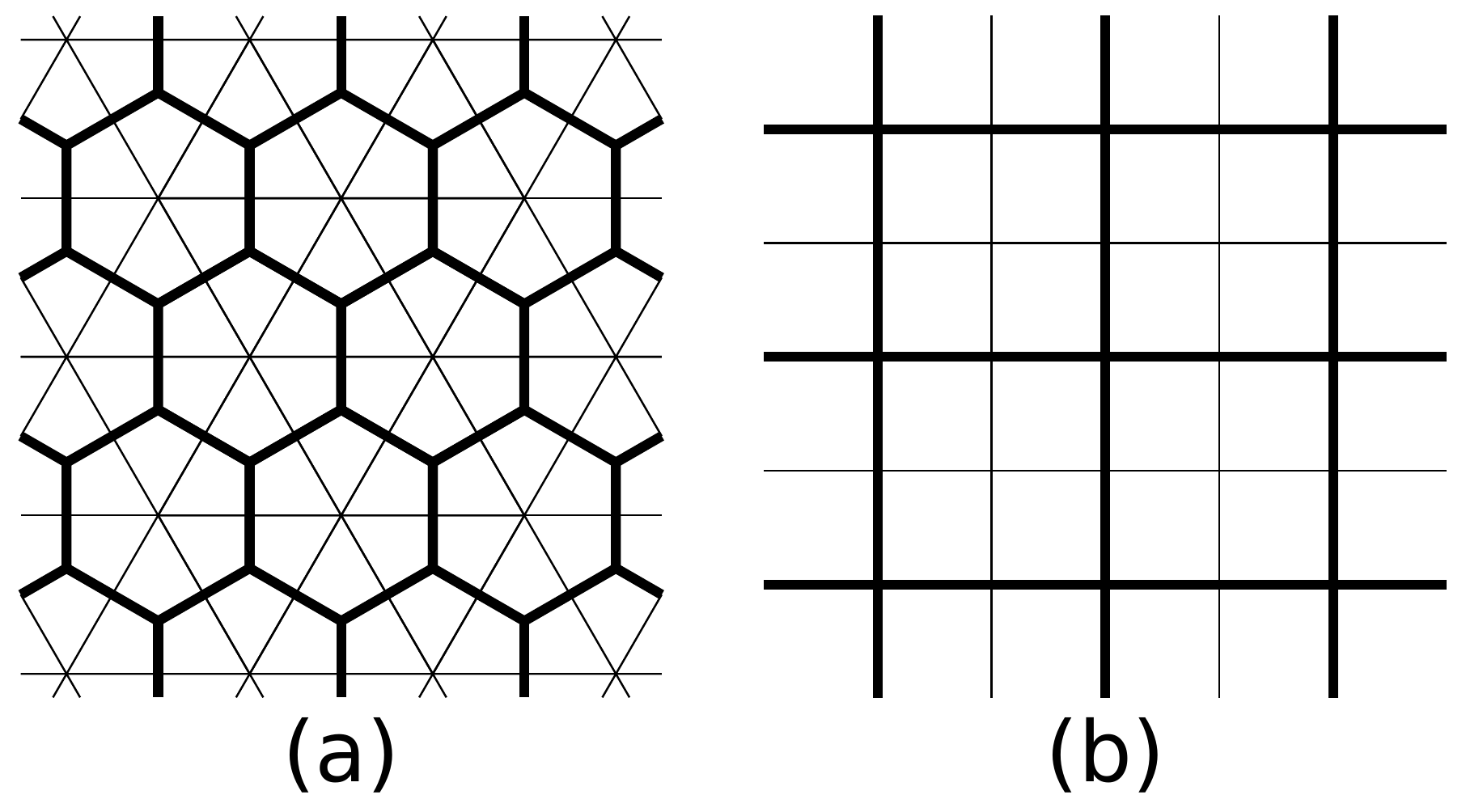}
	\caption{Duals of the regular tessellations of the Euclidean plane: (a) $\left\lbrace 6,3 \right\rbrace \stackrel{dual}{\longleftrightarrow}\left\lbrace 3,6 \right\rbrace$, (b) $\left\lbrace 4,4 \right\rbrace \stackrel{dual}{\longleftrightarrow}\left\lbrace 4,4 \right\rbrace$. See also Fig. \ref{fig:tessel_plg}.}
\label{fig:tessellation_dual}
\end{figure}

\begin{figure}[!h]
	\centering
	\includegraphics[width=0.45\textwidth]{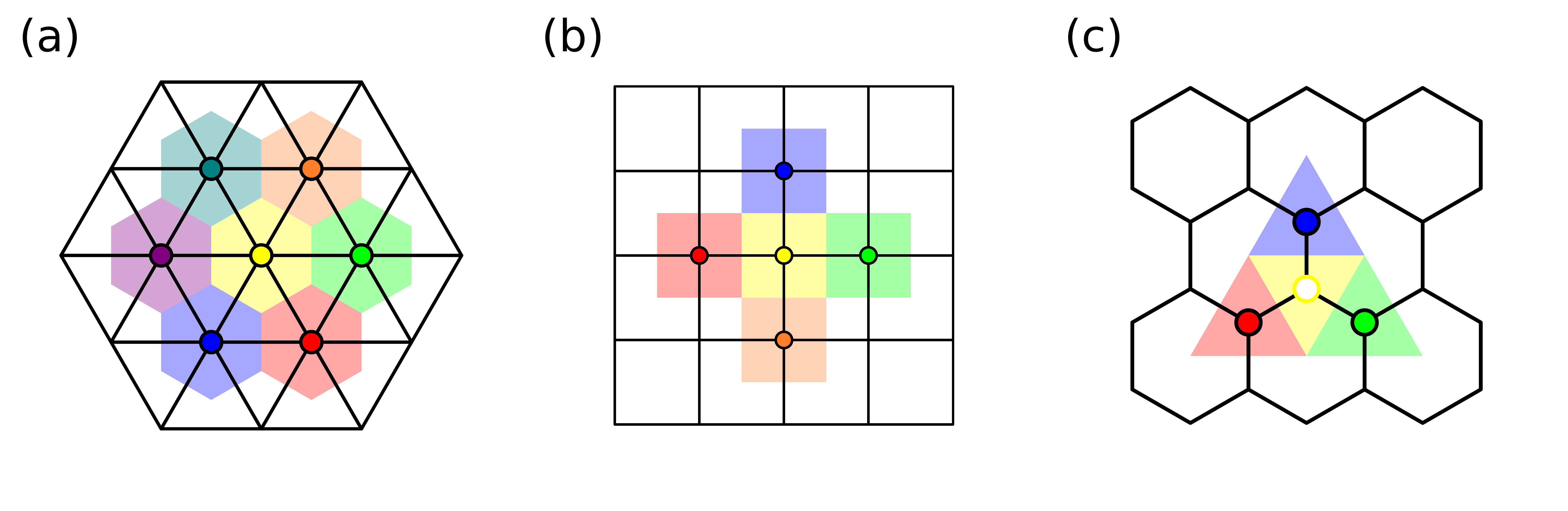}
	\caption{The expedient adopted to represent the maps of the populations of the vertices for the (a) triangular, (b) square, and (c) honeycomb lattice graph: the value of the population in a vertex (colored circle) of the lattice graph is assigned to the corresponding patch of its dual lattice. In doing so, the degree of the vertex is recovered.}
\label{fig:maps_legend}
\end{figure}

Here we report how we plot the maps representing the time evolution of the probability density (population of the vertices). The idea is to assign to each vertex a patch colored according to the corresponding value of the population. The patch must have a shape reproducing the degree of the vertex, so that the adjacent patches really represent its NNs. The \textit{dual} of $\left\lbrace p, q \right\rbrace$ is the tessellation whose edges are the perpendicular bisectors of the edges of $\left\lbrace p, q \right\rbrace$ (Fig. \ref{fig:tessellation_dual}). Thus the dual of $\left\lbrace p, q \right\rbrace$ is $\left\lbrace q, p \right\rbrace$, and \textit{vice versa}; the vertices of either are the centers of the faces of the other \cite{coxeter1969introduction}. Adopting the dual of the tessellation of interest provides patches exactly meeting our needs. So, the proper way of representing and interpreting the maps of the populations of the vertices is shown in Fig. \ref{fig:maps_legend}. Be aware that a map made of hexagonal patches refers to triangular lattice graph, whereas a map made of triangular patches to honeycomb lattice graph, being one the dual of the other; instead, a map made of square patches refers to square lattice graph, since the dual of $\left\lbrace 4, 4 \right\rbrace$ is an equal $\left\lbrace 4, 4 \right\rbrace$.

\subsection{Indexing and coordinates of vertices}
\label{subapp:index_coords}

\subsubsection{Indexing}
\label{subsubapp:index}
\begin{figure}[!h]
	\centering
	\includegraphics[width=0.45\textwidth]{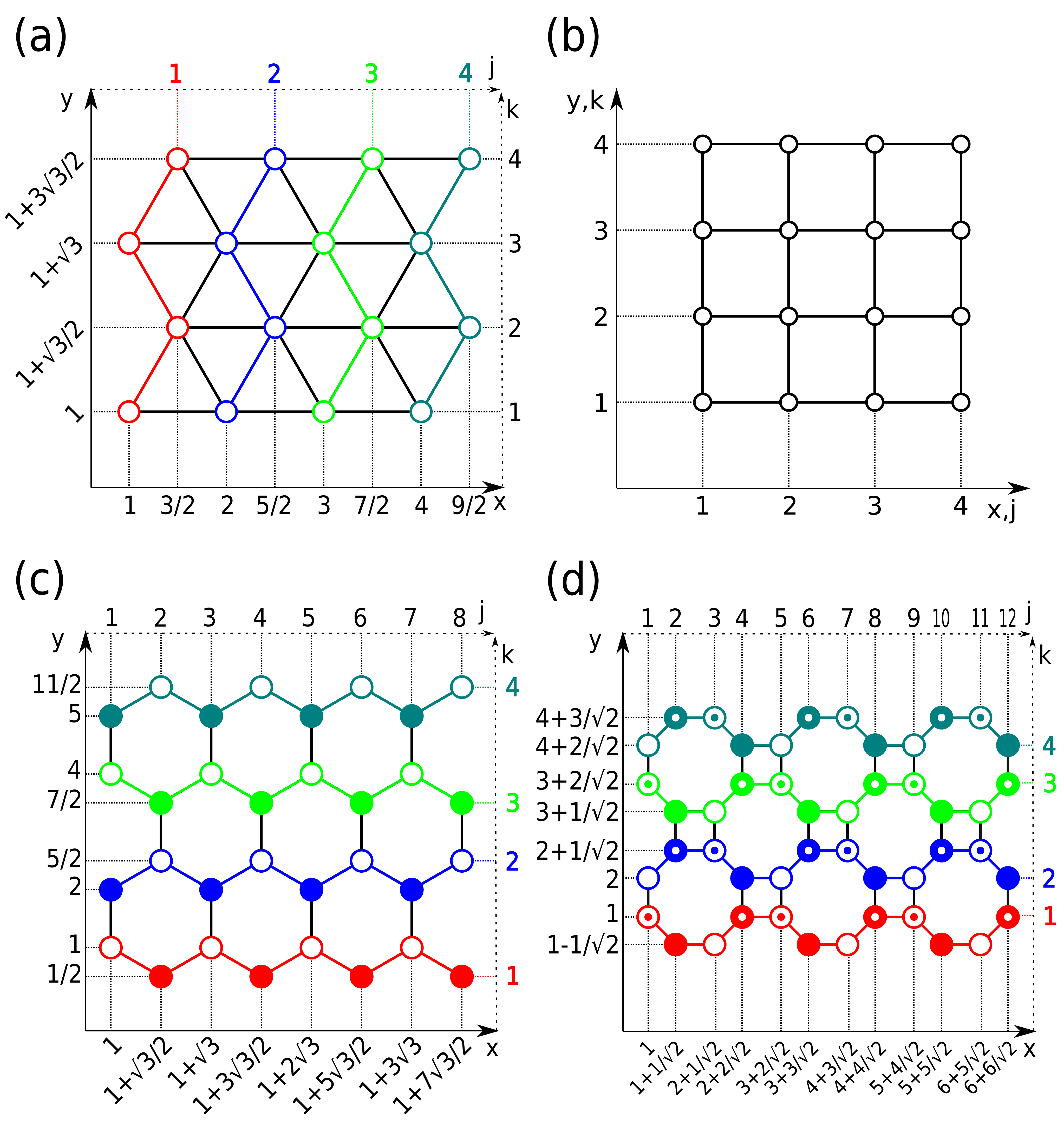}
	\caption{Labeling of vertices with a couple of indices $(j,k)$ and corresponding $(x,y)$ coordinates for the (a) triangular, (b) square, (c) honeycomb, and (d) truncated square lattice graph. Lattice parameter $a=1$.}
\label{fig:labeling_plg}
\end{figure}

When numerically dealing with PLGs (in particular the non-square ones), the first issue is how to label each vertex of the graph with a couple of indices $(j,k)$, with $j=1,\ldots,N_j$ and $k=1,\ldots,N_k$ (the lattice graph is finite). We adopt the indexing shown in Fig. \ref{fig:labeling_plg}. While for the square lattice graph the $x$ and $y$ directions provide the natural framework in which defining the couple of indices $(j,k)$, for the non-square PLGs we have to define \textit{polylines} of vertices referring to the same $x$-index, denoted as $j$, or to the same $y$-index, denoted as $k$. So, we denote by $N_j$ and $N_k$ the total number of vertices along the $j$- and $k$-polyline, respectively. Notice, when implementing the system, that the NNs of a vertex may be differently labeled depending on the location of such vertex.\\

\paragraph*{Triangular.} In a triangular lattice graph, according to our indexing (Fig. \ref{fig:labeling_plg}(a)), we have to distinguish between \textit{even} ($k \bmod 2=0$) and \textit{odd} ($k \bmod 2=1$) $y$-indices, $k$. Consider, e.g., the vertex $V=(j,k)=(2,1)$: if we move one step along the unit vector $(1/2, \sqrt{3}/2)$ we reach the vertex $(2,2)=(j,k+1)$, but if we do the same starting from $V'=(j',k')=(2,2)$ we reach the vertex $(3,3)=(j'+1,k'+1)$, not $(2,3)=(j',k'+1)$.\\

\paragraph*{Square.} In a square lattice graph (Fig. \ref{fig:labeling_plg}(b)) the coordinates of a vertex are integer multiples of the lattice parameter $a=1$, so they provide the indices $(x_j,y_k)=(j,k)$.\\

\paragraph*{Honeycomb.} In a honeycomb lattice graph there are two classes of non-equivalent vertices $\{\circ,\bullet\}$. According to our indexing (Fig. \ref{fig:labeling_plg}(c)), a vertex $V=(j,k)$ belongs to either class according to the following rule:
\begin{equation}
V=
\begin{cases}
(V,\circ) & \text{if } (j+k)\bmod 2=0\,,\\
(V,\bullet) & \text{if } (j+k)\bmod 2=1\,.
\end{cases}
\end{equation}

\paragraph*{Truncated square.} In a truncated square lattice graph there are four classes of non-equivalent vertices $\{\whitecirc,\blackcirc,\whitecircdot,\blackcircdot\}$. According to our indexing (Fig. \ref{fig:labeling_plg}(d)), a vertex $V=(j,k)$ belongs to one of the different classes according to the following rule:
\begin{equation}
V=
\begin{cases}
(V,\whitecircdot) & \text{if } (k\bmod2=1 \wedge j\bmod4=1)\\
&\vee (k\bmod2=0 \wedge j\bmod4=3)\,,\\
(V,\blackcirc) & \text{if } (k\bmod2=1 \wedge j\bmod4=2)\\
&\vee (k\bmod2=0 \wedge j\bmod4=0)\,,\\
(V,\whitecirc) & \text{if } (k\bmod2=1 \wedge j\bmod4=3)\\
&\vee (k\bmod2=0 \wedge j\bmod4=1)\,,\\
(V,\blackcircdot) & \text{if } (k\bmod2=1 \wedge j\bmod4=0)\\
&\vee (k\bmod2=0 \wedge j\bmod4=2)\,.
\end{cases}
\end{equation}
Notice that being this lattice graph symmetric under rotation of $\pi/2$, we may adopt the same indexing along the $y$ axis as well as along the $x$ axis. However, this would generate a virtual logically rectangular grid where we should discard the vertices not corresponding to the actual ones: e.g., we might pick up a virtual vertex inside the octagon which actually does not exist. Adopting an indexing analogous to that of the honeycomb ensures that any couple of indices $(j,k)$ is associated to an actual vertex.\\

In conclusion, after labeling each vertex with a couple of indices $(j,k)$ (which also label the vertex state $\ket{j,k}$), from the computational point of view it is worth indexing vertices with a single index $l$. This is accomplished, for instance, as follows:
\begin{equation}
(j,k) \longrightarrow l=N_k (j-1)+k\,,
\end{equation}
where $N_k$ denotes the number of vertices along the $k$-polyline, which plays the role of the $y$ axis.

\subsubsection{Coordinates}
\label{subsubapp:coords}
Here we show how to restore the coordinates $(x_V,y_V)$ of a vertex $V$ given its indices $(j_V,k_V)\in[1,N_j]\times[1,N_k]$ (lattice parameter $a=1$).\\

\paragraph*{Triangular.} In a triangular lattice graph, the coordinate $y_V$ is an integer multiple of $\sqrt{3}/2$, whereas the coordinate $x_V$ is integer or half-integer depending on the parity of the index $k_V$:
\begin{algorithmic}[1]
\setstretch{1.35}
\State $x_V=j_V+(1-\operatorname{mod}(k_V,2))/2$;
\State $y_V=\sqrt{3}\ast k_V/2$;
\end{algorithmic}
\
\paragraph*{Square.} In a square lattice graph, the coordinates $x_V$ and $y_V$ of a vertex coincide with the indices $j_V$ and $k_V$, respectively:
\begin{algorithmic}[1]
\State $x_V=j_V$;
\State $y_V=k_V$;
\end{algorithmic} 
\
\paragraph*{Honeycomb.} In a honeycomb lattice graph, the coordinate $x_V$ of a vertex is an integer multiple of $\sqrt{3}/2$, whereas the coordinate $y_V$, with respect to the index $k_V$, requires a correction depending on the parity of both the indices $(j_V,k_V)$ and a shift by $\Delta=\Delta(k_V)$:
\begin{algorithmic}[1]
\State $x_V=\sqrt{3}\ast j_V/2$;
\State $\Delta=\operatorname{floor}((k_V-1)/2)$;
\State $y_V=k_V+\Delta+(1-\operatorname{mod}(j_V,2))\ast(1/2-\operatorname{mod}(k_V,2))$;
\end{algorithmic}
where the expression $(1-\operatorname{mod}(j_V,2))\ast(1/2-\operatorname{mod}(k_V,2))$ adjusts the value $k_V+\Delta$ by $0$ or $\pm1/2$ according to the parity of the indices.
\\
\paragraph*{Truncated square.} In a truncated square lattice graph, along the $x$ and $y$ axes the coordinate increases by $1$ or $1/\sqrt{2}$. Because of the indexing analogous to the honeycomb lattice graph, the coordinate $x_V=x_V(j_V)$, whereas the coordinate $y_V$, with respect to the index $k_V$, requires a correction depending on the parity of both the indices $(j_V,k_V)$ and a shift by $\Delta=\Delta(k_V)$:
\begin{algorithmic}[1]
\State $x_V=\operatorname{floor}((j_V+1)/2)+\operatorname{floor}(j_V/2) /\sqrt{2}$;
\State $\Delta=\sqrt{2}\ast\operatorname{floor}((k_V-1)/2)$;
\State $y_V=k_V+\Delta+\sqrt{2}(1/2-\operatorname{mod}(k_V,2))\ast(\operatorname{mod}(j_V,2)\neq \operatorname{mod}(j_V,4))$;
\end{algorithmic}
where the expression $\operatorname{mod}(j_V,2)\neq \operatorname{mod}(j_V,4)$ is to be understood as the (logical) value $1$ if true and $0$ if false.

\subsection{Boundary conditions}
\begin{figure}[!h]
	\centering
	\includegraphics[width=0.45\textwidth]{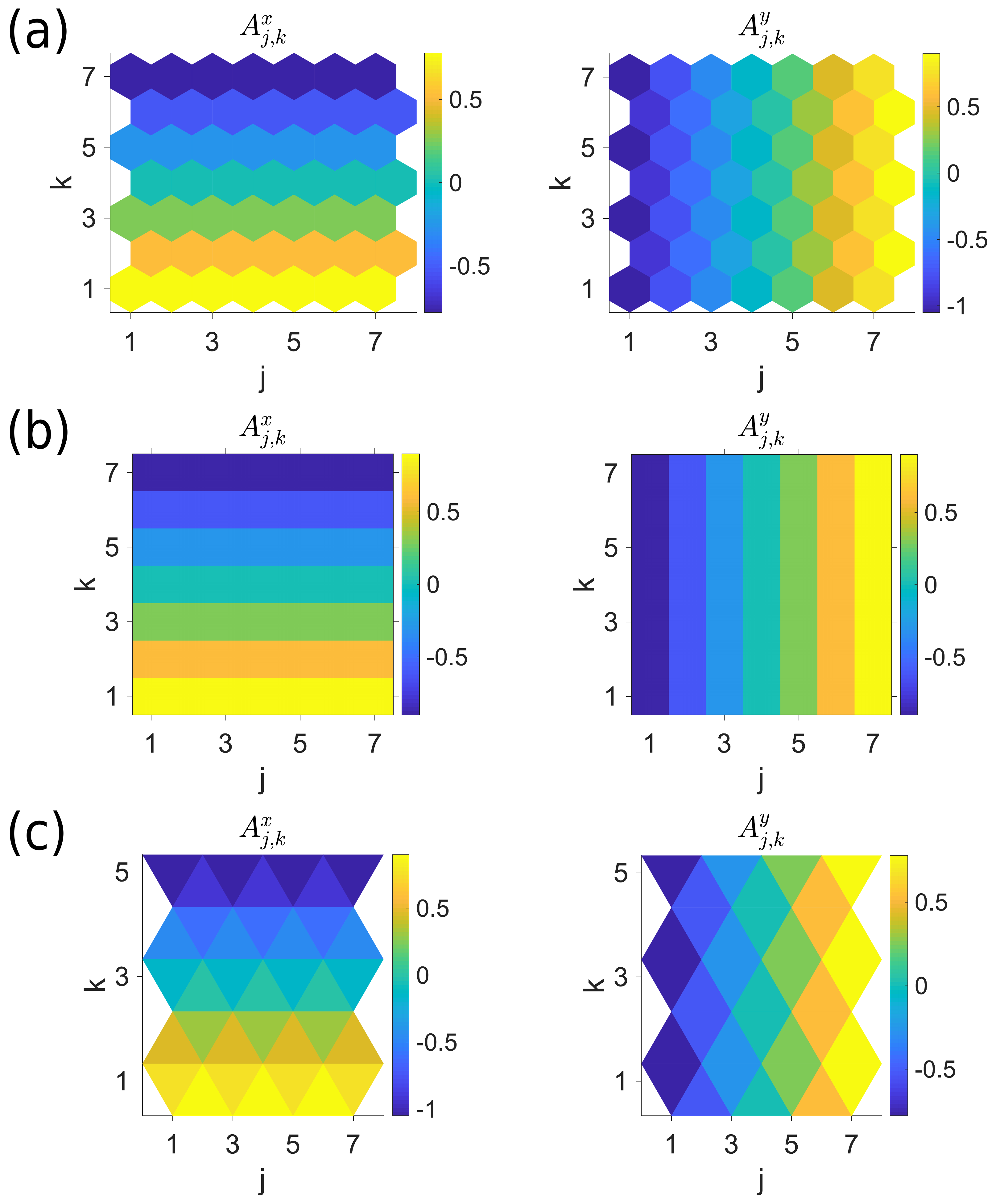}
	\caption{The components $A^x$ and $A^y$ of the vector potential (symmetric gauge, $B=0.6$) computed in a (a) $7\times 7$ triangular, (b) $7\times 7$ square, and (c) $7\times 5$ honeycomb lattice graph. The vector potential is centered in $(4,4)$ in the triangular and square lattice graph, in $(4,3)$ in the honeycomb one.	Such components show a discontinuity at the boundaries. So, since the hopping terms in the Hamiltonian may involve both the components, periodic boundary conditions are not appropriate. However, since in the square lattice graph the hopping along the $x$($y$) direction only involves $A^{x(y)}$, periodic boundary conditions can be assumed.}
\label{fig:A_PBC}
\end{figure}

When dealing with the magnetic field, since the hopping terms depend on the vector potential evaluated in both the initial and final vertex, boundary conditions may raise some issues. In Fig. \ref{fig:A_PBC} the components $A^x$ and $A^y$ of the vector potential (symmetric gauge) computed in the different PLGs are reported and they show a discontinuity at the boundaries. Since the hopping terms in the Hamiltonian may involve both the components, e.g. in the triangular (see Eqs. \eqref{eq:H_TRI_plg_peierls} and \eqref{eq:H_TRI_plg}) and honeycomb (see Eqs. \eqref{eq:H_HON_plg_peierls} and \eqref{eq:H_HON_plg}) lattice graph, periodic boundary conditions are not appropriate. On the other hand, in the square lattice graph the hopping occurs along the orthogonal directions $x$ and $y$ and it respectively only involves $A^x$ and $A^y$ (see Eqs. \eqref{eq:H_SQU_plg_peierls} and \eqref{eq:H_SQU_plg}). Thus, in this case, periodic boundary condition can be assumed.
\label{subapp:bc}

\section{Units}
\label{app:units}
The CTQW Hamiltonian of the system has some characteristic parameters, such as the electric charge $q$, the mass $m$ of the particle, and the lattice parameter $a$. In order to perform numerical simulations we have to declare them. We consider it appropriate to design a computation whose character is as general as possible. To this end we set the lattice parameter, the reduced Planck's constant and the elementary electric charge equal to 1, i.e. $a=\hbar=e=1$, and so these quantities are adimensional in the resulting system of units, which we refer to as \textit{QW units}. The physical quantities we treat in the present work derive from the fundamental ones (in the SI): mass (M), length (L), time (T), and electric current (I). Setting $a=\hbar=e=1$ means that:
\begin{enumerate}[(i)]
\item length is measured in units of $a$ (lattice parameter);
\item angular momentum (its modulus $\vert \vb{L}\vert$) is measured in units of $\hbar$ (reduced Planck's constant);
\item electric charge $q$ is measured in units of $e$ (elementary electric charge).
\end{enumerate}
Let $X$ be a physical quantity. Then its dimensions read as follows:
\begin{align}
\left[ X \right] &= \overbrace{M^\alpha L^\beta T^\gamma I^\delta}^{\textit{SI base quantities}} \nonumber \\
&=\underbrace{M^A L^B \left[ \vert \vb{L}\vert \right]^C \left[ q \right]^D}_{\textit{QW base quantities}} = M^{A+C} L^{B+2C} T^{-C+D} I^D\,,
\end{align}
since $\left[ \vert \vb{L}\vert \right] = M L^2 T^{-1}$ and $\left[ q \right]= I T$. This means that
\renewcommand\arraystretch{1.5}
\begin{eqnarray}
\left\lbrace
\begin{array}{lll}
\alpha &=& A+C\,,\\
\beta &=& B+2C\,,\\
\gamma &=& -C+D\,,\\
\delta &=& D\,,
\end{array}
\right.
& \Rightarrow &
\left\lbrace
\begin{array}{lll}
A &=& \alpha + \gamma - \delta\,,\\
B &=& \beta + 2\gamma - 2\delta\,,\\
C &=& -\gamma + \delta\,,\\
D &=& \delta\,,
\end{array}
\right.
\end{eqnarray}
from which
\begin{equation}
\left[ X \right] = M^{\alpha + \gamma - \delta} L^{\beta + 2\gamma - 2\delta} \left[ \vert \vb{L}\vert \right]^{-\gamma + \delta} \left[ q \right]^\delta\,. 
\end{equation}
Then we have
\begin{equation}
\left[ X \right]\vert_{QW} = M^{\alpha + \gamma - \delta}\,,
\end{equation}
since $a \vert_{QW} =  \hbar \vert_{QW}= e \vert_{QW}=1$, and so the relation between QW units and the SI ones is:
\begin{equation}
 X \vert_{QW} = x \vert_{QW}\, kg^{\alpha + \gamma - \delta} = x \, kg^\alpha \, m^\beta \, s^\gamma \, A^\delta\,,
\end{equation}
where $x=x \vert_{QW} \,a^{\beta + 2\gamma - 2\delta}\, \hbar^{-\gamma + \delta} \,e^\delta$. In the present work the relevant quantities are:
\begin{enumerate}[(i)]
\item space coordinates, for which $\alpha=\gamma=\delta=0$, $\beta=1$;
\item time, for which $\alpha=\beta=\delta=0$, $\gamma=1$;
\item modulus of magnetic field, for which $\alpha=1$, $\beta=0$, $\gamma=-2$, $\delta=-1$;
\item hopping amplitude (see, e.g., Eq. \eqref{eq:J_SQU_plg}), for which $\alpha=1$, $\beta=2$, $\gamma=-2$, $\delta=0$.
\end{enumerate}

\begin{table*}[t]
\center
\begin{ruledtabular}
\begin{tabular}{ccccc@{\hspace{2em}}cccc}
\renewcommand\arraystretch{1.75}
\multirow{2}{*}{\textbf{Base quantity}}& \multicolumn{3}{c}{\textbf{Dimensions}} & &\multirow{2}{*}{\textbf{Derived quantity}}& \multicolumn{3}{c}{\textbf{Dimensions}} \\\cline{2-4}\cline{7-9}
 & \textbf{SI} & \textbf{SI$\,\to\,$QW}	& \textbf{QW} && & \textbf{SI} & \textbf{SI$\,\to\,$QW}	& \textbf{QW} \\\cline{1-4}\cline{6-9}
mass & $M$ & $M$ & $M$ && time & $T$ & $M L^2 \left[\vert \vb{L} \vert\right]^{-1}$ & $M$ \\
length & $L$ & $L$ & adim.	&& magnetic field & $M T^{-2}I^{-1}$ & $ L^{-2} \left[\vert \vb{L} \vert\right] \left[q\right]^{-1}$ & adim. \\
angular momentum & $M L^2 T^{-1}$ & $\left[\vert \vb{L} \vert\right]$ & adim. && hopping amplitude & $M L^{2}T^{-2}$ & $M^{-1} L^{-2} \left[\vert \vb{L} \vert\right]^{2}$ & $M^{-1}$\\
electric charge	& $IT$ & $\left[q\right]$ & adim.
\end{tabular}
\end{ruledtabular}
\caption{Dimensional analysis of the QW base quantities and the derived quantities in different systems of units: in the International System of Units (SI), after redefining the base quantities (SI$\,\to\,$QW), and in the QW system of units (QW), for which $a=\hbar=e=1$.}
\label{tab:QW_quantities}
\end{table*}

The definition of this system of units is consistent with the SI. In particular, if we consider the four fundamental quantities - length, time, mass and electric current -, length is redefined accordingly to $a=1$, whereas time and electric current to $\hbar=e=1$. In this way the dimensions of all the physical quantities are expressed in terms of mass (hence $J$), which becomes the only characteristic physical quantity of the system (see Table \ref{tab:QW_quantities}).

\newpage
\bibliography{qwplgbib}

\end{document}